\documentclass[printer]{tADP2e}
\usepackage{color}

\def\st[#1]{| #1 \rangle}

\def\ffrac#1#2{\textstyle{#1\over#2}\displaystyle}

\newcommand{\e}{\mathrm{e}}
\newcommand{\etal}{{\it et al\/}\ }

\def\Ladder{
\unitlength=0.50mm
\begin{picture}(160.,50.)
\put(0,0){\line(1,0){150.}}
\put(0,30){\line(1,0){150.}}
\put(155.,0.){\makebox(0.,0.){$S$}}
\put(155.,30.){\makebox(0.,0.){$T$}}
\put(-7.,0.){\makebox(0.,0.){$g_s$}}
\put(-7.,30.){\makebox(0.,0.){$g_t$}}
\put(30,0){\line(0,1){30.}}
\put(60,0){\line(0,1){30.}}
\put(90,0){\line(0,1){30.}}
\put(120,0){\line(0,1){30.}}
\put(30.,0.){\makebox(0.,0.){$\bullet$}}
\put(30.,30.){\makebox(0.,0.){$\bullet$}}
\put(60.,0.){\makebox(0.,0.){$\bullet$}}
\put(60.,30.){\makebox(0.,0.){$\bullet$}}
\put(90.,0.){\makebox(0.,0.){$\bullet$}}
\put(90.,30.){\makebox(0.,0.){$\bullet$}}
\put(120.,0.){\makebox(0.,0.){$\bullet$}}
\put(120.,30.){\makebox(0.,0.){$\bullet$}}
\put(63.,14.){\makebox(0.,0.){$J_{\perp}$}}
\put(74.,25.){\makebox(0.,0.){$J_{\parallel}$}}
\put(74.,2.){\makebox(0.,0.){$J_{\parallel}$}}
\put(84.,14.){\makebox(0.,0.){$J_{\perp}$}}
\end{picture}}

\begin{document}
\doi{10.1080/00018730xxxxxxxxxxxx}
 \issn{1460-6976}
\issnp{0001-8732} \jvol{00} \jnum{00} \jyear{2007} \jmonth{February}

\markboth{M.T. Batchelor, X.-W. Guan, N. Oelkers and Z. Tsuboi}
{Integrable quantum spin ladders: comparison between theory and experiment}

\title{Integrable models and quantum spin ladders:\\ 
comparison between theory and experiment\\ for the strong coupling ladder compounds}

\author{M.T. BATCHELOR$^{1\dagger}$\footnote{$\dagger$ Murray.Batchelor@anu.edu.au}, 
X.W. GUAN$^{1\ddagger}$\footnote{$\ddagger$ xwe105@rsphysse.anu.edu.au}, 
N. OELKERS$^{2\S}$\footnote{$\S$ oelkers@maths.uq.edu.au}, 
Z. TSUBOI$^{3\P}$\footnote{$\P$ zengo\_tsuboi@pref.okayama.jp}
\thanks{\vspace{6pt}\newline\centerline{\tiny{ {\em Advances in Physics} ISSN 0001-8732 print/ ISSN 1460-6976 online
\textcopyright 2005 Taylor \& Francis Ltd}}
\newline\centerline{\tiny{ http://www.tandf.co.uk/journals}}
\newline \centerline{\tiny{DOI:10.1080/00018730xxxxxxxxxxxx}}}
\newline $^1$ Department of Theoretical Physics, Research School of Physical
Sciences \& Engineering and Mathematical Sciences Institute,
Australian National University, Canberra ACT 0200,  Australia
\newline $^2$ Centre for Mathematical Physics, 
University of Queensland, Brisbane QLD 4072, Australia
\newline $^3$ Okayama Institute for Quantum Physics, 
1-9-1 Kyoyama, Okayama City 700-0015, Japan}
\received{February 2007}

\maketitle

\begin{abstract}
This article considers recent advances in the investigation of the thermal and magnetic properties of
integrable spin ladder models and their applicability to the physics of strong coupling ladder compounds. 
For this class of compounds the rung coupling $J_\perp$ is much stronger than the 
coupling $J_\parallel$ along the ladder legs.
The ground state properties of the integrable two-leg spin-$\frac{1}{2}$ and the mixed 
spin-($\frac{1}{2},1$) ladder models at zero temperature are analysed by means of the
Thermodynamic Bethe Ansatz (TBA).  
Solving the TBA equations yields exact results for the critical fields and critical behaviour.
The thermal and magnetic properties of the models are discussed in terms of the 
recently introduced High Temperature Expansion (HTE) method, which is reviewed in detail.
In the strong coupling region the integrable spin-$\frac{1}{2}$ ladder model  
exhibits three quantum phases: 
(i) a gapped phase in the regime $H<H_{c1}=J_{\perp}-4J_{\parallel}$, 
(ii) a fully polarized phase for $H>H_{c2}=J_{\perp}+4J_{\parallel}$, and 
(iii) a Luttinger liquid magnetic phase in the regime $H_{c1}<H<H_{c2}$. 
The critical behaviour in the vicinity of the critical points $H_{c1}$ and $H_{c2}$ is of Pokrovsky-Talapov type. 
The temperature-dependent thermal and magnetic properties are directly evaluated from the 
exact free energy expression and compared to known
experimental results for the strong coupling ladder compounds 
(5IAP)$_2$CuBr$_4\cdot 2$H$_2$O,
Cu$_{2}$(C$_5$H$_{12}$N$_2$)$_2$Cl$_4$, (C$_5$H$_{12}$N)$_2$CuBr$_4$, 
BIP-BNO and [Cu$_2$(C$_2$O$_2$)(C$_{10}$H$_8$N$_2$)$_2$)](NO$_3$)$_2$.
Similar analysis of the mixed spin-($\frac{1}{2},1$) ladder model reveals a
rich phase diagram, with a $\frac13$ and a full saturation magnetization plateau within
the strong antiferromagnetic rung coupling regime. 
For weak rung coupling, the fractional magnetization plateau is diminished and a
new quantum phase transition occurs. 
The phase diagram can be directly deduced from the magnetization curve obtained from
the exact result derived from the TBA and HTE. 
The results are applied to the mixed ferrimagnetic ladder compound PNNBNO.
The thermodynamics of the spin-orbital model with
different single-ion anisotropies is also investigated. 
For this model single-ion anisotropy can trigger different quantum phase
transitions within the spin and orbital degrees of freedom, with 
magnetization plateaux arising from different spin and orbit Land\'{e} $g$-factors. 
\end{abstract}

\section{Introduction}

The subject of integrable or exactly solved models in statistical mechanics 
has inspired a number of profound developments in both physics and mathematics.
There are several key solvable models in low dimensions, the analysis of 
which has developed to an extraordinary level of mathematical sophistication. 
This article is concerned with the calculation of physical properties of the 
integrable quantum spin ladders and their applicability
to experimental ladder compounds.
By integrable it is meant that the ladder models are exactly solvable in the 
Yang-Baxter sense \cite{McG, Yang, Baxter}, with infinitely many conserved quantities 
and an underlying Bethe Ansatz solution.
Most importantly,  this entails access to the thermodynamic properties \cite{QISMbook,Tbook}. 
In this way the theory of integrable models, 
which has developed for over 75 years since Bethe's \cite{Bethe,Batchelor} pioneering solution of the
$su(2)$ Heisenberg chain,  
has recently been applied \cite{HTE1}  to make direct contact with experiments 
on spin ladder compounds.
The integrable ladder models, for example, the two-leg spin-$\frac{1}{2}$ ladder 
and the mixed spin-$(\frac{1}{2},1)$ ladder, are variants of the $su(4)$ and $su(6)$ 
models based on the general family of integrable $su(N)$ permutator 
models \cite{McG,Yang,Uimin, Lai, Suth}.
Their two-dimensional classical lattice model counterparts are the $N$-state $A_{N-1}$ vertex
models \cite{VVB,Jimbo}, which are special cases of the Perk-Schultz model \cite{Perk}.

\subsection{Physical motivation}

Over the last few decades there has been considerable theoretical and
experimental interest in low-dimensional quantum spin systems. 
One  focus of attention is the  spin liquid behaviour.
It is well established  that the critical behaviour of the one-dimensional Heisenberg spin chain 
in the absence of frustration depends on the spin magnitude $S$, i.e., on whether $2S$ is 
an even or odd integer.
In one dimension the quantum fluctuations are so
strong that the long range N\'eel order no longer survives.  
For $2S$ odd the spin chain has low-lying gapless excitations 
with a power law spin-spin correlation \cite{LSM}. 
On the other hand, for $2S$ even the spin-spin correlations decay exponentially
due to the Haldane gap \cite{Haldane1,Haldane2}. 
In two-dimensional quantum spin systems the quantum fluctuations compete with the 
thermal fluctuations for domination of the critical behaviour at low temperatures. 
Here long range N\'eel order is possibly stable against quantum fluctuations.

Spin ladder systems, which intermediate between one-dimensional
and two-dimensional spin systems, exhibit rich and novel phase transitions
and a diverse range of quantum magnetic effects at low temperatures
\cite{ladder1,exp0,exp1,Spin-liquid}.
The long range N\'eel order phase emerges when the interaction between
the adjacent two-leg ladders is greater than some critical value \cite{phase1}. 
However, if the two-leg ladder structures are well
isolated from each other, the intra- and inter-chain interactions
induce rich spin liquid phases, such as gapped dimerised and gapless phases. 
The existence of a spin gap, magnetization plateaux,
quantum critical points and superconductivity under hole doping are
examples of key physical properties observed in the ladder compounds. 
The physics of the spin ladder systems reveals a significant pairing effect, 
namely the Heisenberg spin ladders have an energy gap in the spin excitation 
spectrum if the number of ladder legs is even,
whereas there is no gap for an odd number of legs \cite{ladder1,exp0,exp1}.
This similarity with the above mentioned Haldane conjecture for spin-$S$ chains arises
from the fact that the ladder model can be mapped onto a
one-dimensional counterpart with on-site anisotropy or
next-nearest-neighbour interaction, at least in the strong coupling limit \cite{White,ortiz}.  
In addition, the ladder systems have been
argued to provide simple models relevant to the mechanism for high
temperature superconductivity in layered cuprate superconductors \cite{exp1}.
The challenge for theorists is to calculate and to predict the properties of 
the ladder systems.

\subsection{Experiments and theories for ladders}

The experimental study of ladder systems was initiated from the observation that
the compound (VO)$_2$P$_2$O$_7$ has a singlet ground state with
an energy gap in the spectrum of spin excitations \cite{Johnston}. 
Subsequently, the theoretical study of ladder systems began when the hole-doped 
two-leg $t$--$J$ ladder was found to exhibit a finite spin gap \cite{Dagotto92}. 
In this model, at each site the ions are either in a spin-$\frac{1}{2}$ Cu$^{2+}$ state
or a Cu$^{3+}$ hole state. 
Hole ions are bound along rungs so that the spin-$\frac{1}{2}$ Cu$^{2+}$ ions form rung singlets. 
This configuration leads to superconducting pairing correlations \cite{Dagotto92,exp1}. 
In this context, there have been many theoretical papers focussing on hole-doped ladders, 
see Refs. \cite{exp1,numers4}. 
Experimental superconducting ladder compounds, e.g.,
(La,Sr,Ca)$_{14}$Cu$_{24}$O$_{41}$ \cite{McCarron,Siegrist} and other
doped ladder compounds, e.g., LaCuO$_{2.5}$ with Sr doping \cite{Hiroi,numers4}, 
provide experimental evidence of superconductivity. 
In Cu-based ladders, the hole doping provides a ``d-wave'' channel for a 
superconducting phase \cite{Dagotto92,exp1}.
In this article we focus on undoped or pure spin ladder materials for
comparisons with the theoretical results.

The Heisenberg antiferromagnetic spin-$\frac{1}{2}$ ladder model was first proposed
in Ref. \cite{ladder1}. 
It is known that the Heisenberg coupling $J_{\perp}$ along the rungs can open a 
gapped phase for arbitrary coupling strength. 
If $J_{\perp}=0$, the ladder decouples into two non-interacting 
spin-$\frac{1}{2}$ Heisenberg chains, with no gap to spin excitations \cite{LSM}. 
The decoupled chains are in critical states exhibiting no long range order. 
In the presence of weak interchain interactions (small $J_{\perp}$), 
the lowest spin excitation can be separated from
the ground state by an energy gap of magnitude 
$\Delta \approx {J_{\perp}}/{2}$ \cite{exp1,numers4,White1,Gopalan,Shelton}.
Essentially, the weakly coupled ladder models may exhibit antiferromagnetic 
long-range order at finite temperatures.
For example, the trellis layer structure of the Sr(Cu$_{1-x}$Zn$_x$)$_2$O$_3$ 
lattice with nonmagnetic isoelectronic Zn$^{+2}$ doping \cite{Azuma2,Fujiwara} 
was found to induce long-range N\'{e}el order. 
It was verified experimentally that the undoped two-leg ladder compound
SrCu$_{2}$O$_3$ exhibits a spin gap, but the three-leg ladder
compound Sr$_2$Cu$_{3}$O$_5$ does not \cite{Azuma,Kojima,Ishida}. 
In the strong coupling limit, the ground state of the two-leg ladder
compound is dimerised along the rungs.
The intrachain interaction between neighbouring magnetic moments is 
quenched by the formation of a rung singlet state. 
In order to create a spin excitation the rung singlet must be replaced by a rung triplet.  
Breaking the rung singlet causes an energy gap
$\Delta \approx J_{\perp}-J_{\parallel}$ which separates the lowest
excited state from the dimerised ground state \cite{exp1,phase1,phase2}.  
Here $J_{\parallel}$ is the (intrachain) coupling along the ladder legs.
The number of legs significantly affects the magnitude of the spin gap. 
For an even number of spin-$\frac{1}{2}$ ladder legs the gap decreases 
as the number of legs grows \cite{Azuma,Kojima}, as one would expect 
as the ladder becomes more like a two-dimensional antiferromagnet.
The ground state is a spin singlet with zero total spin. 
For an odd number of legs with strong rung coupling, the ground state is a magnetic
degenerate state with total spin $S=\frac{1}{2}$, with 
algebraically decaying spin-spin correlations and 
no gap to spin excitations. 
The qualitatively different behaviour between even- and odd-leg spin 
ladders has been discussed at length \cite{Gopalan,Rice93,Sigrist,exp1}.

\begin{figure}
\centerline{\includegraphics[width=.30\linewidth]
{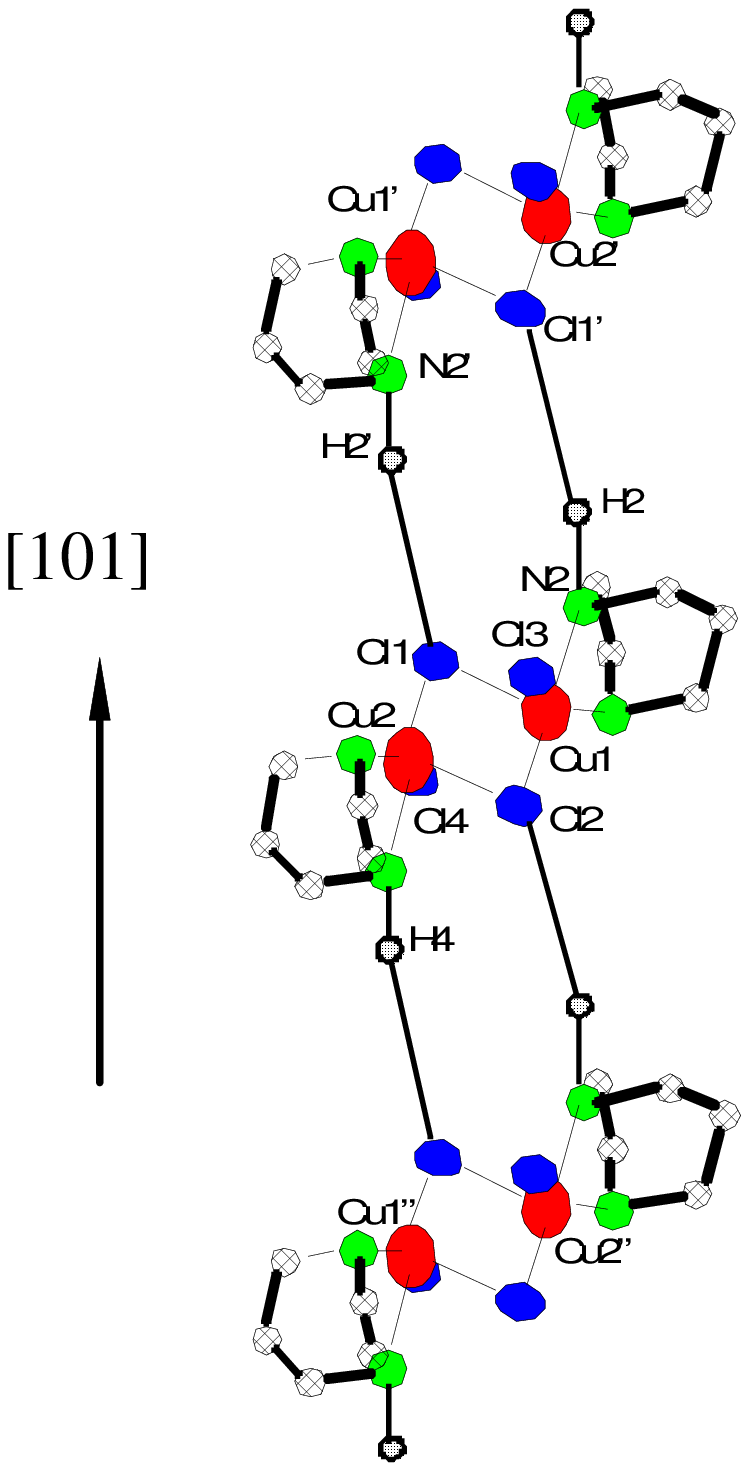}}
\caption{
The schematic structure of the two-leg spin ladder compound
  Cu$_2$(C$_5$H$_{12}$N$_2$)$_2$Cl$_4$. The Cu-Cu binuclear units
  along the chain direction [101] form the intrachain interaction. The
  superexchange paths between two Cu ions along each rung form the interchain interaction.
Reproduced from Ref.~\cite{Chaboussant2-2}.
}
\label{fig:B5i2aTSZ-structure}
\end{figure}

Of particular interest are the strongly coupled ladders as their
energy gaps are experimentally accessible. 
Many organic spin ladders with strong rung coupling have been synthesised. 
In particular, (5IAP)$_2$CuBr$_4$$\cdot 2$H$_2$O \cite{5IAP}
Cu$_2$(C$_5$H$_{12}$N$_2$)$_2$Cl$_4$
\cite{compound1,Chaboussant1,Chaboussant2,Chaboussant2-2} and
(C$_5$H$_{12}$N)$_2$CuBr$_{4}$ \cite{Watson} have a well isolated ladder structure. 
The ladder structure for the compound
Cu$_2$(C$_5$H$_{12}$N$_2$)$_2$Cl$_4$ is shown in figure \ref{fig:B5i2aTSZ-structure}.  
This compound is made up of effective two-leg spin-$\frac12$ ladders
with strong antiferromagnetic rung interaction through superexchange
between adjacent molecular units.  
The interchain coupling between Cu-Cu ions is parametrized by the rung coupling $J_{\perp}$.  
The dimeric units along each rung are weakly linked to the nearest neighbour ones
through hydrogen bonds along the [101] chain direction. 
The exchange between  two neighbouring rungs is much weaker than the intradimer path. 
This intrachain coupling is denoted by the parameter $J_{\parallel}$. 
It is believed that the family of compounds KCuCl$_3$, TlCuCl$_3$, NH$_4$CuCl$_3$ and
KCuBr$_3$ also exhibit such double-spin-chain 
structure \cite{KCL1a,KCL1b,KCL2,KCL3,KCL4a,KCL4b,KCL4c,KCL5a,KCL5b,TLCL1a,TLCL1b,Hagiwara,TLCL2,F1,F2}.
These double-chain compounds have been studied extensively
from both the theoretical and experimental points of view.

The Heisenberg ladder model, with antiferromagnetic spin-spin interactions
along the legs and rungs, describes these compounds rather well. 
The strong coupling compounds have a singlet ground
state for magnetic fields less than a critical value $H_{c1}$. 
The theoretical prediction for the energy gap is 
$\Delta \approx J_{\perp}-J_{\parallel}$ \cite{ladder2,FT1,exp1}, 
which is in excellent agreement with the experimental results, obtained
from Nuclear Magnetic Resonance measurements \cite{Chaboussant1}, along with
high field magnetic properties \cite{Hagiwara,5IAP,Chaboussant3,KCL2}. 
In addition, the
susceptibility $\chi \propto e^{-\Delta/T}/\sqrt{T}$ with energy gap
$\Delta$ obtained from the quantum transfer matrix method \cite{Troyer} 
generally provides a good fit for the strongly coupled two-leg ladders.  
For magnetic field $H  > H_{c1}$ the triplet rung state enters into the ground state,
with a gapless Luttinger phase in the regime $H_{c1} < H < H_{c2}$. 
The critical field $H_{c2}$ indicates a
quantum phase transition separating the Luttinger liquid magnetic phase
from a fully polarized state. 
It was suggested \cite{FT2} that the gapless phase can be viewed as 
a Bose-Einstein condensation of magnons \cite{AffleckA,AffleckB,BEC1,BEC3,BEC4,BEC2}.
The quantum fluctuation of low-lying excitations for these quantum spin
systems are  characterized by spin waves which reveal bosonic
signature.  
In this way the strongly coupled spin ladder with magnon excitations in high magnetic
field can be mapped onto a one-dimensional $XXZ$ Heisenberg model with
an effective magnetic field. 
{}From this mapping, the critical fields are given by 
$H_{c1}\approx J_{\perp}-J_{\parallel}$ and 
$H_{c2}\approx J_{\perp}+2J_{\parallel}$ \cite{ladder2,FT1}, which 
agree rather well with experiments \cite{exp1}.

Considerable interest has been generated by fractional magnetization plateaux,
which have been found in Shastry-Sutherland dimer systems \cite{FPL2,FPL3,FPL} and 
some spin organic ferrimagnetic ladders \cite{mixl1,mixl2a,mixl2b}.
Theoretical studies and numerical results suggest that magnetization 
plateaux exist in a mixed spin ladder \cite{mixl5,mixl4,mixl3} and the mixed
spin-$(\frac{1}{2},1)$ chains \cite{mixc1a,mixc1b,mixc2a,mixc2b,mixc3a,mixc3b,mixc4,mixc5}.
It was concluded that for certain strong rung coupling values, there
exist magnetization plateaux at $M^z=0.5$ and at $M^z=1$, but no
plateaux for negative (ferromagnetic) rung coupling values. 
In the strong coupling limit, the mixed spin ladder can be viewed as a
ferrimagnetic chain which displays unusual quantum effects at low
temperatures.

\subsection{Integrable ladder models} 

Haldane's conjectured universal properties of the spin-$S$ Heisenberg
antiferromagnet depending on whether $2S$ is odd or even have been
recognized in many disordered spin liquid systems, such as in spin
ladders, mixed spin ladders and alternating spin chains.  
The spin liquid systems have a finite correlation length and a spin gap at zero
temperature. The ground state is a spin singlet. The lowest spin
triplet excitations have an energy gap due to the finite correlation
length.  In this context, Nersesyan and Tsvelik \cite{ladder3}
predicted a different gapped phase for a generalised two-leg
spin-$\frac{1}{2}$ ladder model with a strong four-spin interaction
term (the biquadratic interaction strength $|V|$ is very large \cite{ladder3}). 
This model includes the conventional two-leg
Heisenberg spin-$\frac{1}{2}$ ladder and the integrable spin ladder
model based on $su(4)$ algebra symmetry \cite{I-ladder1} as special
cases, i.e.,  $V=0$ and $|V|=4J_{\parallel}$, respectively.  
In the case of weak biquadratic interaction ($|V|$ very small), the
generalised Heisenberg ladder model has the same Haldane spin liquid
phase as that of the conventional spin ladder.  However, for strong
enough biquadratic interaction this generalized ladder model exhibits
a spontaneous dimerization phase which is different from the Haldane spin
liquid gapped phase.
The physical importance of the model has been subsequently noted  \cite{ladder4}.

{}From the theoretical perspective, these ladder models cannot be solved exactly, 
in contrast to the spin-$\frac12$ Heisenberg chain. 
Thus their thermal and magnetic properties are not accessible via exact methods.
In order to get a handle on the physical properties, some integrable extensions of 
the Heisenberg ladders have been constructed 
\cite{I-ladder1,I-ladder1-2a,I-ladder1-2b,I-ladder2a,I-ladder2b,I-ladder2c,I-ladder2d,I-ladder3a,I-ladder3b,
I-ladder4a,I-ladder4b,I-ladder4c,I-ladder5a,I-ladder5b,I-ladder5c,I-ladder5d,I-ladder5e,I-ladder5f,
I-ladder5g,I-ladder5h,I-ladder5i,I-ladder5j,I-ladder5k}.
These models include an integrable ladder based on the $su(4)$ algebra
\cite{I-ladder1} and integrable $n$-leg ladder models
\cite{I-ladder2a,I-ladder2b,I-ladder2c,I-ladder2d,I-ladder3a,I-ladder3b} among a number of variants.
At first glance, due to the complicated spin-spin interactions along the
legs and rungs, it appears that none of these models can be 
used to predict physical properties for direct comparison with experimental data. 
Indeed, the application of the integrable ladder models to
ladder materials should be taken with caution. 
In this article we focus on the calculation and study of the exact thermal and magnetic
properties of the integrable ladder models and attempt to fit these properties to the
experimental data for some known strong coupling ladder compounds by setting 
aside the issues regarding the applicability of the integrable Hamiltonian to those materials, 
in particular the effect of the biquadratic interaction term.
We demonstrate that under certain conditions, namely
\begin{equation}
J_{\perp} \gg J_{\parallel} \quad \mbox{\rm and} \quad K_BT > J_{\parallel}
\end{equation}
the integrable ladder models may be used to study the thermal and magnetic
properties of ladder materials with strong rung coupling. 
However, at low temperatures the biquadratic interaction term in the 
integrable models may lead to discrepancies with the physics of the 
ladder materials.  
These discrepancies are minimal if $J_{\perp} \gg J_{\parallel}$.  
We point out that the good agreement between the theoretical
predictions and the experimental results for macroscopic quantities such as the 
magnetization, the susceptibility and the specific heat do not imply that the
materials considered have the integrable Hamiltonian structure. 
For determining the precise structure of the materials more physical analysis, 
such as dynamical structure and dispersion in low-lying excitations,
should be carried out in addition to the study of the thermal and
magnetic properties.  
At high temperatures,
the thermodynamics of the compounds can also be modelled by a
rung-dimer system \cite{MU03}. 
However, the dimer model gives poor agreement with
experimental results at low temperatures. 
For the dimer model, it appears necessary to include Dzyaloshinskii-Moriya interaction
terms in order to better explain the experimental behaviour \cite{Capponi}.

The integrable ladder model based on the $su(4)$ algebra \cite{I-ladder1} 
appears to be a good candidate \cite{HTE1,TBAladder1a,TBAladder1b} for 
studying the ladder compounds, 
since its Hamiltonian is that of the standard Heisenberg ladder 
with an additional biquadratic spin interaction. 
The physical interpretation of the biquadratic spin interaction term is 
of various kinds. 
In the continuum limit it represents an interchain
coupling of spin-dimerization fields which therefore influence the
low-energy properties of the spin ladder \cite{ladder3}.  
In the strong coupling limit, the effective biquadratic spin-spin coupling arises
from virtual electron transitions between neighbouring magnetic atoms. 
This coupling is in general very small in spin chain systems \cite{Milla-Zhang}, 
smaller than for example the quadratic interaction term.
In principle, the strength of the biquadratic interaction in ladder materials 
might be accurately determined through calculation of exchange integrals.

For strong rung coupling, i.e., $4J_{\parallel} \ll J_{\perp}$, 
the spin-spin exchange interaction is dominated by the strong rung interaction. 
It follows that the contribution from the interaction along the ladder legs,
 including the biquadratic term, to the low-temperature physics is much less 
 than the contribution from the rung interaction.
However, the leg interactions and the biquadratic spin interactions do 
cause a shift in the critical field values. 
This kind of shift in the critical field values can be
balanced by adjusting the interaction strengths.
We shall see clearly that the values of the coupling constants $J_{\parallel}$ 
for the various compounds determined from the integrable model
are always smaller than the values fixed by the standard Heisenberg spin ladder,
as a result of including the biquadratic interaction term. 
The integrable spin ladder model provides a consistent and elegant way to
obtain the ground state and thermodynamic properties of
the strongly coupled ladder compounds, i.e., with $4J_{\parallel} \ll
J_{\perp}$.  
We emphasise again that we set aside any issues regarding the reality of the 
biquadratic interaction term. 
The advantage of the integrable model approach is that it
opens up the exact calculation of finite temperature thermodynamic
properties via the well established methods developed for 
integrable systems \cite{QISMbook,Tbook}, including the more 
recently developed high-temperature expansion method reviewed in this article.

The interaction $J_{\parallel}$ plays an essential role in  
opening up a Luttinger liquid magnetic phase between the
singlet and fully-polarized phases for $H_{c1}<H<H_{c2}$. 
Although the physical properties are not sensitive to the value of
$J_{\parallel}$ at large temperatures, $J_{\parallel}$ has a 
significant influence on the low temperature behaviour.  
It is argued that for the integrable model  the collective effect 
from the biquadratic interaction and the exchange interaction 
along the ladder legs are equivalent to the pure Heisenberg type 
interaction in the standard Heisenberg spin ladder model 
for the strong coupling region $4J_{\parallel} \ll  J_{\perp}$.
The coupling constants can be fixed by an overall fit of the
critcal fields and thermodynamic properties with the experimental results.  
In this way the integrable model is seen to exhibit similar critical behaviour, 
i.e., in the same universality class,  
in the correlation functions, magnetization, susceptibility, specific heat etc 
to the non-integrable Heisenberg ladder, which has so far been used to examine 
spin ladder materials \cite{exp1}.

The Thermodynamic Bethe Ansatz (TBA) \cite{TBA1,TBA2} has been widely 
applied to many physical problems, e.g., to the Heisenberg magnet \cite{TBA2,DV95}, 
the Hubbard model \cite{Takahashi,Hubbardbook}, the supersymmetric $t$--$J$ 
model \cite{Schlottmanna,Schlottmannb,EK92,Angela} and the Kondo problem \cite{Kondo}.  
It has most recently evolved into the exact High Temperature Expansion (HTE)
method \cite{HTE2,ZT1,ZT2,RST02,BEU00}.  
The key ingredients of the HTE method are the Quantum Transfer 
Matrix (QTM) \cite{Suzuki,QTMrefsa,QTMrefsb,QTMproof,SI87,K87,SAW90,JKS98,FK99,KLNP} 
and a functional relation called the
$T$-system \cite{KunibaFusiona,KunibaFusionb,KRfusion}, from which one derives nonlinear
integral equations which can be solved in an exact perturbative fashion. 
To date this approach has been applied to the Heisenberg
model \cite{HTE2}, the $osp(1|2s)$ model \cite{ZT1}, 
the $su(N)$ Uimin-Lai-Sutherland model \cite{ZT2}, the higher spin
Heisenberg model \cite{ZT3}, the $su(N)$ Perk-Schultz model \cite{ZT4} 
and the $su(m|n)$ Perk-Schultz model \cite{ZT5}.  
In particular, in this way, i.e., via the HTE expansion, it has been demonstrated that
the integrable $su(4)$ ladder model can be used to study the thermodynamics and magnetic
properties of the strongly coupled ladder compounds \cite{HTE1}. 

For strongly coupled ladders, the rung coupling $J_{\perp}$ can overwhelm the 
intrachain interaction $J_{\parallel}$ such that a pair of spin-$S$ magnetic ions along 
each rung forms an effective spin-$2S$ ion, allowing the
strongly coupled ladders to be mapped to one-dimensional spin chains. 
The integrable ladder models based on $su(N)$ symmetry 
provide a tunable way to describe experimental compounds for which the rung 
coupling dominates the exchange interaction along the legs. 
In a certain basis, the rung interaction terms are diagonal while the basis maintains 
the $su(N)$ permutator exchange interaction along the legs. 
More precisely, the integrable ladder models based on $su(N)$ symmetry can be mapped 
to one-dimensional $su(N)$ \cite{HTE1,I-ladder2a,I-ladder2b,I-ladder2c,I-ladder2d} 
chains with additional chemical potentials.  
Consequently, the thermal and magnetic properties can be systematically derived from 
the exact TBA and HTE approaches.

The TBA method is seen to be most convenient for predicting critical fields
of relevance to quantum phase transitions in ladder compounds 
\cite{TBAladder1a,TBAladder1b,mix,YRFC,ying2}, spin chains \cite{BGN,BGNF} with strong single-ion anisotropies
\cite{orendaca,orendacb}  and spin-orbital models \cite{yinga,yingb}.
It has been shown \cite{TBAladder1a,TBAladder1b,YRFC} that in the strong coupling
region the integrable two-leg spin-$\frac{1}{2}$ ladder model exhibits three
quantum phases: 
\begin{enumerate}
\item  a gapped phase in the regime $H< H_{c1}=J_{\perp}-4J_{\parallel}$, 
\item a fully polarized phase for $H>H_{c2}=J_{\perp}+4J_{\parallel}$, 
\item a Luttinger liquid magnetic phase in the regime $H_{c1}<H<H_{c2}$. 
\end{enumerate}
The critical behaviour in the vicinity of the critical points $H_{c1}$ and $H_{c2}$ 
is seen to be of the Pokrovsky-Talapov  type.  
On the other hand, the HTE method gives the temperature-dependent free energy, 
from which physical properties such as the magnetic susceptibility and the specific 
heat can be derived \cite{HTE1,ying2}.  
Indeed, as we shall discuss here, in certain regimes, the integrable two-leg ladder with strong 
rung coupling \cite{HTE1,ying2} describes some experimental compounds, e.g., 
(5IAP)$_2$CuBr$_4$$\cdot 2$H$_2$O \cite{5IAP},
Cu$_2$(C$_5$H$_{12}$N$_2$)$_2$Cl$_4$ \cite{Chaboussant1,Chaboussant2},
(C$_5$H$_{12}$N)$_2$CuBr$_{4}$ \cite{Watson}, as well as the double spin
chain systems KCuCl$_3$ 
\cite{KLC1a,KCLb,KCL2,KCL3} TlCuCl$_3$
\cite{KCL2,TLCL1a,TLCL1b,TLCL2}, organic polyradical ladders BIP-BNO
\cite{BIPa,BIPb}, [Cu$_2$(C$_2$O$_2$)(C$_{10}$H$_8$N$_2$)$_2$)](NO$_3$)$_2$
\cite{SC2} and the ferrimagnetic mixed spin ladder PNNBNO
\cite{mixl1,mixl2a,mixl2b}.

\subsection{Key results and outline}

In this article we review the calculation of the physical properties of
an integrable two-leg spin-$\frac{1}{2}$ ladder and an integrable mixed spin-$(\frac{1}{2},1)$ ladder. 
For the integrable spin-$\frac{1}{2}$ ladder with strong
rung coupling the singlet ground state is separated from the
lowest magnon (triplet) excitation by a finite energy gap.
Notably, the values for  the energy gap, $\Delta=J_{\perp}-4J_{\parallel}$, and
the critical points, $H_{c1}=J_{\perp}-4J_{\parallel}$ and
$H_{c2}=J_{\perp}+4J_{\parallel}$, derived from the TBA are in excellent
agreement with the available experimental results for strong coupling
ladder compounds.
The compounds discussed are 
(5IAP)$_2$CuBr$_4$$\cdot 2$H$_2$O \cite{5IAP}, 
Cu$_2$(C$_5$H$_{12}$N$_2$)$_2$Cl$_4$ \cite{Chaboussant1,Chaboussant2}, 
(C$_5$H$_{12}$N)$_2$CuBr$_{4}$ \cite{Watson}, 
the organic polyradical ladders BIP-BNO \cite{BIPa,BIPb} and
[Cu$_2$(C$_2$O$_2$)(C$_{10}$H$_8$N$_2$)$_2$)](NO$_3$)$_2$ \cite{SC2}.
The finite temperature thermal and magnetic properties are examined
for these compounds via the exact HTE, with good overall agreement between 
the theoretical predictions and the experimentally measured thermal and magnetic 
properties.

For the integrable mixed spin-$(\frac{1}{2},1)$ ladder, the ground
state lies in a gapless phase and a one-third saturation
magnetization plateau occurs as the rung coupling exceeds
$J_{\perp}>\frac{4}{3}J_{\parallel}\ln 2$. 
In the absence of a magnetic field the model exhibits three quantum 
phases associated with doublets, quadruplets and $su(6)$ symmetry. 
The fractional plateau corresponding to a fully-polarized doublet state 
opens at the critical field $H_{c1}$ and vanishes at the critical field $H_{c2}$. 
If the magnetic field is greater than $H_{c3}$, the quadruplets are fully polarized.
The phase diagram  is reminiscent of the spin-$\frac{3}{2}$ chain \cite{BGN, BGNF}. 
The magnetic properties of the organic mixed ferrimagnetic ladder compound PNNBNO 
are examined via the HTE.

The ground state properties of two integrable spin-orbital models are
also treated via the TBA method, from which it is seen that the 
single-ion anisotropy and orbit splitting field can trigger a magnetization plateau 
with respect to
different Land\'{e} factors with spin and orbital degrees of freedom \cite{yinga,yingb}. 
The magnetization for spin and orbital degrees of freedom
is also calculated numerically from the TBA equations.

The article is organised as follows. 
In section \ref{sec:ISL}, we review the two integrable spin ladder models --
a two-leg spin-$\frac{1}{2}$ ladder and a mixed spin-($\frac12,1)$ ladder -- 
along with their Bethe Ansatz solutions. 
In section \ref{sec:TBA}, we discuss the TBA equations for the two-leg spin-$\frac{1}{2}$ 
ladder and the investigation of the ground state properties via the TBA equations at 
zero temperature. 
The exact and resulting numerical calculations for the magnetization 
and susceptibility are treated in detail. 
The Quantum Transfer Matrix method is reviewed in section \ref{QTM-NLIE}, 
along with the derivation of the nonlinear integral equations in terms of the $T$-system. 
It is then shown how the free energy is derived from the HTE. 
The results are then applied in section \ref{sec:Exp_comp}  
to examine the thermal and magnetic properties of the
strong coupling ladder compounds 
(5IAP)$_2$CuBr$_4$$\cdot 2$H$_2$O, 
Cu$_2$(C$_5$H$_{12}$N$_2$)$_2$Cl$_4$,
(C$_5$H$_{12}$N)$_2$CuBr$_{4}$, 
and the organic polyradical ladders BIP-BNO
and [Cu$_2$(C$_2$O$_2$)(C$_{10}$H$_8$N$_2$)$_2$)](NO$_3$)$_2$.
In section \ref{mixHTE}, the same approach is applied to the mixed spin-($\frac{1}{2},1$) 
ladder with a discussion of the ladder compound PNNBNO. 
Section \ref{sec:SO} is devoted to the study of the magnetic properties of the spin-orbital model. 
A concluding discussion is given in section \ref{sec:concl}.

It should be noted that readers more interested in the physical properties of the ladder
compounds, rather than their detailed calculation, may choose to skip sections 
\ref{sec:TBA} and \ref{QTM-NLIE}. 
In any case the relevant results are collected at the beginning of section \ref{sec:Exp_comp} 
on the comparison of the thermal and magnetic properties with the experimental data.
Conversely, readers more interested in the mathematical methods may prefer to concentrate 
on sections \ref{sec:TBA} and \ref{QTM-NLIE}. 


\section{Integrable spin ladder models}
\label{sec:ISL}

\begin{figure}[ht]
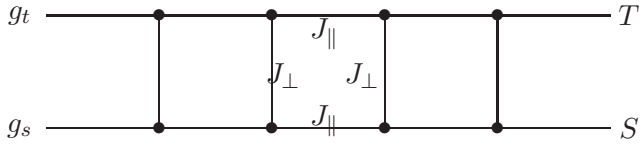

\begin{center}
\Ladder \\
\end{center}
\caption{The two-leg spin ladder.
In general, sites on the lower (upper) leg carry spin-$S$ ($T$) with $g_s$ ($g_t$) 
the Land\'{e} factor along the leg.
$J_{\parallel}$ is the intrachain or leg coupling and 
$J_{\perp}$ is the interchain or rung coupling.}
\label{fig:Figmix1}
\end{figure}

The general two-leg ladder configuration is depicted in figure \ref{fig:Figmix1}.
The two-leg Heisenberg spin-$\frac{1}{2}$ ladder model is defined by the 
Hamiltonian \cite{ladder1,exp1, ladder2} 
\begin{equation}
{\cal H}=J_{\parallel}\sum_{j=1}^{L}(\vec{S}_j\cdot \vec{S}_{j+1}+\vec{T}_j\cdot \vec{T}_{j+1})+
J_{\perp}\sum_{j=1}^{L}\vec{S}_j\cdot \vec{T}_j-\mu_{{\rm B}}gH\sum_{j=1}^{L}(S_j^z+T_j^z). \qquad
\label{eq:Ham1}
\end{equation}
Here $\vec{S}_j=(S_j^x,S_j^y,S_j^z)$ and $\vec{T }_j=(T_j^x,T_j^y,T_j^z)$ are
spin-$\frac{1}{2}$ operators acting on site $j$.
For example, $\vec{S}_j=\frac12  \vec{\sigma}_j$, where 
$\vec{\sigma}_j=(\sigma_j^x, \sigma_j^y, \sigma_j^z)$ are Pauli matrices.
The essential parameters are the rung coupling $J_{\perp}$, the intrachain coupling 
$J_{\parallel}$ and the external magnetic field $H$.  
The Bohr magneton is $\mu_{{\rm B}}$ and $g=g_s=g_t$ is the Land\'{e} factor.  
There are $L$ rungs and periodic boundary conditions, $\vec{S}_{L+1}=\vec{S}_1$ and 
$\vec{T}_{L+1}=\vec{T}_1$, are imposed.

This model has been extensively studied via numerical methods 
\cite{ladder1,numers1,numers2,numers3,numers4},
perturbative field theory \cite{ladder2,FT1,FT2,FT3,Sierra} 
and transfer matrix methods \cite{Troyer}. 
Some essential physics of the model can be obtained from the strong coupling region
$J_{\perp} \gg J_{\parallel}$.
Consider the more general spin-$S$ ladder.
Then in this limit the ladder effectively becomes a spin-$2S$ chain with single-ion anisotropy. 
The physics of the ladder/chain equivalence holds even for weak coupling, but this 
is by no means obvious. 
The gapped phase persists for arbitrary rung coupling, with  
the spin-spin correlation decaying exponentially. 
The analogy between the physics of the spin chain with $2S$ even/odd 
and the spin ladder with even/odd number of legs is well described by the quantum field theoretic 
approach \cite{QF-ladd-1,QF-ladd-2,Sierra}. 
Both the antiferromagnetic Heisenberg spin chain and the ladder can be mapped to the  
nonlinear sigma models. 
The field theory approach thus gives a unified description of the Heisenberg chains 
and ladders. 
In this framework the gapped and gapless phases of the spin chains and ladders 
are governed by the topological term $\theta =0$ or $\pi$,  respectively, 
in the nonlinear sigma model \cite{Affleck1,Sierra}.

\subsection{The integrable two-leg spin-$\frac{1}{2}$ ladder}

The simplest integrable spin ladder is defined by the Hamiltonian \cite{I-ladder1} 
\begin{equation}
{\cal H}=J_{\parallel}{\cal H}_{{\rm leg}}+J_{\perp}\sum_{j=1}^{L}\vec{S}_j\cdot \vec{T}_j-
\mu_{{\rm B}}gH\sum_{j=1}^{L}(S_j^z+T^z_j)
\label{eq:Ham2}
\end{equation}
where
\begin{equation}
{\cal H}_{{\rm leg}}=\sum_{j=1}^{L}\left(\vec{S}_j\cdot \vec{S}_{j+1}+\vec{T}_j
\cdot \vec{T}_{j+1}+4(\vec{S}_j\cdot \vec{S}_{j+1})(\vec{T}_j\cdot \vec{T}_{j+1})\right).
\label{eq:intra} 
\end{equation}
Periodic boundary conditions, $\vec{S}_{L+1} = \vec{S}_1$, $\vec{T}_{L+1} = \vec{T}_1$, 
are applied.
This integrable Hamiltonian contains a biquadratic term, which minimises the
Haldane phase \cite{ladder3,TBAladder1a,TBAladder1b} and causes a shift of the critical value of
the rung coupling $J_{\perp}$ at which the gap vanishes (recall that the gap
vanishes at zero rung coupling for the Heisenberg ladder). 
The other key feature of the model is that it has the same magnetic field and rung interaction
terms as the standard Heisenberg ladder (\ref{eq:Ham1}).
In the absence of symmetry breaking potentials, the dispersion of the low-lying excitations 
is entirely caused by the leg interaction.
However, symmetry breaking potentials, such as rung interaction or external field, 
result in different band fillings such that the lowest spin excitation is gapped.
The two-body and biquadratic interactions along the legs are effectively 
quenched by the strong rung singlet. 
The model thus lies in the same phase as the standard Heisenberg ladder.

The critical behaviour of the model is determined by the 
competition between the rung and leg couplings and the 
magnetic field $H$.  
A key factor in the integrability of this Hamiltonian is that ${\cal H}_{{\rm leg}}$ 
is (up to a constant) simply the permutation operator corresponding to the 
$su(4)$ algebra \cite{I-ladder3a,I-ladder3b}.  
This can be seen by performing a change of basis to the states
\begin{eqnarray}
\begin{array}{ll}
\st[1] = \frac{1}{\sqrt{2}}\left(\st[\! \uparrow
\downarrow] -\st[\! \downarrow \uparrow ] \right), & \st[2]
=\st[\! \uparrow\uparrow]  \\
\st[3] =
\frac{1}{\sqrt{2}}\left(\st[\! \uparrow \downarrow] +\st[\! \downarrow
\uparrow ] \right),& \st[4] =\st[\! \downarrow\downarrow]
\end{array}
\label{eq:su4basis}
\end{eqnarray}
where state $\st[1]$ denotes the rung singlet and the remaining three
states denote the triplet components.  
Here we have set 
$\st[\!\! \uparrow] =   {1 \choose 0}$ and $\st[\!\! \downarrow] =  {0 \choose 1}$.
Thus $| \!\! \uparrow \rangle \langle \uparrow \!\! | = \frac12 (1+\sigma^z)$, 
$| \!\! \downarrow \rangle \langle \downarrow \!\! | = \frac12 (1-\sigma^z)$,
$| \!\! \uparrow \rangle \langle \downarrow \!\! | = \sigma^+$ and 
$| \!\! \downarrow \rangle \langle \uparrow \!\! | = \sigma^-$. 
Note that in this new basis the leg part remains of the same form while the rung term 
becomes diagonal. 
The rung term reduces the $su(4)$ symmetry of ${\cal H}_{{\rm leg}}$ 
to $su(3)\oplus u(1)$ symmetry under the action of the local chemical potential 
$J_\perp | 1 \rangle_{j\,j} \langle 1 |$.
Switching on the magnetic field further breaks this symmetry due to Zeeman splitting. 
Moreover, the $su(3)\oplus u(1)$ symmetry does not survive if we consider
different $g$ factors along the two legs.
Let $g_t$ and $g_s$ be these Land\'{e} factors along the upper and lower legs, respectively. 
Then the Hamiltonian reads \cite{yinga,yingb}
\begin{equation}
{\cal H}=J_{\parallel}{\cal H}_{{\rm leg}}+J_{\perp}\sum_{j=1}^{L}\vec{S}_j\cdot \vec{T}_j-
\mu_BH\sum_{j=1}^L(g_sS^z_j+g_tT^z_j)
\end{equation}
where the local eigenbasis on the $j$th rung is given by
\begin{eqnarray} 
\phi_0
&=&\frac{1}{\sqrt{1+(y^{(-)})^2}}\left(\st[\! \uparrow\downarrow]+
y^{(-)}\st[\! \downarrow \uparrow ]\right),\quad \phi_1 = \st[\! \uparrow \uparrow ]\nonumber \\
\phi_2
&=&\frac{1}{\sqrt{1+(y^{(+)})^2}}\left(\st[\! \uparrow\downarrow]+
y^{(+)}\st[\! \downarrow \uparrow ]\right),\quad 
\phi_3 = \st[\! \downarrow \downarrow ]
\label{ebasis2}
\end{eqnarray}
with 
\begin{equation}
y^{(\pm)}=(g_s-g_t)\mu_BH/J_{\perp}\pm \sqrt{1+(g_s-g_t)^2\mu_B^2H^2/J_{\perp}^2}.
\end{equation}

In this procedure it is important to note that the leg content, 
${\cal H}_{{\rm leg}}$, equation (\ref{eq:intra}), of the
Hamiltonian is invariant under any change of rung eigenbasis.
It is also invariant under any choice of reference state for the application of the
algebraic Bethe ansatz, while the rung and the field terms are altered by these
changes. 
The validity for reordering the basis arises from the fact
that the intrachain interaction is the permutation operator, which is
independent of the choice of basis for the vector space 
$V = V_1 \oplus V_2$.
On the other hand, the rung interaction and magnetic field vary the
energy levels of the  eigenbasis states
following from equations (\ref{eq:su4basis}) and (\ref{ebasis2}).

Generally speaking, the difference between $g_t$ and $g_s$ is often very
small and even negligible for the isotropic two-leg ladder model, but this may
be different for the spin-obital models \cite{SOtheo1,SOtheo5a,SOtheo5b}.  
For simplicity,  unless otherwise stated, we consider here the isotropic ladder 
model (\ref{eq:Ham2}) for the application of the TBA and HTE methods.  

In section \ref{QTM-NLIE} it is shown that the integrable Hamiltonian (\ref{eq:Ham2}) 
can be derived, up to an unimportant constant, 
by exploiting the spectrum parameter $v$ and the row-to-row transfer matrix 
$t(v)={\rm tr}_0 R_{0,L}(v)R_{0,L-1}(v )\cdots R_{0,2}(v)R_{0,1}(v)$
via the relation  
\begin{equation}
{\cal H}=J_{\parallel}\frac{d}{dv}\ln t
(v)|_{v=0} - J_{\perp}\sum_{j=1}^L \st[1]_j\langle 1|_j -
\mu_B g H \sum_{j=1}^L \left(\st[2]_j\langle 2|_j {\scriptstyle -}\st[4]_j\langle 4|_j\right)
\label{eq:H-T1}
\end{equation}
in terms of the underlying $su(4)$ $R$-matrix, which is simply $R(v)=v I+{\cal P}$,
where ${\cal P}$ is the $16 \times 16$ permutation operator (cf. equation (\ref{eq:R-mat})).
After some algebra, the energy spectrum can be obtained from relation (\ref{eq:H-T1}) 
as \cite{TBAladder1a,TBAladder1b}
\begin{equation}
{\cal E}=-J_{\parallel}\sum^{M_1}_{i=1}
\frac{1}{(v_i^{(1)})^2+\frac{1}{4}}-J_{\perp}N_0-\mu_Bg(N_+-N_-)H
\label{spec1}
\end{equation}
in which $N_0$ denotes the number of rung singlets and $N_+$ and $N_-$
denotes the number of the triplet components $\st[ 2] $ and $\st[ 4]$,
respectively.

The number $M_1$ appearing in the energy spectrum (\ref{spec1}) 
is the number of Bethe roots $v_i^{(1)}$  
which satisfy the $r=3$ case of the general $su(r+1)$ Bethe equations
\begin{equation}
 \prod_{i=1}^{M_{k-1}} \frac{v_j^{(k)}-v_i^{(k-1)}+\ffrac12{\mathrm{i}}}
{v_j^{(k)}-v_i^{(k-1)}-\frac12{\mathrm{i}}}=\prod^{M_k}_{\stackrel{\scriptstyle l=1}{l\neq j}}
\frac{v_j^{(k)}-v_l^{(k)}+\mathrm{i}}{v_j^{(k)}-v_l^{(k)}-\mathrm{i}}
\prod_{l=1}^{M_{k+1}}\frac{v_j^{(k)}-v_l^{(k+1)}-\frac12{\mathrm{i}}}
{v_j^{(k)}-v_l^{(k+1)}+\frac12{\mathrm{i}}}
\label{eq:BE}
\end{equation}
obtained by application of the nested Bethe Ansatz \cite{Yang,Uimin,Lai,Suth}.
Here $k=1, 2, \ldots, r$ and $j=1,\ldots, M_k$, with the conventions
$v_j^{(0)}=v_j^{(r+1)}=0,\, M_{r+1}=0$ and $M_0=L$.

If $g_s \neq g_t$, the corresponding energy is given by \cite{yinga,yingb} 
\begin{eqnarray}
{\cal E} &=&-J_{\parallel}\sum^{M_1}_{i=1}\frac{1}{(v_i^{(1)})^2+\frac{1}{4}}
-\ffrac12 J_{\perp} \left[1+\sqrt{1+(g_s-g_t)^2h^2}\right]N_{\phi_0}\nonumber\\
& &-\ffrac{1}{2}\mu_{{\rm B}}(g_s+g_t)HN_{\phi_1}+\ffrac{1}{2}\mu_{{\rm B}}(g_s+g_t)HN_{\phi_3}\nonumber\\
& &-\ffrac12 J_{\perp} \left[1-\sqrt{1+(g_s-g_t)^2h^2}\right]N_{\phi_2}
\label{spec2}
\end{eqnarray}
with $h=\mu_{{\rm B}}H/J_{\perp}$.
In this case $N_{\phi_a}$ ($a \in \{0, 1, 2, 3 \}$) is the number of $\phi_a$ states.

\subsection{The integrable mixed spin-$(\frac{1}{2},1)$ ladder}
\label{subsec:TBA}

Recalling figure  \ref{fig:Figmix1}, the 
integrable mixed spin-$(\frac{1}{2},1)$ ladder model is defined by the 
Hamiltonian \cite{I-ladder3a,I-ladder3b,mix}
\begin{eqnarray}
~~{\cal H}&=&J_{\parallel}  {\cal H}_{{\rm leg}} +J_{\perp}
\sum_{j=1}^{L}\vec{T}_j \cdot \vec{S}_j -\mu_{{\rm
B}}H\sum_{j=1}^{L}\left( g_tT^z_j+g_sS^z_j\right)
\label{eq:Ham3}
\\
{\cal H}_{{\rm leg}}&=&\sum_{j=1}^{L}\left(\ffrac{1}{2}+2\, \vec{T}_j \cdot \vec{T}_{j+1}\right)
\left(-1+\vec{S}_j \cdot \vec{S}_{j+1}+(\vec{S}_j \cdot \vec{S}_{j+1})^2\right).
\label{eq:intra1}
\label{eq:Ham3all}
\end{eqnarray}
For this model $\vec{T}_j$ remain spin-$\frac{1}{2}$ operators but now 
$\vec{S}_j$ are spin-$1$ operators and we consider general Land\'{e} factors $g_t$ and $g_s$.
Periodic boundary conditions are again imposed and $L$ is the number of rungs.

By construction ${\cal H}_{{\rm leg}}$ defined in (\ref{eq:intra1}) is the permutation operator
with $su(6)$ algebraic symmetry.  
In this case the rung term in (\ref{eq:Ham3}) breaks the $su(6)$ symmetry 
into an $su(4) \oplus su(2)$ symmetry, i.e., quadruplet plus doublet symmetry.
Now changing the fundamental basis  of $su(6)$ 
into rung quadruplet and doublet states (via Clebsch-Gordan
decomposition) results in the six-dimensional state space splitting into the
direct sum of quadruplets and doublets with regard to the rung interaction.
The projectors onto the doublet and quadruplet subspace are given by
\begin{equation} 
{\cal P}_{{\rm d}}=-\ffrac{2}{3}(\vec{T} \cdot \vec{S}-\ffrac{1}{2}), \qquad 
{\cal P}_{{\rm q}}=\ffrac{2}{3}(\vec{T} \cdot \vec{S}+1).
\end{equation}
 
Although the magnetic field preserves the integrability of the leg part of the
Hamiltonian, the different $g$-factors on each leg break the 
doublet/quadruplet basis for the Hamiltonian  (\ref{eq:Ham3}) due to Zeeman splitting.
Fortunately, there is another basis,
\begin{eqnarray}
\psi^{(\pm)}_{\frac{1}{2}}&=&\frac{1}{\sqrt{1+(y^{(\pm)}_{\frac{1}{2}})^2}}
\left(\st[1,-\ffrac{1}{2}]
+y^{(\pm)}_{\frac{1}{2}}\st[0,\ffrac{1}{2}]\right)\nonumber\\
\psi^{(\pm)}_{-\frac{1}{2}}&=&\frac{1}{\sqrt{1+(y^{(\pm)}_{-\frac{1}{2}})^2}}
\left(\st[-1,\ffrac{1}{2}]
+y^{(\pm)}_{-\frac{1}{2}}\st[0,-\ffrac{1}{2}]\right) \\
\psi_{\frac{3}{2}}&=&\st[1,\ffrac{1}{2}]
\nonumber \\
\psi_{-\frac{3}{2}}&=&\st[-1,-\ffrac{1}{2}] \label{eq:mixbasis}
\end{eqnarray}
that diagonalises the rung and magnetization terms simultaneously.
Here the quantities $y^{(\pm)}_{a}$ are given by 
\begin{eqnarray}
 y^{(\pm)}_{a}=a\sqrt{2}[(g_s-g_t)h^{'}+a] \pm
\sqrt{1+\ffrac{1}{2}(g_sh^{'}-g_th^{'}+a)^2}
\end{eqnarray}
where $a=\pm \frac{1}{2}$ and $h^{'}=\mu_B H/J_{\perp}$.
Further
\begin{equation}
\st[1] = \left( \begin{array}{c} 1 \cr 0 \cr 0  \end{array} \right), \,
\st[0] = \left( \begin{array}{c} 0 \cr 1 \cr 0  \end{array} \right), \,
\st[-1] = \left( \begin{array}{c} 0 \cr 0 \cr 1  \end{array} \right), \,
\st[\ffrac12] = \left( \begin{array}{c} 1 \cr 0 \end{array} \right), \,
\st[-\ffrac12] = \left( \begin{array}{c} 0 \cr 1 \end{array} \right) 
\end{equation}
with $\st[b,a] = \st[b] \otimes \st[a]$.
If $g_s=g_t$, the basis states $\psi^{(-)}_{\frac{1}{2}}$
and $\psi^{(-)}_{-\frac{1}{2}}$ reduce to the doublet, with
the other states reducing to the quadruplet.
It follows that  the leg and rung parts of the Hamiltonian (\ref{eq:Ham3all})
are given by the relation 
\begin{equation}
{\cal H}=J_{\parallel}\frac{d}{dv}\ln t (v)|_{v=0}+\sum _{j=1}^{L}\sum_{\alpha,\beta }\mu_{\beta}^{(\alpha) }
\st[ \psi^{(\alpha)}_{\beta}] _j\langle \psi^{(\alpha)}_{\beta}\mid_j + {\rm~constant}
\label{eq:H-T}
\end{equation}
associated with the row-to-row transfer matrix. 
Here $\alpha =+,0,-$ and $\beta =\pm \frac12, \pm \frac32$, with $\psi_\beta^{(0)}=\psi_\beta$. 
The chemical potentials $\mu_{\beta}^{(\alpha) }$ are given in (\ref{eq:energy}) below.

The eigenspectrum is obtained from relation (\ref{eq:H-T}) and the Bethe Ansatz to be \cite{mix}
\begin{equation}
{\cal E}=-J_{\parallel}\sum^{M_1}_{i=1}\frac{1}{{v_i^{(1)}}^2+\frac{1}{4}}+{\cal E}_{\perp+h}
\end{equation}
where the Bethe roots $v_i^{(1)}$ satisfy the Bethe equations (\ref{eq:BE}) with $r=5$.
The energy contribution from the rung interaction and
the magnetic field terms is given  by
\begin{eqnarray}
{\cal E}_{\perp+h} &=&\left[-\ffrac14 J_{\perp} -\ffrac{1}{2} g_s\mu_BH-\ffrac{1}{\sqrt 2} J_{\perp} 
\sqrt{1+\ffrac{1}{2}(g_sh^{'}-g_th^{'}+\ffrac{1}{2})^2}\right]N^{(-)}_{\frac{1}{2}}\nonumber\\
& &
+\left[-\ffrac14 J_{\perp} +\ffrac{1}{2}g_s\mu_BH-\ffrac{1}{\sqrt 2} J_{\perp} 
\sqrt{1+\ffrac{1}{2}(g_sh^{'}-g_th^{'}-\ffrac{1}{2})^2}\right]N^{(-)}_{-\frac{1}{2}}\nonumber\\
& &+\left[\ffrac12 J_{\perp} -(\ffrac{1}{2}g_t+g_s)\mu_BH\right]N_{\frac{3}{2}}\nonumber\\
& &+\left[-\ffrac14 J_{\perp} -\ffrac{1}{2}g_s\mu_BH+\ffrac{1}{\sqrt 2} J_{\perp} 
\sqrt{1+\ffrac{1}{2}(g_sh^{'}-g_th^{'}+\ffrac{1}{2})^2}\right]N^{(+)}_{\frac{1}{2}}\nonumber\\
& &+\left[-\ffrac14 J_{\perp}+\ffrac{1}{2}g_s\mu_BH+\ffrac{1}{\sqrt 2} J_{\perp} 
\sqrt{1+\ffrac{1}{2}(g_sh^{'}-g_th^{'}-\ffrac{1}{2})^2}\right]N^{(+)}_{-\frac{1}{2}}\nonumber\\
& &+\left[\ffrac12 J_{\perp}+(\ffrac{1}{2}g_t+g_s)\mu_BH\right]N_{-\frac{3}{2}}
\label{eq:energy}
\end{eqnarray}
In the above equation, $N^{(\pm)}_{\pm \frac12}$ and $N^{(\pm)}_{\pm \frac32}$
are the occupation numbers of the corresponding states and $h'=\mu_B H/J_{\perp}$.

\section{Thermodynamic Bethe Ansatz}
\label{sec:TBA}

\subsection{Derivation of the TBA equations }

In this section we turn to the study of the ground state properties and the
phase diagram of the integrable two-leg spin-$\frac{1}{2}$ ladder (\ref{eq:Ham2}) 
by means of the Thermodynamic Bethe Ansatz (TBA). 
In the thermodynamic ($L \rightarrow \infty$) limit, the roots of the
Bethe equations for the ground state are real. 
At zero temperature there are no holes in the ground state root distribution.
Yang and Yang \cite{TBA1} introduced the TBA in treating the thermodynamics 
of the one-dimensional boson model with delta function interaction, where the
elementary excitations involve moving quasimomenta out of the Dirac sea.
In general complex conjugate pairs of Bethe roots form bound states of two magnons. 
The string hypothesis has been developed to classify the 
many permissible bound state solutions to the Bethe equations,
originally for  the Heisenberg chain\cite{TBA2},  the
one-dimensional Hubbard model \cite{Takahashi}, 
the supersymmetric $t-J$ model \cite{Schlottmanna, Schlottmannb,EK92,Angela}, 
Kondo problems \cite{Kondo} and 
the higher-spin Heisenberg chain \cite{Babujian} among others.
Recall that the Bethe equations (\ref{eq:BE}) for the ladder model (\ref{eq:Ham2}) 
consist of a set of three coupled equations with Bethe Roots of three different colours, 
$v_i^{(k)}$, $k=1, 2, 3$.
Taking the thermodynamic limit \cite{TBA1,TBA2,Kirillov} results in the string solutions
\begin{eqnarray}
{v^{(k)}}^n_{\alpha_k j}&=&{v^{(k)}}^n_{\alpha_k}+\ffrac12 \mathrm{i} (n+1-2j)
\end{eqnarray}
where $j=1,\ldots,n$ and $\alpha_k =1,\ldots,N_n^{(k)}$.
Here ${v^{(k)}}^n_{\alpha_k},\,k=1, 2, 3$, 
are the positions on the real axis of the centre of an $n$-string, 
where $n$ is the string length in each colour. 
There are $N_n^{(k)}$ strings of the $n$-string form. 
The  number of $n$-string $N_n^{(k)}$  satisfy the relation
$N^{(k)}=\sum_{n=1}^\infty nN_n^{(k)}$,  where $N^{(k)}$
is the total number of Bethe roots of colour $k$.

It is assumed that the distributions of Bethe roots are dense along the real axis. 
The missing roots of each colour $k$ are the holes, with positions  
${v^{(k)h}}^n_{\alpha_k}$ on the real axis.
The distribution densities of $n$-string roots, $\rho^{(k)}_n(v)$, and holes, 
$\rho^{(k)h}_n(v)$, are defined to be
\begin{eqnarray}
\rho^{(k)}_n(v)&=& \lim _{L\rightarrow\infty} \frac{1}{L({v^{(k)}}^n_{\alpha_k +1}-{v^{(k)}}^n_{\alpha_k})}\\
\rho^{(k)h}_n(v)&=& \lim _{L\rightarrow\infty} \frac{1}{L({v^{(k)h}}^n_{\alpha_k +1}-{v^{(k)h}}^n_{\alpha_k})}
\end{eqnarray}
for the three colours $ k=1,2,3$. 

Adopting the standard notation \cite{Babujian}
\begin{eqnarray}
A_{nm}(\lambda)&=& \delta(\lambda)\delta_{nm}+(1-\delta_{nm})a_{|n-m|}(\lambda)
+a_{n+m}(\lambda)\nonumber \\
& &+\,\, 2\sum^{{\rm Min}(n,m)-1}_{l=1}a_{|n-m|+2l}(\lambda)\\
a_{nm}(\lambda)& =& \sum^{{\rm Min}(n,m)}_{l=1}a_{n+m+1-2l}(\lambda)
\end{eqnarray}
with $a_n(\lambda)=\frac{1}{2\pi}\frac{n}{n^2/4+\lambda ^2}$, the
Bethe equations (\ref{eq:BE}) can be transformed after a long though standard 
calculation into the form
\begin{eqnarray}
\rho^{(1)h}_n(v)&=&a_n(v) -\sum_{m}(A_{nm}*\rho^{(1)}_m)(v)
+\sum_{m}(a_{nm}*\rho^{(2)}_m)(v)\label{eq:1st}\\
\rho^{(2)h}_n(v)&=&-\sum_{m}(A_{nm}*\rho^{(2)}_m)(v)+\sum_{m}(a_{nm}*(\rho^{(1)}_m+\rho^{(3)}_m))(v)\\
\rho^{(3)h}_n(v)&=&-\sum_{m}(A_{nm}*\rho^{(3)}_m)(v)+\sum_{m}(a_{nm}*\rho^{(2)}_m)(v)
\label{eq:Tbethe1}
\end{eqnarray}
where $*$ denotes the convolution integral
$(f*g)(\lambda) = \int_{-\infty}^\infty f(\lambda-\lambda') g(\lambda') d\lambda'$.

In order to find the equilibrium state of the system at a fixed
temperature $T$ and external magnetic field, the free energy $F=E-TS-\mu_{{\rm B}}g M^z$ 
is minimised with respect to the densities. 
To this end, the energy per site can be written as
\begin{equation}
{{\cal E}}/{L}=-J_{\parallel}\sum_{n=1}^{\infty}\int_{-\infty}^{\infty}dv \,a_n(v) \rho_n^{(1)}(v)-J_{\perp}N_{0}
\end{equation}
in terms of the densities of roots and holes.
Following \cite{TBA1} the total entropy of the system is given by
\begin{eqnarray}
S&=&\sum_{n=1}^{\infty}\int_{-\infty}^{\infty}dv\left(\sum_{k=1}^3(\rho_n^{(k)}(v)+
\rho^{(k)h}_n(v))\ln(\rho_n^{(k)}(v)+\rho^{(k)h}_n(v))\right.\nonumber\\
& &
\left.-\rho_n^{(k)}(v)\ln\rho_n^{(k)}(v)-\rho^{(k)h}_n(v)\ln\rho^{(k)h}_n(v)\right)
\end{eqnarray}
with magnetization
\begin{equation}
M^z=\mu_{{\rm B}}g(N_+-N_-).
\end{equation}

Minimizing the free energy, i.e., $\delta F=0$, and making use of the equations (\ref{eq:1st})-(\ref{eq:Tbethe1}) gives
the TBA equations
\begin{eqnarray}
\ln (1+\eta _n^{(k)}(v)) &=&
\frac{1}{T}g^{(k)}_n(v)+\sum_{m=1}^{\infty}A_{nm}*\ln(1+{\eta _m^{(k)}}^{-1})(v)\nonumber\\
&&-\sum_{m=1}^{\infty}a_{nm}*\ln(1+{\eta _m^{(k-1)}}^{-1})(v)\nonumber\\
&&-\sum_{m=1}^{\infty}a_{nm}*\ln(1+{\eta _m^{(k+1)}}^{-1})(v)
\label{eq:tba}
\end{eqnarray}
in terms of the functions $\eta_n^{(k)}(v)=\rho^{(k)h}_n(v)/\rho_n^{(k)}(v)$. 
Henceforth for notational simplicity we will omit the argument $v$.
The driving terms $g^{(k)}_n(v)=nx^{(k)}-\delta_{1,k}2\pi a_n(v)$ appearing in the TBA equations
(\ref{eq:tba}) play a key role to be discussed further below. 
Here $x^{(k)}$ is the chemical potential for the colour $k$-branch.

The TBA equations are often written in terms of the dressed energies 
$\epsilon^{(k)}_n(v)$, defined via 
$\eta _n^{(k)}(v) =\exp(\epsilon^{(k)}_n(v)/T)$ for $k=1, 2, 3$ with 
$\epsilon^{(0)}_n(v)=\epsilon^{(4)}_n(v)=0$.  
Specifically, 
\begin{eqnarray}
\epsilon^{(k)}_1&=&g_1^{(k)}+Ta_2*\ln(1+\e^{-{\epsilon^{(k)}_1}/{T}})+
T(a_0+a_2)\sum_{m=1}^{\infty}a_m*\ln(1+\e^{-{\epsilon^{(k)}_{m+1}}/{T}})\nonumber\\
&&-T\sum_{m=1}^{\infty}a_m*\left(\ln(1+\e^{-{\epsilon^{(k-1)}_m}/{T}})+
\ln(1+\e^{-{\epsilon^{(k+1)}_m}/{T}})\right)
\label{eq:TBAe1}
\end{eqnarray}
with
\begin{eqnarray}
\epsilon^{(k)}_n&=&g_n^{(k)}+Ta_1*\ln(1+\e^{-{\epsilon^{(k)}_{n-1}}/{T}}+
Ta_2*\ln(1+\e^{-{\epsilon^{(k)}_n}/{T}}))\nonumber\\
& &+T(a_0+a_2)\sum_{m=n}^{\infty}a_{m-n}*\ln(1+\e^{-{\epsilon^{(k)}_m}/{T}})  \nonumber \\
& &-T\sum_{m=n}^{\infty}a_{m-n+1}*\left(\ln(1+\e^{-{\epsilon^{(k-1)}_m}/{T}})+
\ln(1+\e^{-{\epsilon^{(k+1)}_m}/{T}})\right)
\label{eq:TBAe2}  
\end{eqnarray}
for $n  > 1$.
The dressed energies play the role of
excitation energies measured from the Fermi energy level for each colour, 
and satisfy a Fermi-Dirac-like distribution. 
This can be seen from the ratio of the number of occupied vacancies to the total number of 
vacancies in the interval $v $ to $v +dv$ for colour $k$.
Namely
\begin{equation}
\frac{\rho^{(k)}_n(v)}{\rho^{(k)}_n(v)+\rho^{(k)h}_n(v)}=
\frac{1}{1+\e^{\epsilon^{(k)}_n(v)}}.
\end{equation}
The dressed energies $\epsilon ^{(k)}_n(v)$ for $k=1,2,3$ are monotonically
increasing functions of $|v|$ and even, i.e., $\epsilon
^{(k)}_n(v)=\epsilon ^{(k)}_n(-v)$ with 
$$
\lim_{n\to\infty}\frac{\epsilon_n^{(k)}(v)}{n}=2x^{(k)}.
$$

In the above equations the driving terms $g^{(k)}_1$ and $g^{(k)}_n$
depend on the choice of the basis order due to the energy level
splitting among the singlet and the triplet components with respect to the
competition between the rung interaction and the magnetic field, see
(\ref{spec1}) and (\ref{spec2}).
It is worth noting that for the ladder Hamiltonian (\ref{eq:Ham2}) the singlet rung state is
energetically favoured when $J_{\perp} > 0$ (antiferromagnetic coupling), while for  
$J_{\perp} < 0$ (ferromagnetic coupling) the triplet rung state is energetically more favourable. 
Here we consider only the antiferromagnetic coupling regime, i.e., with 
$J_{\parallel} > 0$ and $J_{\perp} > 0$.
Although the physics of the model doesn't depend on the basis order,
the driving terms vary the Fermi surfaces of each branch in different ways. 
It is most convenient to order the basis according to the energy levels of the 
singlet and triplet components which provides a natural order with respect to the
band filling.  
For strong rung coupling there are only two components, 
the singlet and the lowest component of the triplet, which compete in the ground state. 
The other components induce a gap above the ground state.
In this natural basis order, the TBA equations for the
ground state reduce to one band with finite Fermi points.

Explicitly, in the strong coupling regime,
i.e., $J_{\perp}\geq 4 J_{\parallel}$ and  $J_{\perp}>H$, 
the basis order is chosen as $(\st[ 1], \st[ 2 ], \st[ 3 ], \st[ 4 ])$, 
corresponding to the Bethe states with number $N_1$ of state $\st[ 1]$, 
up to number $N_4$ of state $\st[ 4]$.
Here $N_1=L - M_1, N_2 = M_1 - M_2, N_3 = M_2 - M_3$ and $N_4 = M_3$ (see equation (\ref{eq:BE})).
The driving terms are thus given by
\begin{equation}
g^{(1)}_1=-J_{\parallel}2\pi a_1+J_{\perp}-\mu_{{\rm B}}gH,\quad
g^{(2)}_1= g^{(3)}_1=\mu_{{\rm B}}gH.
\label{eq:basis1}
\end{equation}
The higher order driving terms are $g_n^{(1)}=n\mu_{{\rm B}}gH$,
$g_n^{(2)}=ng_1^{(2)}$ and $g_n^{(3)}=ng_1^{(3)}$. 
Subsequently, up to a constant, the free energy is 
\begin{equation}
f(T,H)=-T\int_{-\infty}^{\infty}\sum_{n=1}^{\infty}a_n(v)\ln(1+\e^{-{\epsilon^{(1)}_n(v)}/{T}})dv.  
\label{eq:FE1}
\end{equation}
However, if $H>J_{\perp}$, the basis order can be taken to be
$(\st[ 2], \st[ 1 ], \st[ 3 ], \st[ 4 ])$ for a simpler analysis of
the TBA equations.  The driving terms are then
\begin{equation}
g^{(1)}_1=-J_{\parallel}2\pi a_1- J_{\perp}+\mu_{{\rm B}}gH,\quad
g^{(2)}_1=  J_{\perp} ,\quad      g^{(3)}_1  =\mu_{{\rm B}}gH
\label{eq:basis2}
\end{equation}
with the higher order driving terms remaining unchanged.
In this case the free energy is 
\begin{equation}
f(T,H)=-\mu_{{\rm B}}gH-T\int_{-\infty}^{\infty}\sum_{n=1}^{\infty}a_n(v)
\ln(1+\e^{-{\epsilon^{(1)}_n(v)}/{T}})dv.  \label{eq:FE2}
\end{equation}

\subsection{Ground state properties and phase diagram}

We turn now to the ground state properties and the phase diagram of the
integrable ladder model, as revealed by the analysis of the TBA equations.
In the low-temperature limit $T\rightarrow 0$ only the negative part of the 
dressed energies $\epsilon^{(l)}$, denoted by $\epsilon^{(l)-}=\min \{ \epsilon^{(l)}, 0 \}$,  
contribute to the ground-state energy. 
Here we use the simplified notation $\epsilon^{(l)} = \epsilon^{(l)}_1$.
The TBA equations (\ref{eq:TBAe1}) then become
\begin{eqnarray}
\epsilon^{(1)}&=&g_1^{(1)}-a_2*\epsilon^{(1)-}+a_1*\epsilon^{(2)-}\nonumber\\
\epsilon^{(2)}&=&g_1^{(2)}-a_2*\epsilon^{(2)-}+a_1*\left[\epsilon^{(1)-}+\epsilon^{(3)-}\right]
\label{eq:TBA1}\\
\epsilon^{(3)}&=&g_1^{(3)}-a_2*\epsilon^{(3)-}+a_1*\epsilon^{(2)-}\nonumber
\end{eqnarray}
where $g_1^{(a)},\,a=1,2,3$, are the driving terms with respect to the
different basis orders given in  (\ref{eq:basis1}) and (\ref{eq:basis2}).

\begin{figure}[t]
\begin{center}	
\includegraphics[width=.80\linewidth]{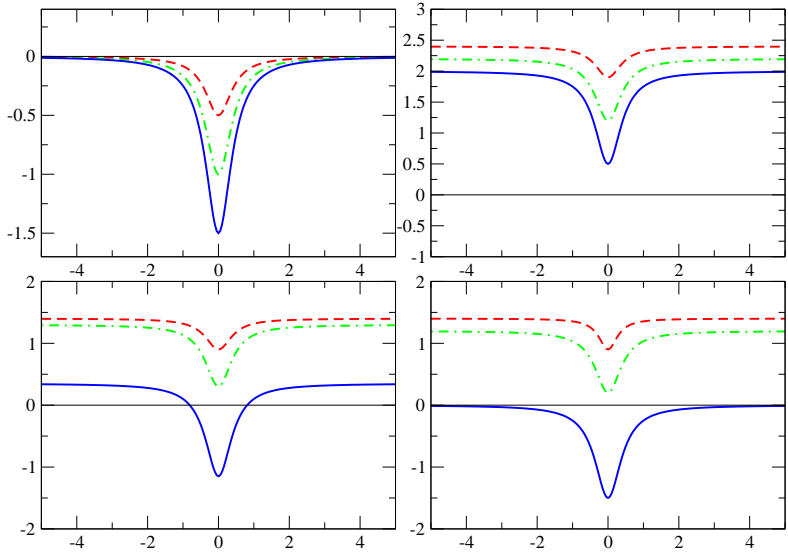}
\end{center}
\caption{Depiction of the dressed energies for the $su(4)$ TBA equations (\ref{eq:TBA1}).
The four panels indicate the quantum phase transitions which occur in the ground state 
of the integrable ladder model (\ref{eq:Ham2}) with strong rung coupling.
The horizontal axis is the spectral parameter $v^{(k)}$ with the origin ($v^{(k)}=0$) at the minima. 
%
%
The upper left panel shows the dressed energies in the absence of the symmetry-breaking 
chemical potential.  
In the presence of the chemical potential, the basis is reordered so that
the Fermi surfaces of the dressed energies remain in the order $\epsilon^{(1)}$ lowest, 
$\epsilon^{(2)}$ second lowest and $\epsilon^{(3)}$ highest (see text).}
\label{fig:TBA-DE}
\end{figure}

In the absence of the symmetry-breaking chemical potential and
magnetic field terms, the Fermi seas of the three bands are completely filled
and have no Fermi surface, as depicted in upper left panel of figure \ref{fig:TBA-DE}. 
The Fermi sea of each band may be raised or lowered by the rung interaction
and the external magnetic field, with spinon excitations created by flipping a spin.
The chemical potential, rung interaction and magnetic field terms vary the  
Fermi surfaces at the bands, leading to  interference effects between the two Fermi points.

In the antiferromagnetic regime ($J_{\perp}>0$) it follows that if $H=0$ the triplet
excitation is massive, with the gap given by $\Delta =J_{\perp}-4J_{\parallel}$. 
This configuration is depicted graphically in the upper right panel of figure \ref{fig:TBA-DE}, 
where all of the dressed energies are gapful. 
The critical point $J^+_c=4J_{\parallel}$ indicates a quantum phase
transition from the three branches of the Luttinger liquid phase to
the dimerised $u(1)$ phase. 
The magnetic field lifts the fermi seas of $\epsilon^{(2)}$ and $\epsilon^{(3)}$,
which can be seen from the TBA equations (\ref{eq:TBA1}).

For $J_{\perp} >J_c^+$, it can be shown that the two components of the
triplet states, $\st[3] $ and $\st[4]$, do not enter into the
groundstate for high magnetic field. 
However, the magnetic field shifts the energy level of the triplet component of $\st[2]$ 
closer to the singlet groundstate.  
The gap can be deduced via the magnetic field $H$.
The first critical field occurs at $H_{c1}$, where $g\mu_BH_{c1}=\Delta$, 
i.e., the effect of the magnetic field is to close the gap. 
The quantum phase transition is from a gapped to a gapless Luttinger phase.
When $H > H_{c1}$ the  triplet component of $\st[2]$ begins to enter into the 
ground state with a finite susceptibility. 
The dressed energy configuration for this case is depicted in the lower left 
panel of figure \ref{fig:TBA-DE}. 
The dressed energies $\epsilon^{(2)}$ and $\epsilon^{(3)}$ are gapful,
whereas $\epsilon^{(1)}$ has two Fermi points. 
The excitation energies are proportional to the momentum of particles.  
If $H > H_{{\rm IP}}=J_{\perp}/\mu_{{\rm B}}g$, the triplet component of $\st[2]$  
becomes the lowest energy level with driving terms (\ref{eq:basis2}). 
At the point $H=H_{{\rm IP}}$, the magnetic field completely diminishes
the rung interaction so that the Fermi sea of the dressed energy
$\epsilon^{(1)}$ is completely filled, whereas the others are still gapful 
(see lower right panel of figure \ref{fig:TBA-DE}).  
In this way we see that the ground state is in a fully-polarized ferromagnetic 
state when the magnetic field is greater than $H_{c2}=J_{\perp}+4J_{\parallel}$.

The TBA predictions for the occurence of critical fields $H_{c1}$ and $H_{c2}$,
along with their values, are in good agreement with the experimental
results for several strong coupling ladder compounds \cite{TBAladder1a,TBAladder1b}, 
as will be discussed further in sections  \ref{sec:mag} and \ref{sec:Exp_comp}.

\subsection{Analytical and numerical solutions of the TBA equations for the magnetization}
\label{sec:mag}

The TBA equations (\ref{eq:TBA1}) may be solved 
numerically \cite{TBAladder1a,TBAladder1b,yinga,yingb,ying2,mix}
to study the magnetization of the integrable spin ladder model in detail.
As a concrete example we take the strong coupling
compound ($5$IAP)$_2$CuBr$_4 \cdot 2$H$_2$O \cite{5IAP} for which the 
coupling constants can be used in the numerical calculations. 
This compound was recognized as a strong coupling ladder compound with 
the critical field values $H_{c1}\approx 8.3$T and $H_{c2}\approx 10.4$T. 
For the integrable spin ladder model (\ref{eq:Ham2}) the critical
field values are $H_{c1}=J_{\perp}-4J_{\parallel}$ and $H_{c2}=J_{\perp}+4J_{\parallel}$.  
In this way reasonable values for the coupling constants $J_{\parallel}$ and $J_{\perp}$ 
can be suggested from the experimental critical field values. 
The $g$-factors are uniquely determined by experiment and are not 
free parameters in this approach.  
However, the coupling constants fixed by the critical fields only lead to a good fit 
with the energy gap. 
In order to make a good overall fit for each of the key thermal and magnetic properties, the
coupling constants should be fixed by the energy gap, the magnetization and the susceptibility. 
Once the coupling constants are fixed, all thermal and magnetic properties follow naturally.
For this compound a fit to the experimental critical fields and the magnetization at different
temperatures suggests the coupling constants $J_{\perp}=13.3\,$K and
$J_{\parallel}=0.2875\,$K, where we have used  $g=2.1$.
The leg coupling constant $J_{\parallel}$ is smaller than for the pure Heisenberg model leg
interaction due to the presence of the biquadratic term, which strengthens the overall leg interaction.
Recall that the leg interaction (\ref{eq:intra}) differs from that for the pure Heisenberg ladder model.
As a result the leg coupling constant $J_{\parallel}$ for compounds determined from the 
integrable model (\ref{eq:intra}) differs from that determined from the Heisenberg ladder model
(\ref{eq:Ham1}). 
In order to compare the difference between $J_{\parallel}$ determined by these two theoretical  
models, a scaling parameter $\gamma$ was introduced \cite{TBAladder1a,TBAladder1b},
where ${J_{\parallel}}/{\gamma}$ characterizes the interaction
strength along the legs for the integrable spin ladder.

{}From the above analysis, when $H\leq H_{c1} =8.6\,$T the ground state is the product of rung
singlets such that the magnetization $M^z=0$. 
If $ H_{c1} < H <H_{{IP}}=9.42\,$T, the dressed energies $\epsilon^{(2)}$ and
$\epsilon^{(3)}$ are both gapful, so that the magnetization is given by
\begin{equation}
M^z=\int_{-Q}^{Q}\rho^{(1)}(v)dv
\end{equation}
where
\begin{equation}
\rho^{(1)}(v)=\frac{1}{2\pi}\frac{1}{v^2+\frac{1}{4}}-\frac{1}{2\pi}\int _{-Q}^{Q}
\frac{2}{1+(v-k)^2}\rho^{(1)}(k)dk.\label{eq:Qbethe}
\end{equation}
In the above equations $Q$ denotes the Fermi boundary determined from 
\begin{equation}
\epsilon ^{(1)}(v)=-2\pi J_{\parallel}a_1(v)+J_{\perp}-\mu_{{\rm B}}gH-
\frac{1}{2\pi}\int _{-Q}^{Q}\frac{2}{1+(v-k)^2}\epsilon^{(1)-}(k)dk. \label{eq:Q1}
\end{equation}
In the region where the magnetic field is greater than $9.42\,$T but less than $H_{c2}=10.2\,$T, 
the Fermi boundary is determined by 
\begin{equation}
\epsilon ^{(1)}(v)=-2\pi J_{\parallel}a_1(v)-J_{\perp}+\mu_{{\rm B}}gH-
\frac{1}{2\pi}\int _{-Q}^{Q}\frac{2}{1+(v-k)^2}\epsilon^{(1)-}(k)dk \label{eq:Q2}
\end{equation}
and the magnetization is 
\begin{equation}
M^z=1-\int_{-Q}^{Q}\rho^{(1)}(v)dv.
\end{equation}

\begin{figure}
\vskip 5mm
\centerline{\includegraphics[width=.70\linewidth]{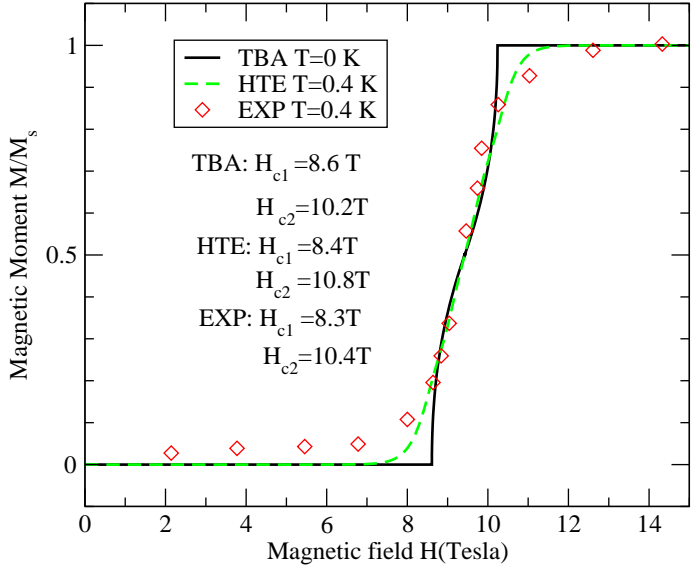}}
\caption{
Comparison between theory and experiment for magnetization versus 
magnetic field $H$ in units of saturation magnetization for the strong coupling ladder
compound ($5$IAP)$_2$CuBr$_4 \cdot 2$H$_2$O \cite{5IAP}.  
The theoretical curves are obtained from the integrable model with the
values $J_{\perp}=13.3\,$K, $J_{\parallel}=0.2875\,$K and $g=2.1$.
The full curve is obtained from the TBA equations at zero temperature.
For further comparison, the dashed curve is obtained using the HTE at $T=0.4\,$K,
as outlined in section \ref{sec:Exp_comp}.
The experimental magnetization at $T=0.4\,$K (diamonds) is shown for comparison.
The discrepancy in the curves at $T=0.4\,$K is due to paramagnetic impurities, 
which result in nonzero magnetization in the singlet ground state.  
At the inflection point, $H=J_{\perp}/\mu_{{\rm B}}g$, the magnetization is $0.5$.  
Both the TBA and HTE curves indicate values of the critical fields 
in excellent agreement with the experimental results.
}
\label{fig:B5i2aTSZ}
\end{figure}

The numerical solution of these equations gives a reasonable
magnetization curve (see figure \ref{fig:B5i2aTSZ}) which passes through an
inflection point midway between $H_{c1}$ and $H_{c2}$.  
The inflection point has a physical meaning in that the probabilities of the
singlet $\st[1 ]$ and the triplet state $\st[2 ]$ in the ground state are equal.  
At zero temperature, in the strong coupling regime, the one
point correlation function, $\langle S_j \cdot T_j\rangle =-\frac34$, 
lies in a gapped singlet groundstate, which indicates an ordered dimer
phase, while $\langle S_j\cdot T_j\rangle =\frac14$  in the
fully-polarized ferromagnetic phase.  
However, in the Luttinger liquid phase, 
$\langle S_j\cdot T_j\rangle =-\frac34+M^z$.  
This indicates that the magnetic field increases the value of the one point 
correlation function.

\subsection{Small Fermi boundary expansion method in the vicinity of the critical fields}

It was shown \cite{TBAladder1a,TBAladder1b} that the magnetization in the
vicinity of the critical fields $H_{c1}$ and $H_{c2}$ depends on the
square root of the field, indicating a Pokrovsky-Talapov transition.
An analytic expression for the field-dependent magnetization in the
vicinity of the critical points $H_{c1}$ and $H_{c2}$ can be obtained
by a small $Q$ approximation. 
The lower left panel of figure \ref{fig:TBA-DE} shows the dressed energies 
$\epsilon^{(k)},\,k=1,2,3$ in the vicinity of the critical point $H_{c1}$.

Performing a small $Q$ approximation in (\ref{eq:Q1}) gives
\begin{eqnarray}
\epsilon ^{(1)}(v)&=&-2\pi J_{\parallel}a_1(v)+J_{\perp}-\mu_{{\rm B}}gH-
\frac{1}{2\pi}\int _{-Q}^{Q}\frac{2}{1+(v-k)^2}\epsilon^{(1)-}(k)dk \nonumber\\
&\approx & -\frac{4J_{\parallel}}{1+4v^2}+J_{{\rm eff}}-
\frac{2Q}{\pi}\frac{1}{1+v^2}\epsilon^{(1)}(0)
\label{eq:Qapprox}
\end{eqnarray}
with $J_{{\rm eff}}=J_{\perp}-\mu_{{\rm B}}gH$. 
It follows that
 \begin{equation}
\epsilon^{(1)}(0)\approx \frac{(J_{{\rm eff}}-4J_{\parallel})\pi}{\pi+2Q}=
-\frac{\mu_{{\rm B}}g(H-H_{c1})\pi}{\pi+2Q}
\end{equation}
with $H_{c1}=(J_{\perp}-4J_{\parallel})/(\mu_Bg)$.
The condition $\epsilon^{(1)}(Q)=0$ then gives
\begin{equation}
Q\approx \sqrt{\frac{\mu_{{\rm B}}g(H-H_{c1})}{16J_{\parallel}-5\mu_{{\rm B}}g(H-H_{c1})}}.
\end{equation}
Similarly, equation (\ref{eq:Qbethe}) becomes
\begin{equation}
\rho^{(1)}(v)\approx \frac{1}{2\pi }\frac{1}{v^2+1/4}-\frac{4Q}{\pi(1+v^2)}\rho^{(1)}(0)
\end{equation}
with  $\rho^{(1)}(0)=\frac{2}{\pi+4Q}$. 
The magnetization for small $Q$ can then be obtained as
\begin{eqnarray}
M^z =\int_{-Q}^{Q}\rho^{(1)}(v)dv\approx \frac{4Q}{\pi}\left(1-\frac{2Q}{\pi}\right).
\end{eqnarray}
It follows that the magnetization in the vicinity of $H_{c1}$ has a square root field-dependent 
critical behaviour of the form
\begin{eqnarray}
M^z \approx \frac{1}{\pi}\sqrt{{\mu_{{\rm B}}g(H-H_{c1})}/{J_{\parallel}}}.\label{TBAmag1}
\end{eqnarray}

The thermodynamics may be calculated from the free energy (\ref{eq:FE1}) and (\ref{eq:FE2}). 
To demonstrate this, we look at the free energy in the vicinity of $H_{c2}$ via the 
TBA equations (\ref{eq:TBA1}). 
When $H>J_{\perp}/\mu_Bg$, the lowest energy level $\st[ 2 ]$ dominates other states as the 
magnetic field increases. 
The basis order is chosen to be $(\st[2],\st[1 ],\st[3 ],  \st[4 ])$. 
In fact, a Luttinger liquid phase lies in the regime $H_{c1}<H<H_{c2}$, 
where the ground state can be described by an integrable $XXZ$ model
with an effective field $H-J_{\perp}/\mu_Bg$, see (\ref{eq:Q2}). 
Similarly, in the vicinity of $H_{c2}$, from (\ref{eq:Q2}), we have 
\begin{eqnarray}
\epsilon^{(1)}(0)&\approx &-\frac{(J_{{\rm eff}}+4J_{\parallel})\pi}{\pi+2Q}=
-\frac{\mu_{{\rm B}}g(H_{c2}-H)\pi}{\pi+2Q}\\
~~~~Q &\approx & \sqrt{\frac{\mu_{{\rm B}}g(H_{c2}-H)}{16J_{\parallel}-5\mu_{{\rm B}}g(H_{c2}-H)}}
\end{eqnarray}
with $H_{c2}=(J_{\perp}+4J_{\parallel})/\mu_Bg$. 
Here the Fermi boundary $Q$ is very small for $H$ near $H_{c2}$.
Subsequently, the free energy can be derived from (\ref{eq:FE2}) as
\begin{eqnarray}
 F(0,H)&=&-\mu_BgH-T\int_{-\infty}^{\infty}a_1(v)\ln(1+\e^{-{\epsilon^{(1)}_1(v)}/{T}})dv \nonumber \\
& = & -\mu_BgH+\int_{-\infty}^{\infty}a_1(v)\epsilon^{(1)-}_1(v)dv
\nonumber \\
& = & -\mu_BgH+\int_{-Q}^{Q}a_1(v)\epsilon^{(1)}_1(v)dv
\nonumber \\
& = & -\mu_BgH+\int_{-Q}^{Q}a_1(v)\left[-2\pi J_{\parallel}a_1(v)+J_{{\rm eff}}^{'} -\frac{2Q\epsilon^{(1)}_1(0)}{\pi}\frac{1}{1+v^2} \right]dv\nonumber \\
& = & -\mu_BgH -\frac{4Q\mu_Bg(H_{c2}-H)}{\pi }\left(1+\frac{2Q}{\pi}\right).
\end{eqnarray}
For very small $Q$ the magnetization is given by
\begin{equation}
M^z \approx \frac{F(0,H_{c2})-F(0,H)}{M_s(H_{c2}-H)}
\end{equation}
with $M_s=\mu_Bg$. 
Thus
\begin{equation}
M^z\approx 1-\frac{1}{\pi}\sqrt{{\mu_{{\rm B}}g(H_{c2}-H)}/{J_{\parallel}}}.\label{TBAmag2}
\end{equation}

The results (\ref{TBAmag1}) and (\ref{TBAmag2}) demonstrate the square root critical behaviour.
They coincide with the numerical results, as applied to the strong coupling ladder 
compound ($5$IAP)$_2$CuBr$_4 \cdot 2$H$_2$O in figure \ref{fig:B5i2aTSZ}.

\subsection{Wiener-Hopf method}
\label{sec:W-H}

The Wiener-Hopf technique has been widely used to solve functional
equations which are obtained by Fourier transformation of certain
integral equations \cite{Kondo,Schlottmanna,Schlottmannb,W-Hbook}. 
In this subsection we demonstrate how to analytically solve the 
TBA equations with a large Fermi surface. 

The higher energy levels of the triplet states $\st[ 3]$ and $\st[ 4]$ 
are always supressed in the ground state for strong rung coupling
($J_{\perp}>4J_{\parallel}$). 
The spin waves are hard-core bosons with massless excitations.  
In the regime $H_{c1}<H<H_{c2}$, the ground state is that of the
integrable $XXZ$ chain with an effective field $H-J_{\perp}/\mu_Bg$. 
This field triggers two pure states, namely $\st[ 0]$ for $H<H_{c1}$ and 
$\st[ 2]$ for $H>H_{c2}$, which effectively correspond to the states 
$\st[\!\! \downarrow ]$ and $\st[\!\! \uparrow ]$ of the $XXZ$ model, respectively. 
The inflection point $H=J_{\perp}/\mu_Bg$ may be considered as the
$su(2)$ point because at this point $J_{\perp}$ is diminished by
external fields. It reduces to the TBA equation withought any chemical
potential for $XXX$ model where the ground state is the $su(2)$
Luttinger liquid.
These two components are degenerate and fully fill the energy band $\epsilon ^{(1)}$. 
All other bands are empty (as per the lower right panel of figure \ref{fig:TBA-DE}).  
When the magnetic field is greater than $J_{\perp}$, i.e., when $H>\frac{J_{\perp}}{\mu_Bg}$, 
the Zeeman energy introduces a large Fermi surface of the dressed energy $\epsilon^{(1)}$ 
with regard to the basis order $(\st[ 2], \st[ 1 ], \st[ 3 ], \st[ 4 ])$. 
Logarithmic corrections are associated with the interference effects between the two 
Fermi boundaries.  

The integral equations (\ref{eq:Q1}) and (\ref{eq:Q2}) exhibit a large Fermi boundary  
for $H$ in the vicinity of ${J_{\perp}}/{\mu_Bg}$ and can be solved by the Wiener-Hopf method. 
We thus consider the case of $H$ in the vicinity of ${J_{\perp}}/{\mu_Bg}$ such that the
ground state is described by the $su(2)$ TBA equation with an effective field.  
With $J_{{\rm eff}}=-J_{\perp}+\mu_BgH$,  
equation (\ref{eq:Q2}) can then be rewritten as
\begin{equation}
\epsilon^{(1)}(v)=-J_{\parallel}2\pi a_1(v)+J_{{\rm eff}}-(a_2*\epsilon^{(1)-})(v).
\end{equation}
Taking Fourier transforms, defined by 
$\tilde{f}(\omega) = \int_{-\infty}^\infty f(\lambda) \e^{{\rm i}\, \omega \lambda} d\lambda$, gives
\begin{equation}
\tilde{\epsilon}^{(1)}(\omega )=-J_{\parallel}2\pi \frac{\e^{-\frac{|\omega|}{2}}}{1+\e^{-|\omega|}}
+\ffrac12 J_{{\rm eff}} \delta(\omega)+\frac{\e^{-|\omega|}}{1+\e^{-|\omega|}}\tilde{\epsilon}^{(1)+}(\omega).
\label{eq:W-H1}
\end{equation} 
Define the function $G(\lambda)$ by
\begin{eqnarray}
G(\lambda)
=\frac{1}{2\pi}\int_{-\infty}^{\infty}\frac{\e^{-|\omega|/2}}{2\cosh \frac{\omega}{2}}
\e^{-{\mathrm i}\lambda \omega}d\omega.\nonumber
\end{eqnarray}
Then the inverse Fourier transform indicates that the   
dressed energy satisfies a Wiener-Hopf type equation of the form
\begin{equation}
\epsilon^{(1)}(\lambda)=-J_{\parallel}\frac{\pi }{\cosh \pi
\lambda}+\ffrac12 J_{{\rm eff}}+\left( \int_{-\infty}^{-B}+\int_{B}^{\infty}\right)
(G*\epsilon^{(1)})(k)dk.\label{eq:W-H2}
\end{equation}
Because $J_{{\rm eff}}$ is very small, the Fermi boundary is very large. 
Let $\lambda \rightarrow \lambda+B$ and $y(\lambda) = \epsilon^{(1)}(\lambda+B)$, 
then equation (\ref{eq:W-H2}) becomes
\begin{eqnarray}
y(\lambda)&=& -J_{\parallel}\frac{\pi }{\cosh \pi
(\lambda+B)}+\ffrac12 J_{{\rm eff}}+\int_{0}^{\infty}G(\lambda +2B-k)y(k)dk\nonumber\\
& &+\int_{0}^{\infty}G(\lambda -k)y(k)dk.
\end{eqnarray}
Now $G(\lambda +2B-k)\rightarrow \e^{-B}$ as $B \gg 1$. 
The driving term can be expanded in powers of $\e^{-B}$.
We can thus take an expansion $y=y_1+y_2+y_3+\cdots $ with respect to the order of 
$\e^{-mB}, \, m=1,2, \ldots $. 
The leading term $y_1$ describes interference between the two Fermi points. 
The energy of the excitation is proportional to the momentum.  
Subsequently, we have 
\begin{eqnarray}
y_1(\lambda)&=& -J_{\parallel}2\pi \, \e^{-(\lambda+B)\pi}+\ffrac12 J_{{\rm eff}}+
\int_{0}^{\infty}G(\lambda -k)y_1(k)dk\label{eq:W-Hkernel}\\
y_2(\lambda)&=&\int_{0}^{\infty}G(\lambda -k)y_2(k)dk+\int_{0}^{\infty}G(\lambda +2B-k)y_1(k)dk\\
y_3(\lambda)&=&\int_{0}^{\infty}G(\lambda -k)y_3(k)dk+\int_{0}^{\infty}G(\lambda +2B-k)y_2(k)dk
\end{eqnarray}
and so on.

Each of these equations is of Wiener-Hopf type and can be solved in an analytic fashion.  
Here we content ourselves with the leading term, i.e., considering the leading pole in the 
decomposed functions. 
The remaining higher terms $y_n$ can be obtained by iteration. 
Introduce the functions $y_{\pm }(\lambda)$ by
\begin{equation}
y_1^+(\lambda)=\left\{\begin{array}{ll}
y(\lambda),&\lambda >0\\
0, & \lambda < 0,
\end{array}\right. \qquad
y_1^-(\lambda)=\left\{\begin{array}{ll}
y(\lambda),&\lambda <0\\
0,& \lambda > 0.
\end{array}\right. \nonumber
\end{equation}
Taking the  Fourier decomposition, we have
$\tilde{y}_1(\omega )=\tilde{y}_1^+(\omega)+\tilde{y}_1^-(\omega)$,
where $\tilde{y}_1^{\pm}(\omega)$ are analytic in the upper and
lower halves of the complex plane, respectively. 
In order to solve the Wiener-Hopf equations, we need to find a decomposition 
of the kernel and the driving term in (\ref{eq:W-Hkernel}). 
To this end, we denote the driving term by
\begin{equation}
\phi (\lambda )= -J_{\parallel}2\pi \, \e^{-(\lambda+B)\pi}+\ffrac12 J_{{\rm eff}} 
\end{equation}
and  define Fourier transform for driving term
\begin{eqnarray}
& &\phi_+ (\lambda )=\left\{
\begin{array}{ll}0, & \,\,\,\lambda <0 \\
\frac{1}{2\pi }\oint \e^{-\mathrm{i}\omega \lambda} \tilde{\phi}_+(\omega)d\omega 
=\phi (\lambda ), & \,\,\, \lambda >0
\end{array}\right. \nonumber\\
& &\phi_- (\lambda )=\left\{
\begin{array}{ll}0, & \,\,\,\lambda >0 \\
\frac{1}{2\pi }\oint \e^{-\mathrm{i}\omega \lambda} \tilde{\phi}_-(\omega)d\omega 
=\phi (\lambda ), & \,\,\, \lambda <0.
\end{array}\right. \nonumber
\end{eqnarray}
Where the first (second) half-circle contour integral in the above equations is carried out in the lower 
(upper) half complex plane.  
The functions $\tilde{\phi}_{\pm}(\omega)$ are analytic in the upper (+) and lower (-)
halves of the complex plane.  
Applying the forward Fourier transformations then gives
\begin{equation}
\frac{\tilde{y}_1^+(\omega)}{a_+(\omega)a_-(\omega)}+\tilde{y}_1^-(\omega)=\tilde{\phi}_{+}(\omega)+\tilde{\phi}_{-}(\omega)\label{eq:W-H3}
\end{equation}
where 
\begin{eqnarray}
a_+(\omega)a_-(\omega)=[1-\tilde{G}(\omega)]^{-1}=\frac{\e^{\frac{|\omega|}{2}}}{2\cosh \frac{|\omega|}{2}}.
\end{eqnarray}
Explicitly,
\begin{eqnarray}
    a_+(\omega)&=&a_-(-\omega)=\sqrt{2}\pi 
    \left(\frac{\eta -\mathrm{i}\omega }{2\pi \e}\right)^{-\frac{\mathrm{i}\omega }{2\pi}}
    \!\!\!/\Gamma \left(\ffrac{1}{2}-\frac{\mathrm{i}\omega }{2\pi}\right)\\
  \phi_+(\omega)&=&\mathrm{i}\left(\frac{J_{{\rm eff}}}{2(\omega+\mathrm{i}\epsilon)}-
  \frac{2J_{\parallel} \, \e^{-B\pi}}{\omega+\mathrm{i}\pi } \pi\right) .
\end{eqnarray}
Here $\Gamma(z)$ is the gamma function. 
The common strip of analyticity for $a_{\pm}(\omega)$ is $\omega \in (-\mathrm{i} \pi, \mathrm{i} \pi)$.

In the above equations,
$a_{\pm}(\omega)$ factorize the kernel and are analytic in the upper and lower halves
of the complex plane, respectively.
{}From  equation (\ref{eq:W-H3}), we obtain 
\begin{equation}
\frac{\tilde{y}_1^+(\omega)}{a_+(\omega)}-\tilde{\Phi}_+(\omega) =\tilde{\Phi}_-(\omega)+\phi_-(\omega) a_-(\omega)-\tilde{y}_1^-(\omega)a_-(\omega).\label{eq:W-H4}
\end{equation}
Here we have used the decomposition
$\tilde{\phi}_+(\omega)a_-(\omega)=\tilde{\Phi}_+(\omega)-\tilde{\Phi}_-(\omega)$,
where $\tilde{\Phi}_+$ is given by (the evaluation of $\tilde{\Phi}_-$ is not necessary in the following)
\begin{equation}
\tilde{\Phi}_+(\omega)=\mathrm{i}\left(\frac{J_{{\rm eff}}a_-(0)}{2(\omega+\mathrm{i}\epsilon)}
-\frac{2J_{\parallel} \, \e^{-B\pi}a_-(-\mathrm{i}\pi)}{\omega+\mathrm{i} \pi }\right) .
\end{equation}
Owing to the analytic properties of both sides of (\ref{eq:W-H4}), 
they should equal some entire function, since both sides tend to zero
as $\omega \rightarrow \infty$.
Here this entire function turns out to be zero.

At the Fermi point 
\begin{equation}
y_1(0)\equiv \epsilon^{(1)}(B)=\frac{1}{2\pi }\int _{0}^{\infty}\tilde{y}_1(\omega)d\omega =0
\end{equation}
where $\tilde{y}_1(\omega)$ is analytic in the upper half of the complex plane. 
Therefore 
$\lim _{|\omega| \rightarrow \infty}\omega \tilde{y}_1(\omega)=0$ holds, which yields
\begin{equation}
e^{-B\pi}=\frac{J_{{\rm eff}} \, a_-(0)}{4\pi J_{\parallel} \, a_-(-\mathrm{i}\pi)}.
\end{equation}
We see that the Fermi boundary decreases monotonically with increasing magnetic field.  
Subsequently, we have
\begin{equation}
\tilde{y}_1^+(\omega)=\tilde{\Phi}_+(\omega)a_+(\omega)=\mathrm{i} \ffrac12 J_{\rm eff} 
\left(\frac{1}{\omega +\mathrm{i}\epsilon}-\frac{1}{\omega +\mathrm{i}\pi}\right)a_-(0)a_+(\omega).
\end{equation} 
Similarly, $\tilde{y}_-(\omega)$ can be obtained from the right hand side of equation (\ref{eq:W-H4}). 
Nevertheless, we only need $\tilde{y}_+(\omega)$ for calculating the free energy.   
{}From equation (\ref{eq:FE2})  we then have 
%
\begin{eqnarray}
F(0,H)&=&-\mu_{{\rm B}}gH-T\int_{-\infty}^{\infty}a_1(v)\ln(1+\e^{-{\epsilon^{(1)}_1(v)}/{T}})dv \nonumber\\
& &+\frac{1}{2\pi}\int_{-\infty}^{\infty}\e^{-{|\omega|}/{2}}
\left[\tilde{G}(\omega)-1\right]\epsilon^{(1)+}(\omega)d\omega \nonumber\\
&=&F(0,0)+\ffrac{1}{2}(J_{\perp}-\mu_BgH)-\frac{1}{2\pi}\int_{-\infty}^{\infty}d\omega 
\frac{\tilde{\epsilon}^{(1)+}(\omega)}{2\cosh \frac{\omega}{2}}\nonumber\\
&=& F(0,0)+\ffrac{1}{2}(J_{\perp}-\mu_BgH)-\left(\frac{J_{{\rm eff}}}{4\pi }\right)^2 
\frac{a_-(0)a_+(0)}{J_\parallel}.
\end{eqnarray}
Here finally $a_-(0)a_+(0)=2$. 
The zero field free energy $F(0,0)=-J_{\parallel}(\Psi (1)-\Psi (\frac{1}{2}))$, where
$\Psi(a)$ is the digamma function.  
It follows that the susceptibility behaves like $\chi \approx \frac{(\mu_Bg)^2}{4\pi ^2 J_\parallel}$, 
indicating $su(2)$ symmetry at this point as the field diminishes the rung interaction.
This susceptibility is consistent with that of the isotropic Heisenberg XXX chain, 
where $J_\parallel = \frac14$ and $h=H/2$ \cite{Tbook}.
It is seen that at the inflection point the model can be mapped onto the 
$su(2)$ isotropic Heisenberg chain with zero field.

\section{The quantum transfer matrix and  nonlinear integral equation approach}
\label{QTM-NLIE}

In this section we review the recently developed high-temperature expansion (HTE) formalism 
for integrable models, including the quantum transfer matrix (QTM) approach 
and the derivation of the temperature-dependant free energy via nonlinear integral 
equations with only a finite number of unknown functions.
This paves the way for the application of this approach to the integrable spin ladder models and
to the physical description of the strong coupling ladder compounds.
The final results can be applied without any knowledge of the QTM and HTE. 
This can be done by inserting the chemical potential terms $\mu_i$ of a given model into the
free energy expansion (\ref{eq:freeenergyHTE}), which gives instant access
to finite temperature physical properties, as shown for a number of ladder compounds 
in section \ref{sec:Exp_comp}.


The QTM relates a $d$-dimensional quantum system at finite
temperature to a $(d+1)$-dimensional classical system on an inhomogeneous lattice via the so-called 
Trotter-Suzuki mapping \cite{QTMrefsa,QTMrefsb,QTMproof,SI87,K87,SAW90}.
In one-dimension, it involves a general relation between the transfer matrix 
$t(u)=\e^{\mathrm{i}\hat{P}+u {\cal H}+O(u^2)}$ and the Hamiltonian $\cal H$ of a quantum system, 
where $\hat{P}$ denotes the momentum operator. 
One also introduces an auxiliary transfer matrix $\tilde{t}(u)$ which is adjoint to $t(u)$ such that
$\tilde{t}(u)=\e^{-\mathrm{i}\hat{P}+u {\cal H}+O(u^2)}$.
In this way the partition function $Z_L$ of the one-dimensional system at finite temperature  
is mapped to that of a staggered two-dimensional vertex model, with  
$Z_L={\rm tr~} \e^{-\beta {\cal H}}=\lim_{N\to \infty} Z_{L,N}$, where $\beta =1/T$ and
$Z_{L,N}= {\rm tr~}[t(-\beta/N)\tilde{t}(-\beta/N)]^{{N}/{2}}$. 
The trace ${\rm tr}$ is carried out on the quantum space, as opposed to the auxiliary space.
In the following we show explicitly that the partition function of the one-dimensional quantum
system is given by $Z_{L,N}= {\rm tr~}[t_{\rm QTM}(0)]^L$, 
where the QTM $t_{\rm QTM}(u)$ is defined on the columns of the two-dimensional lattice. 
The QTM forms a commuting family, i.e.,  $\left[t_{\rm QTM}(u), t_{\rm QTM}(v)\right]=0$, 
by the virtue of the Yang-Baxter equation.
Thus the problem of describing the
thermodynamics of one-dimensional quantum systems reduces to finding
the largest eigenvalue of the QTM of a finite two-dimensional classical lattice model.
In particular, the free energy per unit length is given by
$$
f=-\lim_{L\to \infty}\frac{1}{L\beta}\ln Z_{L,N}=-\frac{1}{\beta}\ln \Lambda _{\rm max}(0). 
$$
In general the next leading eigenvalues of the QTM give the exponential correlation length $\xi$ of the 
equal time correlations.

\subsection{Preliminaries}

Our starting point is the $su(r+1)$ Uimin-Lai-Sutherland, or permutator, model \cite{Uimin,Lai,Suth}, 
for which one can define the QTM \cite{QTMrefsa,QTMrefsb,QTMproof,SI87,K87,SAW90} 
and the $T$-system \cite{KunibaFusiona,KunibaFusionb,KRfusion}. 
A description of the QTM approach for the Uimin-Lai-Sutherland model 
can be found in \cite{JKS98,FK99}. 
This involves working with the two-dimensional classical counterpart of the Uimin-Lai-Sutherland model, 
which is a rational limit of the Perk-Schultz model \cite{Perk}.

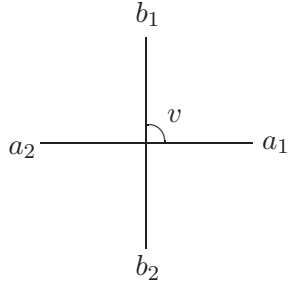
\begin{figure}
    \setlength{\unitlength}{2pt}
    \begin{center}
    \begin{picture}(65,60) 
      \put(30,20){\oval(7,7)[tr]}
      \put(10,20){\line(1,0){40}}
      \put(30,0){\line(0,1){40}}
      \put(52,19){$a_{1}$}
      \put(28,43){$b_{1} $}
      \put(4,18){$a_{2}$}
      \put(28,-5){$b_{2} $}
      \put(34,24){$v$}      
  \end{picture}
  \end{center}
  \caption{The Boltzmann weights $R^{a_{1},b_{1}}_{a_{2},b_{2}}(v)$ of the vertex model.}
  \label{fig:R-mat}
\end{figure}

Let \{$\st[1],\st[2],\dots,\st[r+1] \}$ be an orthonormal basis of an 
$r+1$ dimensional fundamental  representation $V$ of $su(r+1)$, and 
$\{\langle 1 \!\! \mid,\langle 2 \!\! \mid,\dots,\langle r+1 \!\! \mid \}$ be its dual,
with $\langle a\st[b]=\delta_{ab}$. 
The $R$-matrix is given by \cite{Perk} 
\begin{eqnarray}
R(v) \st[a] \otimes \st[b]= \sum_{a^{\prime }=1}^{r+1} 
\sum_{b^{\prime}=1}^{r+1}
\st[a^{\prime}] \otimes \st[b^{\prime }]
R^{a^{\prime } b^{\prime }}_{a b}(v)
\end{eqnarray}
with the matrix element (see figure  \ref{fig:R-mat}) 
\begin{eqnarray}
R^{a^{\prime}b^{\prime}}_{a b}(v) =
\langle a^{\prime} \! \mid \otimes \, \langle b^{\prime} \! \mid \! R(v) \, \st[a] \otimes \st[b]
=v \, \delta_{a a^{\prime}}\delta_{b b^{\prime}}+
 \delta_{a^{\prime }b}\delta_{a b^{\prime }} \label{eq:R-mat}
\end{eqnarray}
where $v \in \mathbb{C}$ and $a,b,a^{\prime},b^{\prime} \in \{1,2,\dots,r+1\}$. 

\begin{figure}
    \setlength{\unitlength}{1.5pt}
    \begin{center}
    \begin{picture}(220,55) 
      \put(38,20){\oval(7,7)[tr]}
      \put(78,20){\oval(7,7)[tr]}
      \put(158,20){\oval(7,7)[tr]}
      \put(198,20){\oval(7,7)[tr]}
      \put(18,20){\line(1,0){80}}
      \put(58,20){\circle*{1.5}}
      \put(98,20){\circle*{1.5}}
      \put(18,20){\circle*{1.5}}
      \put(178,20){\circle*{1.5}}
      \put(218,20){\circle*{1.5}}
      \put(103,20){\line(1,0){6}}
      \put(115,20){\line(1,0){6}}
      \put(127,20){\line(1,0){6}}
      \put(138,20){\circle*{1.5}}
      \put(138,20){\line(1,0){80}}
      \put(38,0){\line(0,1){40}}
      \put(78,0){\line(0,1){40}}
      \put(158,0){\line(0,1){40}}
      \put(198,0){\line(0,1){40}}
      \put(36,43){$\alpha_{1}$}
      \put(76,43){$\alpha_{2}$}
      \put(154,43){$\alpha_{L-1}$}
      \put(194,43){$\alpha_{L}$}
      \put(3,18){$\hat{V}_{2k}$}
      \put(36,-7){$\beta_{1}$}
      \put(76,-7){$\beta_{2}$}
      \put(156,-7){$\beta_{L-1}$}
      \put(196,-7){$\beta_{L}$}
      \put(42,24){$v$}
      \put(82,24){$v$}
      \put(162,24){$v$}
      \put(202,24){$v$}
  \end{picture}
  \end{center}
  \caption{The transfer matrix 
  $t(v)^{\{\alpha_{1},\dots,\alpha_{L}\}}_{\{\beta_{1},\dots,\beta_{L}\}}$. 
  The dots denote summations in definition (\ref{TM-sum}).}
  \label{fig:transfer}
\end{figure}
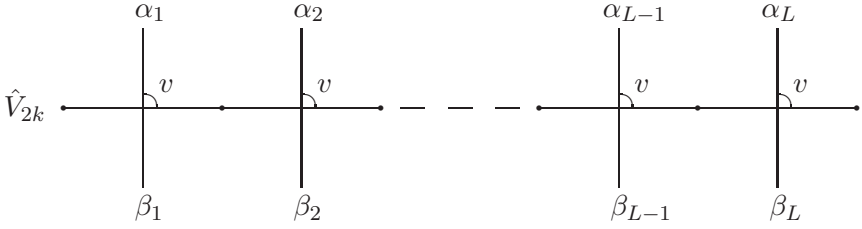

Further let $V_{1},V_{2},\dots,V_{L},  \hat{V}_{1},\hat{V}_{2},\dots,\hat{V}_{N}$ be 
copies of $V$, i.e., $V_{j} = \hat{V}_{j} = V$, where $L,{N}/{2}\in \mathbb{Z}_{\ge 1}$.
The row-to-row transfer matrix with periodic boundary conditions, which acts on the
quantum space $V_{1}\otimes V_{2} \otimes \cdots \otimes V_{L}$, is defined as
\begin{eqnarray}
t(v)={\mathrm{tr}}_{\hat{V}_{2k}}(R_{2k,L}(v)\cdots 
R_{2k,2}(v)R_{2k,1}(v)), 
\quad k=1,2, \dots, {N}/{2}
\end{eqnarray}
where $R_{ab}(v)$ acts on 
$\hat{V}_{1} \otimes \hat{V}_{2} \otimes \cdots \otimes \hat{V}_{N}
\otimes V_{1} \otimes V_{2} \otimes \cdots \otimes V_{L} $ such that 
\begin{eqnarray}
& &
R_{ab}(v)\st[\nu_{1}]\otimes \cdots 
\otimes \st[\nu_{a}] \otimes \cdots 
 \otimes \st[\nu_{N} ]\otimes 
 \st[\alpha_{1}] \otimes \cdots 
 \otimes \st[\alpha_{b} ]\otimes \cdots 
 \otimes \st[\alpha_{L} ]\nonumber \\
& &
\qquad =
 \sum_{\xi_{a}=1}^{r+1}\sum_{\eta_{b}=1}^{r+1}
 \st[\nu_{1}]\otimes \cdots 
\otimes \st[\xi_{a}] \otimes \cdots 
 \otimes \st[\nu_{N}] \otimes 
 \st[\alpha_{1} ]\otimes \cdots \nonumber \\
 &&
\qquad\qquad \qquad \cdots  \otimes \st[\eta_{b} ]\otimes \cdots 
 \otimes \st[\alpha_{L}]
 R^{\xi_{a},\eta_{b}}_{\nu_{a},\alpha_{b}}(v). 
\end{eqnarray}
Here $\nu_{j},\alpha_{j} \in \{1,2,\dots, r+1 \}$ and the matrix element is defined as 
(see figure \ref{fig:transfer})
\begin{equation}
t(v)^{\{\alpha_{1},\dots,\alpha_{L}\}}
  _{\{\beta_{1},\dots,\beta_{L}\}}
=\langle\alpha_{1} \!\! \mid\otimes  \langle\alpha_{2} \!\! \mid \otimes 
\cdots \otimes \langle\alpha_{L} \!\! \mid \! t(v) 
\st[\beta_{1}]\otimes \st[\beta_{2}]\otimes \cdots 
 \otimes \st[\beta_{L}]
\label{TM-sum}
\end{equation} 
where $\alpha_{j},\beta_{j} \in \{1,2,\dots, r+1 \}$.

The Hamiltonian ${\cal H} = {\cal H}_{0} + {\cal H}_{{\rm ch}}$ of the model, which acts on 
$V_{1}\otimes V_{2} \otimes \cdots \otimes V_{L}$, consists of two parts. 
The first part ${\cal H}_{0}$ characterizing the intrachain interaction is given by 
(see also section \ref{sec:ISL})
\begin{eqnarray}
{\cal H}_{0}=J_{\parallel}\frac{d}{dv}\ln t(v) |_{v=0}. 
\end{eqnarray}
The second part is of the form ${\cal H}_{{\rm ch}}=\sum_{j=1}^{L} {\cal H}_{j}$,
where ${\cal H}_{j}$ acts non-trivially on the $j$th site of the quantum space $V_{j}$.
It contains the chemical potential terms which include the rung interaction and magnetic field.
Explicitly,
\begin{equation}
{\cal H}_{{\rm ch}}=J_{\perp}\sum_{j=1}^{L}\vec{S}_j \cdot \vec{T}_j
-\mu_{{\rm B}}gH\sum_{j=1}^{L}(S_j^z+T^z_j)
\end{equation}
for Hamiltonian (\ref{eq:Ham2}).  
As discussed in section \ref{subsec:TBA}, the chemical potential terms can be diagonalised 
in a suitable eigenbasis.  
%
%
The chemical potentials $\mu_{\alpha}$ are given by 
\begin{equation}
\langle \alpha_1 \mid \otimes \cdots \otimes \langle \alpha_L \mid {\cal H}_j \mid \beta_1 \rangle 
\otimes \cdots \otimes \mid \beta_L \rangle = \mu_{\alpha_j} \delta_{\alpha_1 \beta_1} \cdots 
\delta_{\alpha_L \beta_L}
\end{equation}
for $j = 1, \ldots, L$.
For example, for Hamiltonian (\ref{eq:Ham2}), the eigenbasis can be chosen as
given in equation (\ref{eq:su4basis}), for which the  
chemical potentials are given by 
\begin{eqnarray}
\mu_1=-J_{\perp},\,\,\mu_2=-\mu_BgH,\,\,\mu_3=0,\,\,\mu_4=\mu_BgH.
\label{eq:chem}
\end{eqnarray}

\begin{figure}
    \setlength{\unitlength}{2pt}
    \begin{center}
    \begin{picture}(65,60) 
      \put(30,20){\oval(7,7)[tl]}
      \put(10,20){\line(1,0){40}}
      \put(30,0){\line(0,1){40}}
      \put(52,19){$a_{1}$}
      \put(28,43){$b_{1}$}
      \put(4,18){$a_{2}$}
      \put(28,-5){$b_{2}$}
      \put(23,24){$v$} 
  \end{picture}
  \end{center}
  \caption{90 degree rotation  
  $\widetilde{R}^{a_{1},b_{1}}_{a_{2},b_{2}}(v)=R^{b_{1},a_{2}}_{b_{2},a_{1}}(v)$ of $R(v)$.}
  \label{fig:R-mat2}
\end{figure}
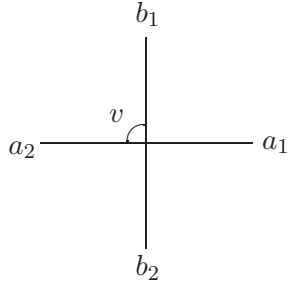

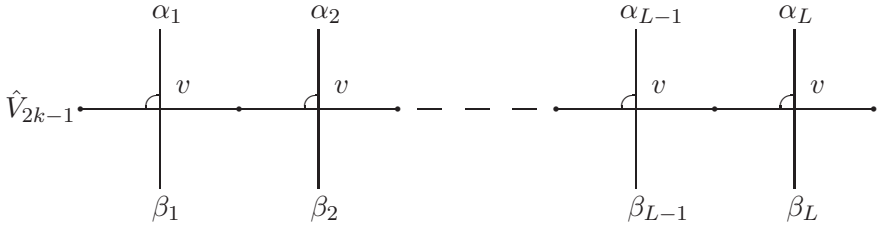
\begin{figure}
    \setlength{\unitlength}{1.5pt}
    \begin{center}
    \begin{picture}(220,55) 
      \put(38,20){\oval(7,7)[tl]}
      \put(78,20){\oval(7,7)[tl]}
      \put(158,20){\oval(7,7)[tl]}
      \put(198,20){\oval(7,7)[tl]}
      \put(18,20){\line(1,0){80}}
      \put(58,20){\circle*{1.5}}
      \put(98,20){\circle*{1.5}}
      \put(18,20){\circle*{1.5}}
      \put(178,20){\circle*{1.5}}
      \put(218,20){\circle*{1.5}}
      \put(103,20){\line(1,0){6}}
      \put(115,20){\line(1,0){6}}
      \put(127,20){\line(1,0){6}}
      \put(138,20){\circle*{1.5}}
      \put(138,20){\line(1,0){80}}
      \put(38,0){\line(0,1){40}}
      \put(78,0){\line(0,1){40}}
      \put(158,0){\line(0,1){40}}
      \put(198,0){\line(0,1){40}}
      \put(36,43){$\alpha_{1}$}
      \put(76,43){$\alpha_{2}$}
      \put(154,43){$\alpha_{L-1}$}
      \put(194,43){$\alpha_{L}$}
      \put(-1,18){$\hat{V}_{2k-1}$}
      \put(36,-7){$\beta_{1}$}
      \put(76,-7){$\beta_{2}$}
      \put(156,-7){$\beta_{L-1}$}
      \put(196,-7){$\beta_{L}$}
      \put(42,24){$v$}
      \put(82,24){$v$}
      \put(162,24){$v$}
      \put(202,24){$v$}
  \end{picture}
  \end{center}
  \caption{The transfer matrix 
  $\tilde{t}(v)^{\{\alpha_{1},\dots,\alpha_{L}\}}_{\{\beta_{1},\dots,\beta_{L}\}}$ defined in 
  (\ref{eq:transfer-tilde}). Summations are taken at the dots.}
  \label{fig:transfer2}
\end{figure}

We also define an auxiliary  transfer matrix with periodic boundary conditions, 
which acts on $V_{1}\otimes V_{2} \otimes \cdots \otimes V_{L}$ as 
\begin{eqnarray}
\widetilde{t}(v)={\mathrm{tr}}_{\hat{V}_{2k-1}}(
\widetilde{R}_{2k-1,L}(v)\cdots \widetilde{R}_{2k-1,2}(v)
\widetilde{R}_{2k-1,1}(v)), \quad k=1,2, \dots, \ffrac{N}{2}.
\label{eq:transfer-tilde}
\end{eqnarray}
Here $\widetilde{R}(v)$ is essentially a 90 degree counterclockwise rotation of $R(v)$,
whose components are given by
$\widetilde{R}^{a_{1},b_{1}}_{a_{2},b_{2}}(v)$
      =$R^{b_{1},a_{2}}_{b_{2},a_{1}}(v)$ (see figure  \ref{fig:R-mat2}). 
$\widetilde{R}_{ab}(v)$ acts on the space 
$\hat{V}_{1} \otimes \hat{V}_{2} \otimes \cdots \otimes \hat{V}_{N}
\otimes V_{1} \otimes V_{2} \otimes \cdots \otimes V_{L} $ such that 
\begin{eqnarray}
& &
\widetilde{R}_{ab}(v)\st[\nu_{1}]\otimes \cdots 
\otimes \st[\nu_{a}] \otimes \cdots 
 \otimes \st[\nu_{N}] \otimes 
 \st[\alpha_{1}] \otimes \cdots 
 \otimes \st[\alpha_{b}] \otimes \cdots 
 \otimes \st[\alpha_{L}]\nonumber\\
& &
\qquad = \sum_{\xi_{a}=1}^{r+1}\sum_{\eta_{b}=1}^{r+1}
 \st[\nu_{1}]\otimes \cdots 
\otimes \st[\xi_{a}] \otimes \cdots \nonumber\\
& & \qquad \qquad
 \otimes \st[\nu_{N}] \otimes 
 \st[\alpha_{1}] \otimes \cdots 
 \otimes \st[\eta_{b}] \otimes \cdots 
 \otimes \st[\alpha_{L}]
 R^{\eta_{b},\nu_{a}}_{\alpha_{b},\xi_{a}}(v)
\end{eqnarray} 
where $\nu_{j},\alpha_{j} \in \{1,2,\dots, r+1 \}$. 
The matrix element is defined as (see figure \ref{fig:transfer2}) 
\begin{eqnarray}
\widetilde{t}(v)^{\{\alpha_{1},\dots,\alpha_{L}\}}
  _{\{\beta_{1},\dots,\beta_{L}\}}
=\langle\alpha_{1}|\otimes \cdots \otimes \langle\alpha_{L}|\widetilde{t}(v) 
\st[\beta_{1}]\otimes \cdots \otimes \st[\beta_{L}]
\end{eqnarray} 
where $\alpha_{j},\beta_{j} \in \{1,2,\dots, r+1 \}$. 

The logarithmic derivative of (\ref{eq:transfer-tilde}) also reproduces ${\cal H}_{0}$.
Explicitly, 
\begin{eqnarray}
&& t(v)=t(0)\{1 + {\cal H}_{0} \,v/J_{\parallel}
 +{\mathcal O}(v^{2})\} \nonumber \\
&& \widetilde{t}(v)=\widetilde{t}(0)\{1 + {\cal H}_{0} \,v/J_{\parallel}
 +{\mathcal O}(v^{2})\}.
\end{eqnarray}
As $t(0)$ and $\widetilde{t}(0)$ are, respectively, left and right translation operators, then
\begin{eqnarray}
t(v)\widetilde{t}(v)=1+ {2}\, {\cal H}_{0} \, v / {J_{\parallel}} +{\mathcal O}(v^{2}).
\label{rtr-two}
\end{eqnarray}

\subsection{Quantum Transfer Matrix method}

The partition function $Z_{L}$ can now be written in terms of the transfer matrices as 
\begin{eqnarray}
Z_{L}&=&{\mathrm{tr}}_{V_{1}\otimes \cdots \otimes V_{L}}
\e^{-\beta {\cal H}}
=
{\mathrm{tr}}_{V_{1}\otimes \cdots \otimes V_{L}}
\e^{-\beta {\cal H}_{\rm ch}} 
\e^{-\beta {\cal H}_{0}} 
\nonumber \\ 
&=&
\lim_{N \to \infty} 
{\mathrm{tr}}_{V_{1}\otimes \cdots \otimes V_{L}}
\e^{-\beta {\cal H}_{\rm ch}}
\left(1- \frac{2 \beta {\cal H}_{0}}{N} \right)^{\frac{N}{2}} 
\nonumber \\ 
 &=&\lim_{N \to \infty}
{\mathrm{tr}}_{V_{1}\otimes \cdots \otimes V_{L}} \e^{-\beta {\cal H}_{\rm ch}}
(t(u_{N})\widetilde{t}(u_{N}))^{\frac{N}{2}}
\end{eqnarray}
where $N$ is the Trotter number, defined by $u_{N}=-{J_{\parallel}\beta}/{ N}$,
with $\beta=1/T$ the inverse temperature.
Here we have set the Boltzmann constant to be unity and have used (\ref{rtr-two}), 
along with commutativity of ${\cal H}_{0}$ and ${\cal H}_{\rm ch}$ 
and the formula $\lim_{N\to \infty}{\mathrm{tr}~}(1-\frac{A}{N})^{N}={\mathrm{tr}~} \e^{-A}$ 
for a suitable operator $A$. 
The free energy per site is
\begin{eqnarray}
f&=&-\lim_{L \to \infty}
\frac{1}{L \beta }\ln Z_{L} \nonumber \\ 
&=&-\lim_{L \to \infty}\lim_{N \to \infty}
\frac{1}{\beta L}\ln
{\mathrm {tr}}_{V_{1}\otimes \cdots \otimes V_{L}} e^{-\beta {\cal H}_{\rm ch}}
(t(u_{N})\widetilde{t}(u_{N}))^{\frac{N}{2}}.
\label{eq:free-trotter}
\end{eqnarray}

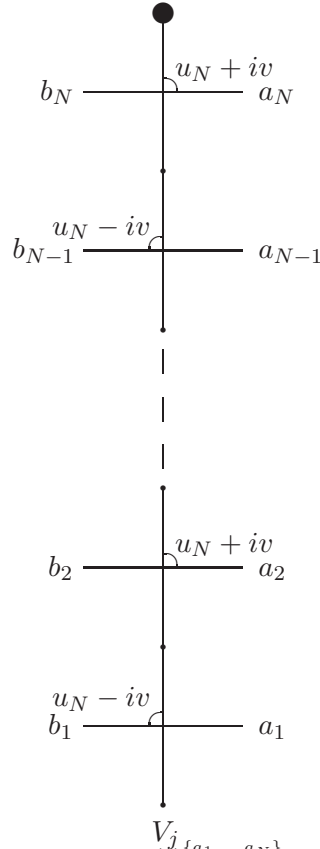
\begin{figure}
    \setlength{\unitlength}{1.5pt}
    \begin{center}
    \begin{picture}(70,203) 
      \put(38,20){\oval(7,7)[tl]}
      \put(38,60){\oval(7,7)[tr]}
      \put(38,140){\oval(7,7)[tl]}
      \put(38,180){\oval(7,7)[tr]}
      \put(38,0){\circle*{1.5}}
      \put(38,40){\circle*{1.5}}
      \put(38,80){\circle*{1.5}}
      \put(38,120){\circle*{1.5}}
      \put(38,160){\circle*{1.5}}
      \put(38,200){\circle*{5}}
      \put(18,20){\line(1,0){40}}
      \put(18,60){\line(1,0){40}}
      \put(18,140){\line(1,0){40}}
      \put(18,180){\line(1,0){40}}
      \put(38,0){\line(0,1){80}}
      \put(38,85){\line(0,1){6}}
      \put(38,97){\line(0,1){6}}
      \put(38,109){\line(0,1){6}}
      \put(38,120){\line(0,1){80}}
      \put(9,18){$b_{1}$}
      \put(9,58){$b_{2}$}
      \put(0,138){$b_{N-1}$}
      \put(7,178){$b_{N}$}
      \put(62,18){$a_{1}$}
      \put(62,58){$a_{2}$}
      \put(62,138){$a_{N-1}$}
      \put(62,178){$a_{N}$}
      \put(10,25){$u_{N}-iv$}
      \put(41,64){$u_{N}+iv$}
      \put(10,144){$u_{N}-iv$} 
      \put(41,184){$u_{N}+iv$} 
      \put(35,-9){$V_{j}$} 
  \end{picture}
  \end{center}
  \caption{The quantum transfer matrix 
  $t_{\rm{QTM}}(v)^{\{a_{1},\dots,a_{N}\}}_{\{b_{1},\dots,b_{N}\}}$.  
 The dots denote summations in definition (\ref{eq:QTM2}). 
 The large dot stands for $\e^{-\beta {\cal H}_{j}}$, $j \in \{1,2,\dots,L\}$.}
  \label{fig:QTM-fig}
\end{figure}

\begin{figure}
    \setlength{\unitlength}{1.5pt}
    \begin{center}
    \begin{picture}(220,208) 
      \put(38,20){\oval(7,7)[tl]}
      \put(38,60){\oval(7,7)[tr]}
      \put(38,140){\oval(7,7)[tl]}
      \put(38,180){\oval(7,7)[tr]}
      \put(78,20){\oval(7,7)[tl]}
      \put(78,60){\oval(7,7)[tr]}
      \put(78,140){\oval(7,7)[tl]}
      \put(78,180){\oval(7,7)[tr]}
      \put(158,20){\oval(7,7)[tl]}
      \put(158,60){\oval(7,7)[tr]}
      \put(158,140){\oval(7,7)[tl]}
      \put(158,180){\oval(7,7)[tr]}
      \put(198,20){\oval(7,7)[tl]}
      \put(198,60){\oval(7,7)[tr]}
      \put(198,140){\oval(7,7)[tl]}
      \put(198,180){\oval(7,7)[tr]}
      \put(38,0){\circle*{1.5}}
      \put(38,40){\circle*{1.5}}
      \put(38,80){\circle*{1.5}}
      \put(38,120){\circle*{1.5}}
      \put(38,160){\circle*{1.5}}
      \put(38,200){\circle*{5}}
      \put(78,0){\circle*{1.5}}
      \put(78,40){\circle*{1.5}}
      \put(78,80){\circle*{1.5}}
      \put(78,120){\circle*{1.5}}
      \put(78,160){\circle*{1.5}}
      \put(78,200){\circle*{5}}
      \put(158,0){\circle*{1.5}}
      \put(158,40){\circle*{1.5}}
      \put(158,80){\circle*{1.5}}
      \put(158,120){\circle*{1.5}}
      \put(158,160){\circle*{1.5}}
      \put(158,200){\circle*{5}}
      \put(198,0){\circle*{1.5}}
      \put(198,40){\circle*{1.5}}
      \put(198,80){\circle*{1.5}}
      \put(198,120){\circle*{1.5}}
      \put(198,160){\circle*{1.5}}
      \put(198,200){\circle*{5}}
      \put(18,20){\circle*{1.5}}
      \put(18,60){\circle*{1.5}}
      \put(18,140){\circle*{1.5}}
      \put(18,180){\circle*{1.5}}
      \put(58,20){\circle*{1.5}}
      \put(58,60){\circle*{1.5}}
      \put(58,140){\circle*{1.5}}
      \put(58,180){\circle*{1.5}}
      \put(98,20){\circle*{1.5}}
      \put(98,60){\circle*{1.5}}
      \put(98,140){\circle*{1.5}}
      \put(98,180){\circle*{1.5}}
      \put(138,20){\circle*{1.5}}
      \put(138,60){\circle*{1.5}}
      \put(138,140){\circle*{1.5}}
      \put(138,180){\circle*{1.5}}
      \put(178,20){\circle*{1.5}}
      \put(178,60){\circle*{1.5}}
      \put(178,140){\circle*{1.5}}
      \put(178,180){\circle*{1.5}}
      \put(218,20){\circle*{1.5}}
      \put(218,60){\circle*{1.5}}
      \put(218,140){\circle*{1.5}}
      \put(218,180){\circle*{1.5}}
      \put(18,20){\line(1,0){80}}
      \put(18,60){\line(1,0){80}}
      \put(18,140){\line(1,0){80}}
      \put(18,180){\line(1,0){80}}
      \put(138,20){\line(1,0){80}}
      \put(138,60){\line(1,0){80}}
      \put(138,140){\line(1,0){80}}
      \put(138,180){\line(1,0){80}}
      \put(38,0){\line(0,1){80}}
      \put(38,85){\line(0,1){6}}
      \put(38,97){\line(0,1){6}}
      \put(38,109){\line(0,1){6}}
      \put(38,120){\line(0,1){80}}
      \put(78,0){\line(0,1){80}}
      \put(78,85){\line(0,1){6}}
      \put(78,97){\line(0,1){6}}
      \put(78,109){\line(0,1){6}}
      \put(78,120){\line(0,1){80}}
      \put(158,0){\line(0,1){80}}
      \put(158,85){\line(0,1){6}}
      \put(158,97){\line(0,1){6}}
      \put(158,109){\line(0,1){6}}
      \put(158,120){\line(0,1){80}}
      \put(198,0){\line(0,1){80}}
      \put(198,85){\line(0,1){6}}
      \put(198,97){\line(0,1){6}}
      \put(198,109){\line(0,1){6}}
      \put(198,120){\line(0,1){80}}
      \put(103,20){\line(1,0){6}}
      \put(115,20){\line(1,0){6}}
      \put(127,20){\line(1,0){6}}
      \put(103,60){\line(1,0){6}}
      \put(115,60){\line(1,0){6}}
      \put(127,60){\line(1,0){6}}
      \put(103,140){\line(1,0){6}}
      \put(115,140){\line(1,0){6}}
      \put(127,140){\line(1,0){6}}
      \put(103,180){\line(1,0){6}}
      \put(115,180){\line(1,0){6}}
      \put(127,180){\line(1,0){6}}
      \put(8,18){$\hat{V}_{1}$}
      \put(8,58){$\hat{V}_{2}$}
      \put(-1,138){$\hat{V}_{N-1}$}
      \put(6,178){$\hat{V}_{N}$}
      \put(35,-8){$V_{1}$}
      \put(75,-8){$V_{2}$}
      \put(155,-8){$V_{L-1}$}
      \put(195,-8){$V_{L}$}
      \put(15,24){$u-iv$}
      \put(41,64){$u+iv$}
      \put(15,144){$u-iv$} 
      \put(41,184){$u+iv$} 
      \put(55,24){$u-iv$}
      \put(81,64){$u+iv$}
      \put(55,144){$u-iv$} 
      \put(81,184){$u+iv$} 
      \put(135,24){$u-iv$}
      \put(161,64){$u+iv$}
      \put(135,144){$u-iv$} 
      \put(161,184){$u+iv$} 
      \put(175,24){$u-iv$}
      \put(201,64){$u+iv$}
      \put(175,144){$u-iv$} 
      \put(201,184){$u+iv$} 
  \end{picture}
  \end{center}
  \caption{
 The QTM lattice structure.
 Summations are taken at dotted points. 
 Large dots stand for $\e^{-\beta {\cal H}_{j}}$,  $j \in \{1,2,\dots,L\}$. 
 The values $u=u_{N}$ and $v=0$ correspond to equation (\ref{eq:ver-hor}). 
 Horizontal lines with $u+iv$ correspond to $t(v)$ while horizontal lines with
 $u-iv$ correspond to $\widetilde{t}(v)$. 
 The vertical lines correspond to $t_{\rm{QTM}}(v)$.} 
 \label{fig:relation-fig}
\end{figure}
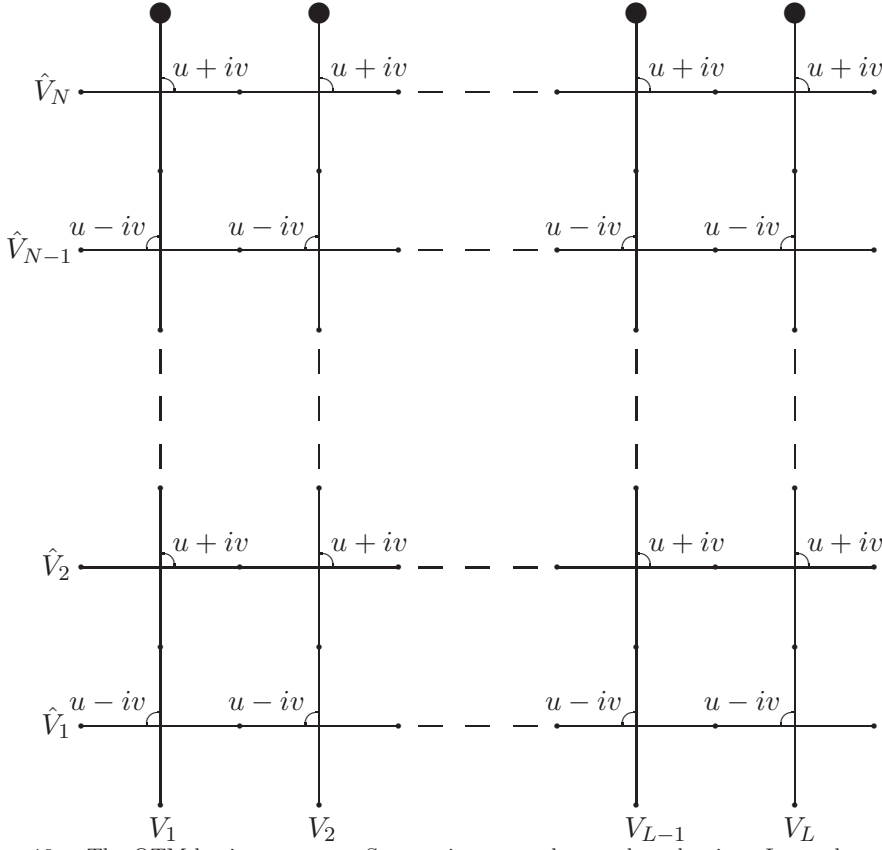

In general, the eigenvalues of $t(u_{N})\widetilde{t}(u_{N})$ 
may be infinitely degenerate in the Trotter limit $N \to \infty$ ($u_{N} \to 0$) 
and taking the trace in (\ref{eq:free-trotter}) is not a trivial operation.
To overcome this difficulty, we define an inhomogeneous transfer matrix 
-- the QTM -- with periodic boundary conditions,
\begin{eqnarray}
&& t_{\mathrm{QTM}}(v)=
{\mathrm{tr}}_{V_{j}} \e^{-\beta {\cal H}_{j}}
R_{N,j}(u_{N}+iv)\widetilde{R}_{N-1,j}(u_{N}-iv)
\cdots  \nonumber \\
&& \hspace{45pt} 
R_{4,j}(u_{N}+iv)\widetilde{R}_{3,j}(u_{N}-iv)
R_{2,j}(u_{N}+iv)\widetilde{R}_{1,j}(u_{N}-iv)
\label{eq:QTM}
\end{eqnarray}
which acts on $\hat{V}_{1}\otimes \hat{V}_{2} \otimes \cdots \otimes \hat{V}_{N}$.
The matrix element of the QTM is defined as
\begin{eqnarray}
t_{\rm_{QTM}}(v)^{\{\alpha_{1},\dots,\alpha_{N}\}}
  _{\{\beta_{1},\dots,\beta_{N}\}}
=\langle\alpha_{1}\mid\otimes \cdots \otimes \langle\alpha_{N}\mid
t_{\rm{QTM}}(v) 
\st[\beta_{1}]\otimes \cdots \otimes \st[\beta_{N}]\nonumber
\end{eqnarray} 
where $\alpha_{j},\beta_{j} \in \{1,2,\dots, r+1 \}$. 
Explicitly, it may be written as (see figure \ref{fig:QTM-fig}) 
\begin{eqnarray}
&& t_{\mathrm{QTM}}(v)^{\{a_{1},\dots, a_{N} \}}
_{\{b_{1},\dots,b_{N} \}} = \nonumber\\
&& \qquad 
\sum_{\{\nu_{k}\}} \e^{-\frac{\mu_{\nu_{1}}}{T}}
\prod_{k=1}^{{N}/{2}}
 R^{a_{2k},\nu_{2k+1}}_{b_{2k},\nu_{2k}}(u_N+iv)
 \widetilde{R}^{a_{2k-1},\nu_{2k}}_{b_{2k-1},\nu_{2k-1}}(u_N-iv)
 \label{eq:QTM2}
\end{eqnarray}
where $\nu_{N+1}=\nu_{1}$ and $\nu_{k} \in \{1,2,\dots, r+1 \}$. 
The QTM is connected to (see also figure \ref{fig:relation-fig})
\begin{eqnarray}
&&{\mathrm{tr}}_{V_{1}\otimes \cdots \otimes V_{L}}
\e^{-\beta {\cal H}_{\rm ch}}
(t(u_{N})\widetilde{t}(u_{N}))^{\frac{N}{2}} 
\nonumber \\
&&={\mathrm {tr}}_{V_{1}\otimes \cdots \otimes V_{L}} 
 \e^{-\sum_{j=1}^{L}\beta {\cal H}_{j}}
\overleftarrow{\prod_{k=1}^{{N}/{2}}} 
{\mathrm {tr}}_{\hat{V}_{2k-1}\otimes \hat{V}_{2k}}
\{
(R_{2k,L}(u_{N})\cdots R_{2k,2}(u_{N}) R_{2k,1}(u_{N}))
\nonumber \\
&& \hspace{60pt} \times 
(\widetilde{R}_{2k-1,L}(u_{N})
 \cdots \widetilde{R}_{2k-1,2}(u_{N}) 
 \widetilde{R}_{2k-1,1}(u_{N}))  \}
\nonumber \\
&&={\mathrm{tr}}_{V_{1}\otimes \cdots \otimes V_{L}} 
 \e^{-\sum_{j=1}^{L}\beta {\cal H}_{j}}
\overleftarrow{\prod_{k=1}^{{N}/{2}}} 
{\mathrm {tr}}_{\hat{V}_{2k-1}\otimes \hat{V}_{2k}}
\{
(R_{2k,L}(u_{N})\widetilde{R}_{2k-1,L}(u_{N}))
\cdots 
\nonumber \\
&& \hspace{60pt} \times 
(R_{2k,2}(u_{N})\widetilde{R}_{2k-1,2}(u_{N}))
(R_{2k,1}(u_{N})\widetilde{R}_{2k-1,1}(u_{N}))
  \}
 \nonumber \\
&& ={\mathrm{tr}}_{\hat{V}_{1}\otimes \cdots \otimes \hat{V}_{N}} 
\overleftarrow{\prod_{j=1}^{L}} 
{\mathrm {tr}}_{V_{j}} \e^{-\beta {\cal H}_{j}}
\overleftarrow{\prod_{k=1}^{{N}/{2}}} 
(R_{2k,j}(u_{N})\widetilde{R}_{2k-1,j}(u_{N}))
\nonumber \\
&& ={\mathrm {tr}}_{\hat{V}_{1}\otimes \cdots \otimes \hat{V}_{N}}
(t_{\mathrm{QTM}}(0))^{L}.
\label{eq:ver-hor}
\end{eqnarray}
Here the ordered product is defined by 
$$
\overleftarrow{\prod_{j=1}^{L}}A(j)=A(L)\cdots A(2)A(1)
$$
for any indexed operator $\{A(j)\}$.

Let $\Lambda_{1},\Lambda_{2},\dots,\Lambda_{(r+1)^{N}}$
be all eigenvalues of $t_{\mathrm{QTM}}(0)$ 
such that $\Lambda_{1} \ge \Lambda_{2} \ge \dots \ge \Lambda_{(r+1)^{N}}$. 
Then the free energy per site (\ref{eq:free-trotter}) is
\begin{eqnarray}
&&f=-\lim_{L \to \infty}\lim_{N \to \infty}
\frac{1}{\beta L}\ln 
\Lambda_{1}^{L} \left\{ 1+
\left(\frac{\Lambda_{2}}{\Lambda_{1}}\right)^{L}
+\left(\frac{\Lambda_{3}}{\Lambda_{1}}\right)^{L}
+\right.\nonumber\\
&&\qquad\qquad\qquad\qquad\qquad\qquad \left.\cdots +
\left(\frac{\Lambda_{(r+1)^{N}}}{\Lambda_{1}}\right)^{L}
\right\} .~~~~~~~
\label{eq:free2}
\end{eqnarray}
It is assumed that there is a finite gap between the largest and the
second largest eigenvalues, $\Lambda_{1}$ and $\Lambda_{2}$, of $t_{\mathrm{QTM}}(0)$, 
at least when $N$ is a finite number. 
This gap is expected to still exist in the Trotter limit if $T>0$.
Moreover, $\Lambda_{1}, \Lambda_{2}, \dots, \Lambda_{(r+1)^{N}}$ 
do not depend on $L$. 
Thus if we exchange \cite{QTMrefsa,QTMrefsb,SI87} the order of the two limits 
$L\to \infty$ and $N\to \infty$ in (\ref{eq:free2}), the free energy per site
can be expressed in terms of only the largest eigenvalue $\Lambda_{1}$ of 
the QTM (\ref{eq:QTM}) at $v=0$.
Namely,
\begin{eqnarray}
f=-T\lim_{N\to \infty}\log \Lambda_{1}
\end{eqnarray} 
a remarkably simple result.

Now the eigenvalue \cite{FK99,JKS98,KW97} $T_{1}^{(1)}(v)$ of the QTM (\ref{eq:QTM}) 
derived by the Bethe Ansatz is 
\begin{eqnarray}
T_{1}^{(1)}(v)=\sum_{d=1}^{r+1}z(d;v).
\label{eq:t11def}
\end{eqnarray}
The functions $\{z(d;v)\}$ are defined as
\begin{equation}
z(a;v)=\psi_{a}(v)
\frac{Q_{a-1}(v-\frac{1}{2}{\rm i}(a+1))Q_{a}(v-\frac{1}{2}{\rm i}(a-2))}
{Q_{a-1}(v-\frac{1}{2}{\rm i}(a-1))Q_{a}(v-\frac{1}{2}{\rm i}a)} 
\label{eq:zdef}
\end{equation}
for $a \in \{1,2,\dots,r+1\}$, with
$$
Q_{a}(v)=\prod_{k=1}^{M_{a}}(v-v_{k}^{(a)})
$$
where $M_{a}\in {\mathbb Z}_{\ge 0}$ and $Q_{0}(v)=Q_{r+1}(v)=1$.  
The vacuum parts $\psi_{a}(v)$ appearing in (\ref{eq:zdef}) are given by
\begin{equation}
\psi_{a}(v)=
 \e^{{-\mu_{a}}/{T}}
 \phi_{+}(v+{\rm i}\, \delta_{a,r+1})\phi_{-}(v-{\rm i}\, \delta_{a,1})
  \label{eq:vac-QTM}
\end{equation}
for $a \in \{1,2,\dots,r+1\}$ and $\phi_{\pm}(v)=(v\pm {\rm i} \, u_{N})^{{N}/{2}}$. 
The complex numbers $v^{(a)}_{k}$ satisfy the Bethe equation 
\begin{equation}
 \frac{\psi_{a}(v^{(a)}_{k}+\frac{1}{2} {\rm i}\, a)}
     {\psi_{a+1}(v^{(a)}_{k}+\frac{1}{2} {\rm i} \, a)}=
-
\frac{Q_{a-1}(v^{(a)}_{k}+\frac{1}{2}{\rm i})Q_{a}(v^{(a)}_{k}-{\rm i})
      Q_{a+1}(v^{(a)}_{k}+\frac{1}{2}{\rm i})}
      {Q_{a-1}(v^{(a)}_{k}-\frac{1}{2}{\rm i})Q_{a}(v^{(a)}_{k}+{\rm i})
      Q_{a+1}(v^{(a)}_{k}-\frac{1}{2}{\rm i})}
     \label{eq:BAE} \\
\end{equation}
for $k\in \{1,2, \dots, M_{a}\}$ and $a\in \{1,2, \dots, r\}$.

All of the above is for general $r$.
Specializing to the $su(4)$ ($r=3$) case of relevance to the integrable ladder model (\ref{eq:Ham2}),
with chemical potential terms (\ref{eq:chem}), we have
\begin{eqnarray}
&& \frac{\phi_{-}(v^{(1)}_{k}-\frac{1}{2}{\rm i})}
     {\phi_{-}(v^{(1)}_{k}+\frac{1}{2}{\rm i})}=
-
\e^{\frac{\mu_{1}-\mu_{2}}{T}}
\frac{Q_{1}(v^{(1)}_{k}-{\rm i})
      Q_{2}(v^{(1)}_{k}+\frac{1}{2}{\rm i})}
      {Q_{1}(v^{(1)}_{k}+{\rm i})
      Q_{2}(v^{(1)}_{k}-\frac{1}{2}{\rm i})}\nonumber\\
&& 
1=-\e^{\frac{\mu_{2}-\mu_{3}}{T}}
\frac{Q_{1}(v^{(2)}_{k}+\frac{1}{2}{\rm i})Q_{2}(v^{(2)}_{k}-{\rm i})
      Q_{3}(v^{(2)}_{k}+\frac{1}{2}{\rm i})}
      {Q_{1}(v^{(2)}_{k}-\frac{1}{2}{\rm i})Q_{2}(v^{(2)}_{k}+{\rm i})
      Q_{3}(v^{(2)}_{k}-\frac{1}{2}{\rm i})}  \label{eq:qtm_BAE4}\\
&& \frac{\phi_{+}(v^{(3)}_{k}+\frac{3}{2}{\rm i})}
     {\phi_{+}(v^{(3)}_{k}+\frac{5}{2}{\rm i})}=
-\e^{\frac{\mu_{3}-\mu_{4}}{T}}
\frac{Q_{2}(v^{(3)}_{k}+\frac{1}{2}{\rm i})Q_{3}(v^{(3)}_{k}-{\rm i})}
      {Q_{2}(v^{(3)}_{k}-\frac{1}{2}{\rm i})Q_{3}(v^{(3)}_{k}+{\rm i})}.\nonumber 
\end{eqnarray}
In the first line, $k\in \{1,2, \dots, M_{1}\}$, in the second, $k\in \{1,2, \dots, M_{2}\}$,
and in the third $k\in \{1,2, \dots, M_{3}\}$.

\subsection{$T$-system}

The next step is to introduce an auxiliary function $T_{m}^{(a)}(v)$, 
whose origin is related to a fusion hierarchy \cite{KRS81} of the QTM 
(cf. the Bazhanov-Reshetikhin formula in \cite{BR90}),
\begin{eqnarray}
T_{m}^{(a)}(v)=\sum_{\{d_{j,k}\}} \prod_{j=1}^{a}\prod_{k=1}^{m}
z(d_{j,k};v-\ffrac{1}{2}\, {\rm i}\,(a-m-2j+2k))
\label{eq:DVF}
\end{eqnarray}
where the summation is taken over 
$d_{j,k}\in \{1,2,\dots,r+1\}$ 
such that $d_{j,k} \prec d_{j+1,k}$ and $d_{j,k} \preceq d_{j,k+1}$ 
($ 1 \prec 2 \prec \cdots \prec r+1$), with
$m \in \mathbb{Z}_{\ge 1}$ and
$a \in \{1,2,\dots,r \}$. 
This general function $T_{m}^{(a)}(v)$ contains the eigenvalue 
$T_{1}^{(1)}(v)$ in (\ref{eq:t11def}) as a special case, namely $(a,m)=(1,1)$. 
The poles of $T^{(a)}_{m}(v)$ are spurious due to the Bethe equations (\ref{eq:BAE}).
The relation (\ref{eq:DVF}) can be interpreted  as Young tableaux with spectral 
parameter $v$ if we set $z(a;v)=\framebox{$a;v$}$.

For example, for the $su(4)$ case ($r=3$), we have 
\begin{eqnarray}
&& T^{(1)}_{1}(u) = 
  \begin{array}{|c|}\hline 
     	1 ;v \\ \hline 
  \end{array}
  +
  \begin{array}{|c|}\hline 
     	2 ;v \\ \hline 
  \end{array}+
  \begin{array}{|c|}\hline 
     	3 ;v \\ \hline 
  \end{array}
  +
  \begin{array}{|c|}\hline 
     	4 ;v \\ \hline 
  \end{array} \label{eq:t11-sl4}
  \\
&& T^{(2)}_{1}(u) = 
  \begin{array}{|c|}\hline 
     	1 ;v-\frac{1}{2}{\rm i} \\ \hline
     	2 ;v+\frac{1}{2}{\rm i} \\ \hline 
  \end{array}
  +
  \begin{array}{|c|}\hline 
     	1 ;v-\frac{1}{2}{\rm i} \\ \hline
     	3 ;v+\frac{1}{2}{\rm i} \\ \hline 
  \end{array}
  +
  \begin{array}{|c|}\hline 
     	1 ;v-\frac{1}{2}{\rm i} \\ \hline
     	4 ;v+\frac{1}{2}{\rm i} \\ \hline 
  \end{array}
  +
  \begin{array}{|c|}\hline 
     	2 ;v-\frac{1}{2}{\rm i} \\ \hline
     	3 ;v+\frac{1}{2}{\rm i} \\ \hline 
  \end{array}
  +
  \begin{array}{|c|}\hline 
     	2 ;v-\frac{1}{2}{\rm i} \\ \hline
     	4 ;v+\frac{1}{2}{\rm i} \\ \hline 
  \end{array}
  \nonumber \\
& & \qquad \qquad
  +
  \begin{array}{|c|}\hline 
     	3 ;v-\frac{1}{2}{\rm i} \\ \hline
     	4 ;v+\frac{1}{2}{\rm i} \\ \hline 
  \end{array}\label{eq:t21-sl4}
  \\
&& T^{(1)}_{2}(u) =
  \begin{array}{|c|c|}\hline 
     	1 ;v+\frac{1}{2}{\rm i}& 1 ;v-\frac{1}{2}{\rm i}\\ \hline 
  \end{array}
  +
  \begin{array}{|c|c|}\hline 
     	1 ;v+\frac{1}{2}{\rm i}& 2 ;v-\frac{1i}{2}{\rm i}\\ \hline 
  \end{array}
  +
  \begin{array}{|c|c|}\hline 
     	1 ;v+\frac{1}{2}{\rm i}& 3 ;v-\frac{1}{2}{\rm i}\\ \hline 
  \end{array}
  \nonumber \\ 
 & &\qquad \qquad +
  \begin{array}{|c|c|}\hline 
     	1 ;v+\frac{1}{2}{\rm i}& 4 ;v-\frac{1}{2}{\rm i}\\ \hline 
  \end{array}
  +
  \begin{array}{|c|c|}\hline 
     	2 ;v+\frac{1}{2}{\rm i}& 2 ;v-\frac{1}{2}{\rm i}\\ \hline 
  \end{array}
  +
  \begin{array}{|c|c|}\hline 
     	2 ;v+\frac{1}{2}{\rm i}& 3 ;v-\frac{1}{2}{\rm i}\\ \hline 
  \end{array}
  \nonumber \\
  & &\qquad \qquad+
  \begin{array}{|c|c|}\hline 
     	2 ;v+\frac{1}{2}{\rm i}& 4 ;v-\frac{1}{2}{\rm i}\\ \hline 
  \end{array}
  +
  \begin{array}{|c|c|}\hline 
     	3 ;v+\frac{1}{2}{\rm i}& 3 ;v-\frac{1}{2}{\rm i}\\ \hline 
  \end{array}
  +
  \begin{array}{|c|c|}\hline 
     	3 ;v+\frac{1}{2}{\rm i}& 4 ;v-\frac{1}{2}{\rm i}\\ \hline 
  \end{array}
  \nonumber \\
  & & \qquad \qquad+
  \begin{array}{|c|c|}\hline 
     	4 ;v+\frac{1}{2}{\rm i}& 4 ;v-\frac{1}{2}{\rm i}\\ \hline 
  \end{array}
  \label{eq:t12-sl4}
\end{eqnarray}
Here we have used the notation
\begin{eqnarray*}
 && z(d_{1,1};v-\ffrac{1}{2}{\rm i})z(d_{2,1};v+\ffrac{1}{2}{\rm i})=\begin{array}{|c|}\hline 
     	d_{1,1} ;v-\frac{1}{2}{\rm i} \\ \hline
     	d_{2,1} ;v+\frac{1}{2}{\rm i} \\ \hline 
  \end{array} \nonumber \\
&& z(d_{1,1};v+\ffrac{1}{2}{\rm i})z(d_{1,2};v-\ffrac{1}{2}{\rm i})=
\begin{array}{|c|c|}\hline 
     	d_{1,1} ;v+\frac{1}{2}{\rm i}& d_{1,2} ;v-\frac{1}{2}{\rm i}\\ \hline 
  \end{array}
\end{eqnarray*} 
In this sense, relation (\ref{eq:DVF}) represents the character of
the $m$-th symmetric and the $a$-th anti-symmetric tensor representation 
of $su(r+1)$ with spectral parameter.  
For $a \in \{1,2,\dots,r\}$ and $m \in {\mathbb Z}_{\ge 1}$, it may be normalised as 
$\widetilde{T}^{(a)}_{m}(v)=T^{(a)}_{m}(v)/\widetilde{{\mathcal N}}^{(a)}_{m}(v)$, 
where
\begin{eqnarray}
\hspace{-30pt} && \widetilde{{\mathcal N}}^{(a)}_{m}(v)=
  \frac{\phi_{-}(v-\frac12 {\rm i}(a+m))\phi_{+}(v+\frac12{\rm i}(a+m))}
  {\phi_{-}(v-\frac12{\rm i}(a-m))\phi_{+}(v+\frac12{\rm i}(a-m))}
  \nonumber \\ 
\hspace{-30pt}  && \hspace{20pt} \times
  \prod_{j=1}^{a}\prod_{k=1}^{m}
  \phi_{-}(v-\ffrac12{\rm i}(a-m-2j+2k))\phi_{+}(v-\ffrac12{\rm i}(a-m-2j+2k)).
  \label{eq:normal}
\end{eqnarray}

One can show that $\widetilde{T}^{(a)}_{m}(v)$ satisfies the functional relation 
\begin{equation}
\widetilde{T}^{(a)}_{m}(v+\ffrac12{\rm i})
\widetilde{T}^{(a)}_{m}(v-\ffrac12{\rm i})
=\widetilde{T}^{(a)}_{m+1}(v)\widetilde{T}^{(a)}_{m-1}(v)
+\widetilde{T}^{(a-1)}_{m}(v)\widetilde{T}^{(a+1)}_{m}(v)
\label{eq:T-system}
\end{equation}
which is called the $T$-system \cite{KunibaFusiona,KunibaFusionb,KRfusion}.
Here $a \in \{1,2,\dots,r\}$ and $m \in {\mathbb Z}_{\ge 1}$, with
\begin{eqnarray}
&& \widetilde{T}^{(a)}_{0}(v)=1,
\quad {\rm for} \quad a \in {\mathbb Z}_{\ge 1}\nonumber \\
&& \widetilde{T}^{(0)}_{m}(v)=
 \frac{\phi_{-}(v+\frac12\,{\rm i}\,{m})\phi_{+}(v-\frac12\,{\rm i}\,m)}
  {\phi_{-}(v-\frac12\,{\rm i}\,{m})\phi_{+}(v+\frac12\,{\rm i}\,{m})},
\quad {\rm for} \quad m \in {\mathbb Z}_{\ge 1} \\
&& \widetilde{T}^{(r+1)}_{m}(v)=
\e^{\frac{-m(\mu_{1}+\mu_{2}+\cdots +\mu_{r+1})}{T}} 
\quad {\rm for} \quad m \in {\mathbb Z}_{\ge 1} \nonumber. 
\end{eqnarray}
It can be verified, for example, that  (\ref{eq:t11-sl4})-(\ref{eq:t12-sl4})
satisfy the $T$-system (\ref{eq:T-system})  
\begin{eqnarray}
&&T^{(1)}_{1}(v+\ffrac{1}{2}{\rm i})
T^{(1)}_{1}(v-\ffrac{1}{2}{\rm i})
=
\sum_{a=1}^{4}\sum_{b=1}^{4}
\begin{array}{|c|c|}\hline 
     	a;v+\frac{1}{2}{\rm i} & b;v-\frac{1}{2}{\rm i} \\ \hline 
\end{array}
\nonumber \\
&&=
\sum_{a\le b}
\begin{array}{|c|c|}\hline 
     	a;v+\frac{1}{2}{\rm i} & b;v-\frac{1}{2}{\rm i} \\ \hline 
\end{array}
+
\sum_{a>b}
\begin{array}{|c|c|}\hline 
     	a;v+\frac{1}{2}{\rm i} & b;v-\frac{1}{2}{\rm i} \\ \hline 
\end{array}
\nonumber \\
&&=
\sum_{a\le b}
\begin{array}{|c|c|}\hline 
     	a;v+\frac{1}{2}{\rm i} & b;v-\frac{1}{2}{\rm i} \\ \hline 
\end{array}
+
\sum_{a>b}
\begin{array}{|c|}\hline 
        b;v-\frac{1}{2}{\rm i} \\ \hline
     	a;v+\frac{1}{2}{\rm i} \\ \hline 
\end{array}
\nonumber \\
&&=T^{(1)}_{2}(v)+T^{(2)}_{1}(v).
\end{eqnarray} 
Then taking note of the relations
\begin{eqnarray}
{\mathcal N}^{(1)}_{1}(v+\ffrac{1}{2}{\rm i})
{\mathcal N}^{(1)}_{1}(v-\ffrac{1}{2}{\rm i})=
{\mathcal N}^{(1)}_{2}(v)=
{\mathcal N}^{(2)}_{1}(v)/ \widetilde{T}^{(0)}_{1}(v)
\end{eqnarray} 
gives the desired result
\begin{eqnarray}
\widetilde{T}^{(1)}_{1}(v+\ffrac{1}{2}{\rm i})
\widetilde{T}^{(1)}_{1}(v-\ffrac{1}{2}{\rm i})
=\widetilde{T}^{(1)}_{2}(v)
+\widetilde{T}^{(0)}_{1}(v)\widetilde{T}^{(2)}_{1}(v).
\end{eqnarray}

\subsection{Nonlinear integral equations}

The thermodynamics of the spin-$\ffrac12$ Heisenberg $XXZ$ chain 
has recently been described by a nonlinear integral equation (NLIE) \cite{Ta01}, 
which was rederived  \cite{TSK01} from the $T$-system \cite{KRfusion} of the QTM. 
It was also derived from a fugacity expansion \cite{KW02}.
There are similar NLIEs for other models \cite{ZT1,ZT2,ZT3,ZT4,ZT5}.
Most importantly, the number of unknown functions for the NLIE is only one, 
corresponding to the rank of $su(2)$. 
In this subsection, we review the derivation of the $su(r+1)$ NLIE (\ref{eq:nlie4})  
from the $T$-system (\ref{eq:T-system}), in which only $r$ unknown functions appear.
This paves the way for the exact high temperature expansion to follow in section \ref{sec:HTE}.

Numerical analysis for finite values of $N$, $u_{N}$ and $r$ reveals the expectation that 
the largest eigenvalue of the QTM (\ref{eq:QTM}) at $v=0$ is given in terms of
a one-string solution (for each colour $a=1,2,\dots, r$) of the Bethe equations (\ref{eq:BAE})
in the sector ${N}/{2}=M_{1}=M_{2}=\dots =M_{r}$.
%
%
{}From now on, we consider only this one-string solution. 
Thus $T^{(1)}_{1}(0)$ should give the largest eigenvalue $\Lambda_{1}$. 
Numerical evidence for this conclusion is given in Appendix C.
We thus state the following conjecture.

\begin{proposition}\label{conj}
For small $u_{N}$ ($|u_{N}|\ll 1$), $a \in \{1,2,\dots,r\}$, 
all zeros $\{\tilde{z}^{(a)}_{1} \}$ of $\widetilde{T}^{(a)}_{1}(v)$ are 
located outside the physical strip $\mathrm{Im} v \in [-\frac{1}{2},\frac{1}{2}]$. 
In particular, they are located near the lines $\mathrm{ Im} v= \pm \frac12(1+a)$, 
at least for the case $\mu_{1}=\mu_{2}=\dots =\mu_{r+1}=0$. 
\end{proposition}

This conjecture is expected to be valid even in the Trotter limit $N \to \infty$. 
Now $\widetilde{T}^{(a)}_{m}(v)$ has poles only at $\pm \tilde{\beta}^{(a)}_{m}$, where 
$\tilde{\beta}^{(a)}_{m}=\frac12 {\rm i} ({m+a})  + {\rm i} u_{N}$, 
whose order is at most $N/2$. 
Moreover, 
\begin{eqnarray}
Q^{(a)}_{1} =
\lim_{|v|\to \infty}\widetilde{T}^{(a)}_{1}(v)
=\sum_{1 \le i_{1} < i_{2} < \ldots < i_{a} \le r+1}
\e^{-\frac{\mu_{i_{1}}}{T}}
\e^{-\frac{\mu_{i_{2}}}{T}}
 \ldots \,
\e^{-\frac{\mu_{i_{a}}}{T}}
\label{eq:Q-sl(r+1)}
\end{eqnarray}
is a finite number.
This quantity is related to a solution of a functional relation called the 
$Q$-system \cite{K89,KR90}, for which 
$(Q^{(a)}_{m})^{2}=Q^{(a)}_{m-1}Q^{(a)}_{m+1}+Q^{(a-1)}_{m}Q^{(a+1)}_{m}$, 
where $m \in \mathbb{Z}_{\ge 1}$ and $a \in \{1,2,\dots, r \}$.

In the $su(4)$ case, we have
\begin{eqnarray}
&& Q^{(1)}_{1}=\e^{-\frac{\mu_{1}}{T}}+
\e^{-\frac{\mu_{2}}{T}}+\e^{-\frac{\mu_{3}}{T}}+
\e^{-\frac{\mu_{4}}{T}} \nonumber \\
&& Q^{(2)}_{1}=\e^{-\frac{\mu_{1}+\mu_{2}}{T}}
+\e^{-\frac{\mu_{1}+\mu_{3}}{T}}+
\e^{-\frac{\mu_{1}+\mu_{4}}{T}}
+\e^{-\frac{\mu_{2}+\mu_{3}}{T}}+
\e^{-\frac{\mu_{2}+\mu_{4}}{T}}+\e^{-\frac{\mu_{3}+\mu_{4}}{T}} \nonumber \\
&& Q^{(3)}_{1}=\e^{-\frac{\mu_{1}+\mu_{2}+\mu_{3}}{T}}
+\e^{-\frac{\mu_{1}+\mu_{2}+\mu_{4}}{T}}+
\e^{-\frac{\mu_{1}+\mu_{3}+\mu_{4}}{T}}
+\e^{-\frac{\mu_{2}+\mu_{3}+\mu_{4}}{T}} 
\nonumber \\ 
&& Q^{(4)}_{1}=\e^{-\frac{\mu_{1}+\mu_{2}+\mu_{3}+\mu_{4}}{T}}. 
\label{eq:Q-sl4}
\end{eqnarray}
It follows that we must put 
\begin{eqnarray}
\widetilde{T}^{(a)}_{1}(v)=Q^{(a)}_{1} +
\sum_{j=1}^{{N}/{2}}
\left\{
  \frac{b^{(a)}_{j}}{(v-\tilde{\beta}^{(a)}_{1})^{j}}
 +\frac{\overline{b}^{(a)}_{j}}{(v+\tilde{\beta}^{(a)}_{1})^{j}}
\right\}
\label{eq:expan}
\end{eqnarray}
where the coefficients $b^{(a)}_{j}, \overline{b}^{(a)}_{j} \in {\mathbb C}$ 
are given by the contour integrals
\begin{eqnarray}
&& b^{(a)}_{j}= \oint_{C^{(a)}} \frac{{\mathrm d} v}{2\pi {\rm i}}
 \widetilde{T}^{(a)}_{1}(v)(v-\tilde{\beta}^{(a)}_{1})^{j-1}\nonumber \\
&& \overline{b}^{(a)}_{j}=
 \oint_{\overline{C}^{(a)}} \frac{{\mathrm d} v}{2\pi {\rm i}}
 \widetilde{T}^{(a)}_{1}(v)(v+\tilde{\beta}^{(a)}_{1})^{j-1}.
 \label{eq:coeff}
\end{eqnarray}
The contour $C^{(a)}$ is a counterclockwise closed loop around 
$\tilde{\beta}^{(a)}_{1}$ which does not encircle $-\tilde{\beta}^{(a)}_{1}$.
Similarly the contour $\overline{C}^{(a)}$ is a counterclockwise 
closed loop around $-\tilde{\beta}^{(a)}_{1}$ not encircling $\tilde{\beta}^{(a)}_{1}$.

Using the $T$-system (\ref{eq:T-system}),  the integrals (\ref{eq:coeff}) can be modified to read
\begin{eqnarray}
b^{(a)}_{j}&=& \oint_{C^{(a)}} \frac{{\mathrm d} v}{2\pi {\rm i}}
 \bigg\{
 \frac{
       \widetilde{T}^{(a)}_{2}(v-\frac{1}{2} {\rm i})}
      {\widetilde{T}^{(a)}_{1}(v-{\rm i})} \nonumber \\
&& +
 \frac{\widetilde{T}^{(a-1)}_{1}(v-\frac{1}{2} {\rm i})
       \widetilde{T}^{(a+1)}_{1}(v-\frac{1}{2} {\rm i})}
      {\widetilde{T}^{(a)}_{1}(v-{\rm i})}
 \bigg\}
 (v-\tilde{\beta}^{(a)}_{1})^{j-1}\nonumber \\
\overline{b}^{(a)}_{j} &=&
 \oint_{\overline{C}^{(a)}} \frac{{\mathrm d} v}{2\pi {\rm i}}
\bigg\{
 \frac{
       \widetilde{T}^{(a)}_{2}(v+\frac{1}{2} {\rm i})}
      {\widetilde{T}^{(a)}_{1}(v+{\rm i})} \nonumber \\
&&+
 \frac{\widetilde{T}^{(a-1)}_{1}(v+\frac{1}{2} {\rm i})
       \widetilde{T}^{(a+1)}_{1}(v+\frac{1}{2} {\rm i})}
      {\widetilde{T}^{(a)}_{1}(v+{\rm i})}
 \bigg\}
 (v+\tilde{\beta}^{(a)}_{1})^{j-1}. 
 \label{eq:coeff2}
\end{eqnarray}
Substitution of (\ref{eq:coeff2}) into (\ref{eq:expan}) and 
performing the summation over $j$ then gives
\begin{eqnarray}
&& \hspace{-20pt}
\widetilde{T}^{(a)}_{1}(v)=Q^{(a)}_{1} \nonumber \\ 
&& +
\oint_{C^{(a)}} \frac{{\mathrm d} y}{2\pi {\rm i}} 
\frac{1-\left(\frac{y}{v-\tilde{\beta}^{(a)}_{1}}\right)^{{N}/{2}}}
 {v-y-\tilde{\beta}^{(a)}_{1}}
 \bigg\{ 
 \frac{
 \widetilde{T}^{(a)}_{2}(y+\tilde{\beta}^{(a)}_{1}-\frac{1}{2}{\rm i})}
 {\widetilde{T}^{(a)}_{1}(y+\tilde{\beta}^{(a)}_{1}-{\rm i})} \nonumber \\
 && \hspace{90pt} +
 \frac{\widetilde{T}^{(a-1)}_{1}(y+\tilde{\beta}^{(a)}_{1}-\frac{1}{2} {\rm i}) 
 \widetilde{T}^{(a+1)}_{1}(y+\tilde{\beta}^{(a)}_{1}-\frac{1}{2} {\rm i})}
 {\widetilde{T}^{(a)}_{1}(y+\tilde{\beta}^{(a)}_{1}-{\rm i})}
 \bigg\} \nonumber \\
&& +
\oint_{\overline{C}^{(a)}} \frac{{\mathrm d} y}{2\pi {\rm i}} 
 \frac{1-\left(\frac{y}{v+\tilde{\beta}^{(a)}_{1}}\right)^{{N}/{2}}}
 {v-y+\tilde{\beta}^{(a)}_{1}}
 \bigg\{ 
 \frac{
 \widetilde{T}^{(a)}_{2}(y-\tilde{\beta}^{(a)}_{1}+\frac{1}{2} {\rm i})}
 {\widetilde{T}^{(a)}_{1}(y-\tilde{\beta}^{(a)}_{1}+{\rm i})} \nonumber \\
 && \hspace{90pt} +
 \frac{\widetilde{T}^{(a-1)}_{1}(y-\tilde{\beta}^{(a)}_{1}+\frac{1}{2} {\rm i}) 
 \widetilde{T}^{(a+1)}_{1}(y-\tilde{\beta}^{(a)}_{1}+\frac{1}{2} {\rm i})}
 {\widetilde{T}^{(a)}_{1}(y-\tilde{\beta}^{(a)}_{1}+{\rm i})}
 \bigg\}  \label{eq:nlie1}
\end{eqnarray}
for $a \in \{1,2,\dots,r\}$ and $m \in {\mathbb Z}_{\ge 1}$.
Here the contour $C^{(a)}$ (resp. $\overline{C}^{(a)}$) 
is a counter\-clockwise closed loop around the origin which does not encircle 
$-2\tilde{\beta}^{(a)}_{1}$ (resp. $ 2 \tilde{\beta}^{(a)}_{1}$).
There are cancellations of singularities in the denominator and 
the numerator in each term in the first curly bracket $\{ \cdots \}$ in (\ref{eq:nlie1}). 
The poles in the first term in the first bracket are $-2 \tilde{\beta}^{(a)}_{1}$ and 
$\tilde{z}^{(a)}_{1}-\tilde{\beta}^{(a)}_{1}+{\rm i}$; 
the poles in the second term in the first bracket 
are $0, -2 \tilde{\beta}^{(a)}_{1}$ and 
$\tilde{z}^{(a)}_{1}-\tilde{\beta}^{(a)}_{1}+{\rm i}$. 
Thus the contribution to the contour integral from the first term in the first bracket vanishes if 
the contour $C^{(a)}$ does not encircle the pole at 
$\tilde{z}^{(a)}_{1}-\tilde{\beta}^{(a)}_{1}+{\rm i}$ 
(cf. Conjecture \ref{conj}). 
As for the second curly bracket in (\ref{eq:nlie1}), 
the poles in the first term are $2 \tilde{\beta}^{(a)}_{1}$ and 
$\tilde{z}^{(a)}_{1}+\tilde{\beta}^{(a)}_{1}-{\rm i}$; 
the poles in the second term are $0, 2 \tilde{\beta}^{(a)}_{1}$ and 
$\tilde{z}^{(a)}_{1}+\tilde{\beta}^{(a)}_{1}-{\rm i}$. 
Thus the contribution to the contour integral from the first term in the second bracket vanishes if 
 the contour $\overline{C}^{(a)}$ does not encircle the pole 
 at $\tilde{z}^{(a)}_{1}+\tilde{\beta}^{(a)}_{1}-{\rm i}$.

It therefore follows that (\ref{eq:nlie1}) reduces to 
\begin{eqnarray}
&&\widetilde{T}^{(a)}_{1}(v)=Q^{(a)}_{1} \nonumber \\ 
&&+
\oint_{C^{(a)}} \frac{{\mathrm d} y}{2\pi {\rm i}} 
\frac{1-\left(\frac{y}{v-\tilde{\beta}^{(a)}_{1}}\right)^{{N}/{2}}}
 {v-y-\tilde{\beta}^{(a)}_{1}}
 \frac{\widetilde{T}^{(a-1)}_{1}(y+\tilde{\beta}^{(a)}_{1}-\frac{1}{2}{\rm i}) 
 \widetilde{T}^{(a+1)}_{1}(y+\tilde{\beta}^{(a)}_{1}-\frac{1}{2}{\rm i})}
 {\widetilde{T}^{(a)}_{1}(y+\tilde{\beta}^{(a)}_{1}-{\rm i})}
 \nonumber \\
&&+
\oint_{\overline{C}^{(a)}} \frac{{\mathrm d} y}{2\pi {\rm i}} 
 \frac{1-\left(\frac{y}{v+\tilde{\beta}^{(a)}_{1}}\right)^{{N}/{2}}}
 {v-y+\tilde{\beta}^{(a)}_{1}}
 \frac{\widetilde{T}^{(a-1)}_{1}(y-\tilde{\beta}^{(a)}_{1}+\frac{1}{2}{\rm i}) 
 \widetilde{T}^{(a+1)}_{1}(y-\tilde{\beta}^{(a)}_{1}+\frac{1}{2}{\rm i})}
 {\widetilde{T}^{(a)}_{1}(y-\tilde{\beta}^{(a)}_{1}+{\rm i})}~~~~~~~~
\label{eq:nlie2}
\end{eqnarray}
for $a \in \{1,2,\dots,r\}$.
Here the contour $C^{(a)}$ (resp. $\overline{C}^{(a)}$) 
is a counterclockwise closed loop around the origin not encircling 
$\tilde{z}^{(a)}_{1}-\tilde{\beta}^{(a)}_{1}+{\rm i}$, 
$-2\tilde{\beta}^{(a)}_{1}$ 
(resp. $\tilde{z}^{(a)}_{1}+\tilde{\beta}^{(a)}_{1}-{\rm i}$, 
$2 \tilde{\beta}^{(a)}_{1}$).
The poles at $y=0$ in 
$\widetilde{T}^{(a-1)}_{1}(y \pm \tilde{\beta}^{(a)}_{1} \mp \frac{1}{2}{\rm i})$
are cancelled by the zeros at $y=0$ 
in $(\frac{y}{v \mp \tilde{\beta}^{(a)}_{1}})^{{N}/{2}}$. 
Thus (\ref{eq:nlie2}) is further simplified to 
\begin{eqnarray}
\widetilde{T}^{(a)}_{1}(v)&=&Q^{(a)}_{1} \nonumber \\ 
&+&
\oint_{C^{(a)}} \frac{{\mathrm d} y}{2\pi {\rm i}} 
 \frac{\widetilde{T}^{(a-1)}_{1}(y+\tilde{\beta}^{(a)}_{1}-\frac{1}{2}{\rm i}) 
 \widetilde{T}^{(a+1)}_{1}(y+\tilde{\beta}^{(a)}_{1}-\frac{1}{2}{\rm i})}
 {(v-y-\tilde{\beta}^{(a)}_{1})\widetilde{T}^{(a)}_{1}(y+\tilde{\beta}^{(a)}_{1}-{\rm i})}
 \nonumber \\
&+&
\oint_{\overline{C}^{(a)}} \frac{{\mathrm d} y}{2\pi {\rm {\rm i}}} 
 \frac{\widetilde{T}^{(a-1)}_{1}(y-\tilde{\beta}^{(a)}_{1}+\frac{1}{2}{\rm i}) 
 \widetilde{T}^{(a+1)}_{1}(y-\tilde{\beta}^{(a)}_{1}+\frac{1}{2}{\rm i})}
 {(v-y+\tilde{\beta}^{(a)}_{1})
 \widetilde{T}^{(a)}_{1}(y-\tilde{\beta}^{(a)}_{1}+{\rm i})}
 \label{eq:nlie3}   
\end{eqnarray}
for $a \in \{1,2,\dots,r\}$.

This is the NLIE for finite Trotter number.
The next important step is to take the Trotter limit $N \to \infty$. 
Define  $\mathcal{T}^{(a)}_{1}(v)=\lim_{N \to \infty} \widetilde{T}^{(a)}_{1}(v)$,
then
\begin{equation} 
\mathcal{T}^{(0)}_{1}(v)=
\exp \left(\frac{J_{\parallel}}{(v^2+\frac{1}{4})T}\right).
\end{equation}
Finally, we have arrived at a system of NLIE which
contains only a {\em finite} number of unknown functions,
$\{{\mathcal T}^{(a)}_{1}(v) \}_{1\le a \le r}$, namely
\begin{eqnarray}
{\mathcal T}^{(a)}_{1}(v)&=&Q^{(a)}_{1} 
+
\oint_{C^{(a)}} \frac{{\mathrm d} y}{2\pi {\rm i}} 
 \frac{\mathcal{T}^{(a-1)}_{1}(y+\beta^{(a)}_{1}-\frac{1}{2}{\rm i}) 
 \mathcal{T}^{(a+1)}_{1}(y+\beta^{(a)}_{1}-\frac{1}{2}{\rm i})}
 {(v-y-\beta^{(a)}_{1})\mathcal{T}^{(a)}_{1}(y+\beta^{(a)}_{1}-{\rm i})}
 \nonumber \\
& &+
\oint_{\overline{C}^{(a)}} \frac{{\mathrm d} y}{2\pi {\rm i}} 
 \frac{\mathcal{T}^{(a-1)}_{1}(y-\beta^{(a)}_{1}+\frac{1}{2}{\rm i}) 
 \mathcal{T}^{(a+1)}_{1}(y-\beta^{(a)}_{1}+\frac{1}{2}{\rm i})}
 {(v-y+\beta^{(a)}_{1})\mathcal{T}^{(a)}_{1}(y-\beta^{(a)}_{1}+{\rm i})}
\label{eq:nlie4}
\end{eqnarray}
for $a \in \{1,2,\dots,r\}$.
Here $\mathcal{T}^{(r+1)}_{1}(v)=\exp(-(\mu_{1}+\cdots +\mu_{r+1})/T)$. 
The contour $C^{(a)}$ (resp. $\overline{C}^{(a)}$) 
is a counterclockwise closed loop around the origin such that 
$y \ne v-\beta^{(a)}_{1}$ (resp. $y \ne v+\beta^{(a)}_{1}$) and it 
does not encircle $z^{(a)}_{1}-\beta^{(a)}_{1}+{\rm i}$, $-2\beta^{(a)}_{1}$ 
(resp. $z^{(a)}_{1}+\beta^{(a)}_{1}-{\rm i}$, 
$2 \beta^{(a)}_{1}$), where 
$z^{(a)}_{1}=\lim_{N\to \infty }{\tilde z}^{(a)}_{1}$.

In particular, for the $su(4)$ case, the NLIEs are 
\begin{eqnarray}
{\mathcal T}^{(1)}_{1}(v)=Q^{(1)}_{1} 
&+&
\oint_{C^{(1)}} \frac{{\mathrm d} y}{2\pi {\rm i}} 
 \frac{\exp \left(\frac{J_{\parallel}}{y(y+{\rm i})T}\right) 
 \mathcal{T}^{(2)}_{1}(y+\frac{1}{2}{\rm i})}
 {(v-y-{\rm i})\mathcal{T}^{(1)}_{1}(y)}
 \nonumber \\
&+&
\oint_{\overline{C}^{(1)}} \frac{{\mathrm d} y}{2\pi {\rm i}} 
 \frac{\exp \left(\frac{J_{\parallel}}{y(y-{\rm i})T}\right) 
 \mathcal{T}^{(2)}_{1}(y-\frac{1}{2}{\rm i})}
 {(v-y+{\rm i})\mathcal{T}^{(1)}_{1}(y)}
 \nonumber \\ 
{\mathcal T}^{(2)}_{1}(v)=Q^{(2)}_{1} 
&+&
\oint_{C^{(2)}} \frac{{\mathrm d} y}{2\pi {\rm i}} 
 \frac{\mathcal{T}^{(1)}_{1}(y+{\rm i}) 
 \mathcal{T}^{(3)}_{1}(y+{\rm i})}
 {(v-y-\frac{3}{2}{\rm i})\mathcal{T}^{(2)}_{1}(y+\frac{1}{2}{\rm i})}
 \nonumber \\
&+&
\oint_{\overline{C}^{(2)}} \frac{{\mathrm d} y}{2\pi {\rm i}} 
 \frac{\mathcal{T}^{(1)}_{1}(y-{\rm i}) 
 \mathcal{T}^{(3)}_{1}(y-{\rm i})}
 {(v-y+\frac{3}{2}{\rm i})\mathcal{T}^{(2)}_{1}(y-\frac{1}{2}{\rm i})}
 \nonumber \\ 
{\mathcal T}^{(3)}_{1}(v)=Q^{(3)}_{1} 
&+&
\e^{-\frac{\mu_{1}+\mu_{2}+\mu_{3}+\mu_{4}}{T}}
\oint_{C^{(3)}} \frac{{\mathrm d} y}{2\pi {\rm i}} 
 \frac{\mathcal{T}^{(2)}_{1}(y+\frac{3}{2}{\rm i})}
 {(v-y-2{\rm i})\mathcal{T}^{(3)}_{1}(y+{\rm i})}
 \nonumber \\
&+&
\e^{-\frac{\mu_{1}+\mu_{2}+\mu_{3}+\mu_{4}}{T}}
\oint_{\overline{C}^{(3)}} \frac{{\mathrm d} y}{2\pi {\rm i}} 
 \frac{\mathcal{T}^{(2)}_{1}(y-\frac{3}{2}{\rm i})}
 {(v-y+2{\rm i})\mathcal{T}^{(3)}_{1}(y-{\rm i})}
 \label{eq:nlie-sl4}
\end{eqnarray}
where $Q^{(a)}_{1}$ are given in (\ref{eq:Q-sl4}).

The free energy per site follows from (\ref{eq:nlie4}) and the relation
\begin{eqnarray}
f=J_{\parallel}-T\ln \mathcal{T}^{(1)}_{1}(0). 
\label{eq:free-en}
\end{eqnarray}

\subsection{High temperature expansion}
\label{sec:HTE}

We now turn to the calculation of the high temperature expansion of 
the free energy (\ref{eq:free-en}) from the NLIE (\ref{eq:nlie4}).  
In the $su(2)$ case (the isotropic Heisenberg chain), 
the high temperature expansion of the free energy 
was calculated from the NLIE \cite{Ta01} up to order 100 \cite{HTE2}.
The first step is to assume the expansion
\begin{eqnarray}
\mathcal{T}^{(a)}_{1}(v)=
 \exp \left(\sum_{n=0}^{\infty}b_{n}^{(a)}(v)
 \left(\frac{J_{\parallel}}{T}\right)^{n} \right), 
 \quad a \in \{1,2,\dots, r\}
 \label{eq:t-expan}
\end{eqnarray}
for large $T$.
In contrast to the $su(2)$ case \cite{HTE2},  poles in the coefficients $\{b_{n}^{(a)}(v)\}$
have to be taken into account. 
This requires the further assumption,
\begin{eqnarray}
b_{n}^{(a)}(v)=\sum_{j=0}^{n-1}
 \frac{c_{n,j}^{(a)}v^{2j}}{(v^2+\frac{(a+1)^2}{4})^n}
\end{eqnarray}
where the coefficients $c_{n,j}^{(a)}$ do not depend on $v$. 
In general, these coefficients are rational functions of $Q^{(a)}_{1}$ given
in (\ref{eq:Q-sl(r+1)}) and thus they are $T$-dependent.
The  coefficients $b_{n}^{(a)}(v)$ are obtained by 
substituting (\ref{eq:t-expan}) into (\ref{eq:nlie4}),
with $b_{0}^{(a)}(v)=\log Q^{(a)}_{1}$.

For the $su(4)$ case relevant to the integrable ladder model
the first few terms of the free energy are given by \cite{ZT2}
\begin{equation}
-\frac{1}{T}f(T,H)=\ln Q^{(1)}_1+c^{(1)}_{1,0}\left(\frac{J}{T}\right)
+c^{(1)}_{2,0}\left(\frac{J}{T}\right)^2+c^{(1)}_{3,0}\left(\frac{J}{T}\right)^3+\cdots
\label{eq:freeenergyHTE}
\end{equation}
The coefficients $c^{(1)}_{n,0}$ are given up to order $n=5$ in Appendix A.
The key point for the ladder model is the quantity $Q^{(a)}_{1}$ given in (\ref{eq:Q-sl4})
which involves the chemical potentials.

The advantage of this approach is that the thermal and magnetic properties 
can be evaluated directly from the exact free energy expression (\ref{eq:freeenergyHTE})
using the standard relations
\begin{equation}
M=- \frac{\partial f(T,H)}{\partial H} 
\!\mid_{T}, \,\,
\chi =-
\frac{\partial^2 f(T,H)}{\partial H^2} 
\!\mid_{T}, \,\,
C= - T\! 
\frac{\partial^2 f(T,H)}{\partial T^2} 
\!\mid_{H} 
\label{eq:Thermo-Relat}
\end{equation}
which define the magnetization, the magnetic susceptibility and the magnetic specific heat.

\section{Thermodynamics of the spin-$\frac{1}{2}$ ladder compounds}
\label{sec:Exp_comp}

Although many two-leg ladders have been synthesised and extensively
studied, the critical points have been measured only  for a few 
strong coupling compounds with accessible gaps.
Theoretical results for the two-leg spin-$\frac12$ ladder model have been
mainly focussed on the thermal and magnetic properties of the
standard Heisenberg ladder. 
According to a celebrated result from perturbation
theory \cite{ladder2,FT1,exp1}, the first-order terms for the zero
temperature gap $\Delta $ and the critical field $H_{c2}$ are given by
$\Delta =J_{\perp}-J_{\parallel}$ and $\mu_BgH_{c2}=
J_{\perp}+2J_{\parallel}$, which are seen to be in good agreement
with the experimental results for some strong coupling compounds \cite{exp1}.
In addition, the susceptibility as a function of temperature, $\chi \propto e^{-\Delta/T}/\sqrt{T}$, 
derived from the quantum transfer matrix algorithm \cite{Troyer} is generally used  to fit the 
susceptibility of the two-leg ladder compounds.

In comparison with perturbation theory or numerical and analytic
methods \cite{White1,numers1,numers2,numers4,numers5} applied to
ladder-like systems, the integrable ladder model (\ref{eq:Ham2})
\cite{I-ladder1,TBAladder1a,TBAladder1b,HTE1,ying2} has a distinct advantage for
studying phase diagrams and quantities such as the high field magnetization, 
specific heat and susceptibility, as they follow directly from the exact free energy.  
Moreover, the ground state properties at zero temperature are now  
understood by means of the TBA \cite{TBAladder1a,TBAladder1b}. 
As demonstrated in section \ref{sec:TBA}, the critical fields,
$ H_{c1}=(J_{\perp}-4J_{\parallel})/\mu_Bg$ and $H_{c2}= (J_{\perp}+4J_{\parallel})/\mu_Bg$, 
also give a good fit for the strong coupling compounds. 
The coupling constants $J_{\perp}$ and $J_{\parallel}$ can be fixed by fitting experimental
results for the susceptibility and the magnetization. 
In general, these coupling constants cannot be fixed from a good
fit just to the susceptibility, as it determines only the amplitude of the gap.  
The way to determine the coupling constants has been discussed at the begining of section \ref{sec:mag}. 
In general the values suggested from only the critical fields are not the best for fitting. 
The coupling constants should provide a good overall fit for each of thermodynamic properties. 
In the following we apply the integrable model approach to the compounds 
(5IAP)$_2$CuBr$_4\cdot 2$H$_2$O \cite{5IAP}, 
bis $5$-iodo-$2$-aminopyridinium tetrabromocuprate (II) dihydrate
(abbreviated B$5$i$2$aT), Cu$_{2}$(C$_5$H$_{12}$N$_2$)$_2$Cl$_4$
(abbreviated CuHpCl) \cite{compound1,Chaboussant1,Chaboussant2}, 
(C$_5$H$_{12}$N )$_2$CuBr$_4$ (abbreviated BPCB) \cite{Watson}, BIP-BNO \cite{BIPa,BIPb} 
and [Cu$_2$(C$_2$O$_2$)(C$_{10}$H$_8$N$_2$)$_2$)](NO$_3$)$_2$ 
(abbreviated CuCON) \cite{SC2}.

Values for the critical fields, obtained from the integrable spin ladder (ISL) via TBA at zero
temperature \cite{TBAladder1a,TBAladder1b}, are compared with the experimental results
for the Heisenberg spin ladder model (HSL) at low temperature in table 1. 
The small discrepancy between the TBA critical fields at
zero temperature and experimental critical fields at low temperature
is a finite temperature effect. 
As the temperature increases the energy gap decreases due to triplet excitations, as the
energy gap is a measure of the energy cost of the first triplet excitation at zero temperature. 
The experimental gap is determined by the first critical field $H_{c1}$ at low temperatures.
It will be seen below that the thermal and magnetic properties derived from the 
HTE are in excellent agreement with the experimental results.

\begin{table}
\begin{center}
\begin{tabular}{|c|c|c|c|c|c|}
\hline 
Compounds&B$5$i$2$aT&Cu(Hp)Cl&BPCB&BIP-BNO&CuCON
\\
\hline 
HSL $J_{\perp}$&$13.0\,$K&$13.1\,$K&$13.3\,$K&$72\,$K&$509\,$K
\\
\hline 
HSL $J_{\parallel}$&$1.15\,$K&$2.62\,$K&$3.8\,$K&$17\,$K&$44\,$K
\\
\hline 
ISL $J_{\perp}$&$13.3\,$K&$14.5\,$K&$15.4\,$K&$83\,$K&$520\,$K
\\
\hline 
ISL $J_{\parallel}$&$0.29\,$K&$1.0\,$K&$1.2\,$K&$7.0\,$K&$10\,$K
\\
\hline 
HSL $H_{c1}$&$8.3\,$T&$7.5\,$T&$6.6\,$T&$38.8\,$T&$323\,$T
\\
\hline 
TBA $H_{c1}$&$8.6\,$T&$7.4\,$T&$7.4\,$T&$41\,$T&$334\,$T
\\
\hline 
HSL $H_{c2}$&$10.4\,$T&$13.0\,$T&$14.6\,$T&&$415\,$T
\\
\hline 
TBA $H_{c2}$&$10.02\,$T&$13.1\,$T&$14.1\,$T&$83\,$T&$389\,$T
\\
\hline 
\end{tabular}
\caption{The coupling strengths, $J_{\perp}$ and  $J_{\parallel}$, and critical field values, 
$H_{c1}$ and $H_{c2}$ 
obtained from previous fits with the Heisenberg spin ladder (HSL) and for 
the integrable spin ladder (ISL) using the TBA approach for some strong coupling ladder 
compounds (see text).} 
\end{center}
\end{table}

\subsection{Note on conversion constants}

Before turning to the study of the thermodynamics of these compounds via the HTE, 
a remark on conversion constants is in order.
Overall conversion constants for the susceptibility, the specific heat
and the magnetization are needed due to the fact that the $T$-system
(\ref{eq:T-system}) is specified up to normalization.
Therefore, in order to match the experimental data for the
unnormalized magnetic properties, a scaling factor needs to be fixed for 
the magnetic properties, such as the magnetization, susceptibility,
specific heat and magnetic entropy.  
In the following, we will see that the conversion constant 
$\chi_{{\rm HTE}}\approx 0.40615 \chi_{{\rm EXP}}$  
for the susceptibility of the ladder compounds is half that of the spin-$1$ chain compounds
\cite{BGN,BGNF} due to there being $2L$ spins in the ladder model.
The specific heat conversion constant is material dependent due to the choice of 
experimental units. 
However, for normalized magnetic properties, 
such as the magnetization ($M/M_s$), the conversion constant plays
no role in fitting the experimental data.

{}From the derivation of the HTE it can be seen that the free energy expansion
\begin{equation}
f_{HTE} (T,H)= c_0 + c_1 \frac{J}{T}
+c_2 \left(\frac{J}{T}\right)^2
+c_3 \left(\frac{J}{T}\right)^3 + \ldots
\label{eq:HTEfree}
\end{equation}
is only fixed up to a normalisation constant, i.e.
\begin{equation}
f_{{\rm phys.units}}= \gamma_{{\rm conv.}} \cdot f_{HTE} (T,H).
\label{eq:HTEfree2}
\end{equation}
Physically this amounts to the fact that the units of energy,
temperature, external field and coupling strength are not unified  in
the theory, i.e., there is no analogue of a `characteristic length'.
Mathematically, it can be traced back to the unit of temperature and
the multiplicativity of the $R$-matrix of the model. 
The scaling factors for different magnetic quantities such as susceptibility,
specific heat and magnetic entropy can be uniquely fixed by their
corresponding experimental curves. 
They depend on their experimental units and are given in the following figure captions.

In the HTE calculations for physical properties this factor has
previously been used to obtain a best fit for experimental data via vertical scaling. 
This procedure is open to critique, as it would offer a new free parameter, 
enabling a better fit than from the pure HTE theory.  
The conversion factor can in fact be calculated, leading to 
predictions of physical properties in physical units (SI) instead of arbitrary HTE units.
One remaining problem is the widespread use of non-standard non-SI units 
in the experimental measurements.
These measurements sometimes even depend on compound-specifics such as 
molecular mass. 
The problem of converting SI units to experimental units, while often not trivial, 
is in no way related to the HTE method and will thus not be further discussed here. 
HTE predictions are made in SI units.

In eqns (\ref{eq:HTEfree}) and (\ref{eq:HTEfree2}) the unknown constant $\gamma_{{\rm conv.}}$ 
is a prefactor, i.e., the same in all orders of the expansion. 
The expansion is generated for interacting spins from the Bethe Ansatz solution. 
For zero order,  the non-interacting case $J=0$,  the expansion reduces to the trivial (non Bethe Ansatz) 
case of $L$ independent spins (here, e.g., a  two site spin-$\frac{1}{2}$ system), for which
\begin{equation}
f_{{\rm 0th\, order}} = \gamma_{{\rm  conv.}} c_0
= -k_B T \ln \left(    {\rm e}^{\frac{\mu_1}{k_B T}}+{\rm e}^{\frac{\mu_2}{k_B T}}     \right).
\end{equation}
The second equality holds for the special point $J=0$, but since $\gamma_{{\rm  conv.}}$ is a constant
it can be extended in eqns (\ref{eq:HTEfree}) and (\ref{eq:HTEfree2}) to all $J>0$.

Once the free energy is thus known in SI units, it is straightforward to obtain all other properties 
derived from it in SI units via
\begin{equation}
M_{{\rm  phys. units.}}=-\frac{\partial}{ \partial H} f_{{\rm phys. units.}}
\end{equation}
The important point here is that the HTE method does make predictions in physical units, 
even if the original $T$-system is only fixed up to normalisation. 
Thus all conversion factors can in theory be calculated using the additional 
information from the non-interacting spin system or directly from
fitting experimental curves.

In practice the HTE scheme is unable to access thermal and magnetic properties 
for temperature scales less than $J_{\parallel}$.
At high temperatures the first few terms in the free energy expansion 
are sufficient to determine the thermodynamic properties.  
The HTE approach works best in practice for small ratios of
${J_\parallel}/{T}$, as only the lowest few terms of the series
are needed to get a good agreement with the experimental results. 
Consequently for high temperatures and materials with a small coupling constant
$J_\parallel$ even the zeroth order HTE suffices to capture the main physical properties. 
This lowest order corresponds to the approximation where $J_\parallel=0$, 
sometimes called the rung-dimer model \cite{MU03}.
In this case the ground state wavefunction is a simple product of rung singlets.  
To this order the partition function of the ladder model consists of only four terms (see
the first term of the $Q$-system for comparison) and can
be evaluated directly without the use of the integrable theory. 
However, not all interesting physical properties can be captured
by this simple dimer model, which is the zeroth order of the HTE series. 
Especially as much of the physics of interest lies in the
lower temperature regime, with the parameter $J_\parallel$ material dependent
and not necessarily small enough for the dimer approximation.

\begin{figure}
\includegraphics[width=.60\linewidth,angle=-90]{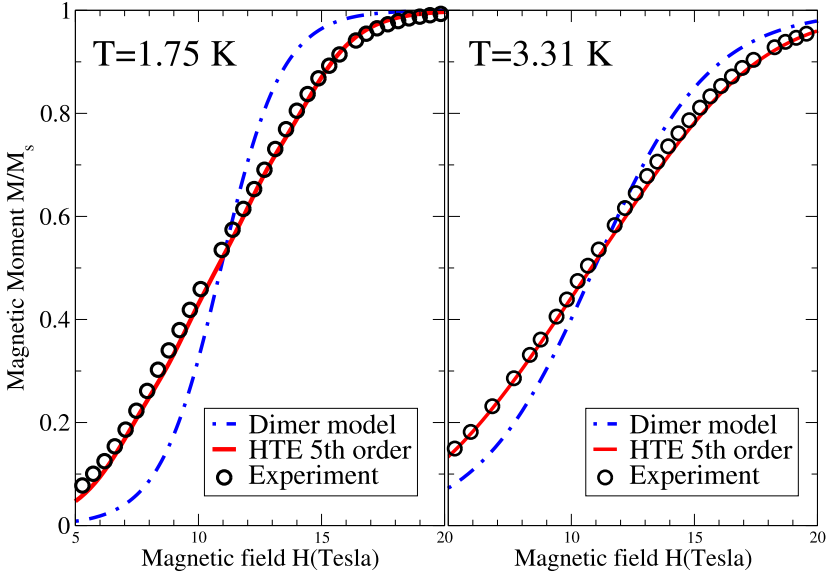}
\caption{
Comparison between the dimer approximation (0th order HTE) and the 5th order HTE result for the
magnetization of the compound (C$_5$H$_{12}$N)$_2$CuBr$_4$.
The dimer approximation does not distinguish the two critical field values, 
for which the experimental difference is $H_{c1}-H_{c1}\approx 7\,$T (see text).
}
\label{fig:dimer-comparison}
\end{figure}

Figure \ref{fig:dimer-comparison} shows a comparison of the magnetization curve at 
two different temperatures for the ladder compound (C$_5$H$_{12}$N)$_2$CuBr$_4$.
It can be seen that the zeroth order approximation of the pure dimer model captures
the qualitative shape of the curve, whereas the HTE series result (discussed further
in figure \ref{fig:BPCBmag}) is in quantitative agreement with the experimental measurement.
The dimer model is obviously a bad approximation for this compound, as the two critical fields,  
namely $H_{c1}=7\,$T and $H_{c2}=14.1\,$T are far apart. 
The difference between these values is proportional to the coupling $J_\parallel$, which the 
dimer model neglects. 

We proceed now to the comparison with the experimental data for a number of compounds.


\subsection{{\rm(5IAP)$_2$CuBr$_4\cdot 2$H$_2$O}}
\label{sec:5IAP}

\subsubsection{Susceptibility} 
\begin{figure}
\begin{center} 
\vskip 5mm
\includegraphics[width=.60\linewidth]{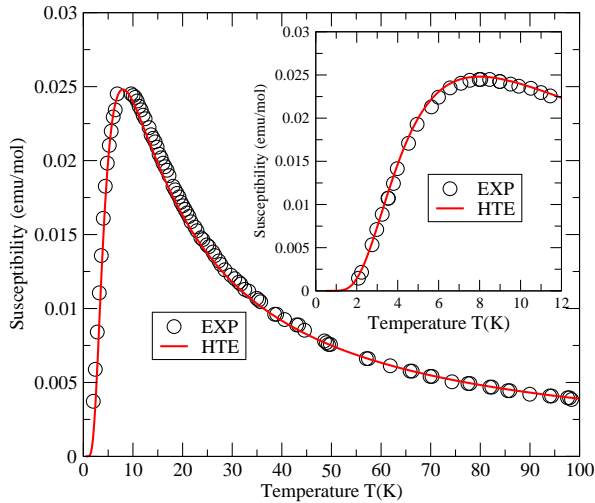}
\caption{
Susceptibility versus temperature for the compound B$5$i$2$aT.
Circles denote the experimental data  \cite{5IAP} and the solid curve
is the susceptibility evaluated directly from the HTE at $H=1\,$T. 
A parameter fit suggests the coupling constants  $J_{\perp}=13.3\,$K and 
$J_{\parallel}=0.2875\,$K with $g=2.1$ and  $\mu_B=0.672\,$K/T. 
The inset shows the same fit in the susceptibility at low temperature. 
The conversion constant is $\chi_{{\rm HTE}}\approx 0.40615 \chi _{{\rm EXP}}$. 
}
\label{fig:B5i2aTsus}
\end{center}
\end{figure}

Measurement of the susceptibility and magnetization of the compound 
B$5$i$2$aT \cite{5IAP} suggests that it lies in the strong coupling regime, 
with coupling constants $J_{\perp}=13.0\,$K, $J_{\parallel}=1.15\,$K  
for the standard Heisenberg ladder. 
{}From the application of the HTE for the Hamiltonian (\ref{eq:Ham2}) we find that
the coupling constants $J_{\perp}=13.3\,$K and $J_{\parallel}=0.2875\,$K
give excellent fits to both the susceptibility and the magnetization.
The temperature dependence of the susceptibility curves is given
in figure  \ref{fig:B5i2aTsus}, where the solid line denotes the
susceptibility derived from the free energy (\ref{eq:freeenergyHTE}) with up to
fifth order of HTE. 
An overall excellent fit with the experimental susceptibility curve is observed.  
The typical rounded peak in the zero magnetic field susceptibility curve, 
characteristic of a low dimensional antiferromagnet, is seen around $8.1\,$K. 
The susceptibility exponentially decreases with further decrease of the 
temperature toward zero.  
The TBA analysis suggests an energy gap $\Delta \approx 8.6\,$T.

The low temperature behaviour of the magnetic susceptibility, 
shown in the inset of figure \ref{fig:B5i2aTsus}, is also in excellent 
agreement with the experimental data. 
We observed that the magnetic field raises the susceptibility 
due to the lower triplet component being energetically favoured 
by the magnetic field at low temperature.
If the magnetic field is much smaller than the temperature, there
is no significant change to the susceptibility.

\subsubsection{High field magnetization} 
\begin{figure}
\begin{center} 
\includegraphics[width=.60\linewidth]{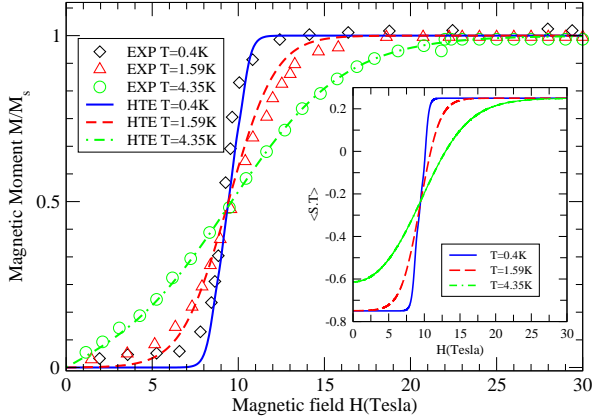}
\caption{
Magnetization versus magnetic field indicating the high field quantum phase diagram 
for the compound B$5$i$2$aT with the same constants as in Fig.\ref{fig:B5i2aTsus}.
The discrepancy in magnetization curves at $T=0.4\,$K and $T=1.59\,$K is due to  
paramagnetic impurities, which result in nonzero magnetization in the  singlet ground state. 
This impurity effect becomes negligible at higher temperatures. 
The inset shows the one point correlation function vs magnetic field -- at low temperature 
the singlet ground state can be clearly identified. 
}
\label{fig:B5i2aTmag}
\end{center}
\end{figure}

The behaviour of the field dependent magnetization curve can lead to the 
prediction of the low temperature phase diagram as well as the magnetization plateaux. 
It follows that the magnetization is a particular interesting quantity for studying 
critical behaviour. 
The high field magnetization curves evaluated from the HTE at different temperatures 
are shown in figure  \ref{fig:B5i2aTmag} in comparison with the experimental curves,
with which there is also excellent agreement.
At very low temperature, provided the magnetic field is less than the critical field $H_{c1}$, 
the rung singlet forms a singlet ground state. 
The antiferromagnetic correlation length is finite while the triplet state is gapful. 
At finite temperature the triplet excitations become involved so that the gap 
decreases with increasing temperature. 
This behaviour is observed in the high field magnetization curves at $T=1.59\,$K and $T=4.35\,$K 
presented in figure \ref{fig:B5i2aTmag}. 
The gap closes at the critical field $H_{c1}=\Delta/\mu_Bg$. 
If the magnetic field is above the critical point $H_{c1}$, the lower component of the triplet
becomes involved in the ground state. 
At zero temperature, one can show rigorously that the other two
(higher) components of the triplet never become involved in the ground state.
Therefore, in this vicinity, the strongly coupled two-leg ladder can be mapped 
to the $XXZ$ Heisenberg chain with an effective magnetic field term, as done
in section \ref{sec:TBA} and in \cite{FT1,FT2}. 
The magnetization increases almost linearly with the field towards the
critical point $H_{c2}$, at which the ground state becomes fully polarized. 
In addition, the inflection point is indicated at 
$H_{{\rm IP}} \approx J_{\perp}/\mu_Bg$ where the magnetization moment
is equal to $0.5$. 
At $T=0.4\,$K, the HTE magnetization curve indicates that $H_{c1} \approx 8.6\,$T 
and $H_{c2} \approx 10.02\,$T.
These values are in excellent agreement with the experimental estimates of
$H_{c1}=8.3\,$T and $H_{c2}=10.4\,$T.  
The experimental magnetization in the singlet ground state at low temperature appears
to be nonzero. 
This is caused by paramagnetic impurities due to the synthesization process.

The inflection point is clearly visible in the experimental
magnetization curves for the compound B$5$i$2$aT \cite{5IAP}.
It follows that for the strong coupling ladder compounds at zero temperature, 
the one point correlation function $\langle S_j \cdot T_j\rangle =-\frac{3}{4}$ 
lies in a gapped singlet groundstate, which indicates an ordered dimer phase, 
while $\langle S_j \cdot T_j\rangle =\frac{1}{4}$ in the fully-polarized ferromagnetic phase.  
However, in a Luttinger liquid phase $\langle S_j\cdot T_j\rangle
=-\frac{3}{4}+S^z$ and the magnetic field increases the one point
correlation function. 
At low temperatures $T<J_{\parallel}$ the one point correlation function is given by 
$\langle S_j \cdot T_j\rangle=\frac{1}{4}+\left(\frac{d}{dJ_{\perp}}f(T,H)\right)_T$. 
The field-induced quantum phase transition can be clearly seen from the
one point correlation function curves at different temperatures 
(see the inset of figure \ref{fig:B5i2aTmag}): 
in the region $H<H_{c1}$, the correlation functions 
$\langle S_j \cdot T_j\rangle=-\frac{3}{4}$ indicate a gapped singlet groundstate; 
in the region $H>H_{c2}$, the correlation functions $\langle S_j \cdot T_j\rangle=\frac{1}{4}$ 
correspond to a fully polarized ferromagnetic state; while for $H_{c1}<H<H_{c2}$, the
one point correlation function $-\frac{3}{4}<\langle S_j \cdot T_j\rangle< \frac{1}{4}$ 
indicates a magnetic Luttinger phase.

\subsection{\rm{Cu$_{2}$(C$_5$H$_{12}$N$_2$)$_2$Cl$_4$}}
\label{sec:CuHpCl}

\subsubsection{Magnetic susceptibility} 
\begin{figure}
\begin{center}
\includegraphics[width=0.60\linewidth]{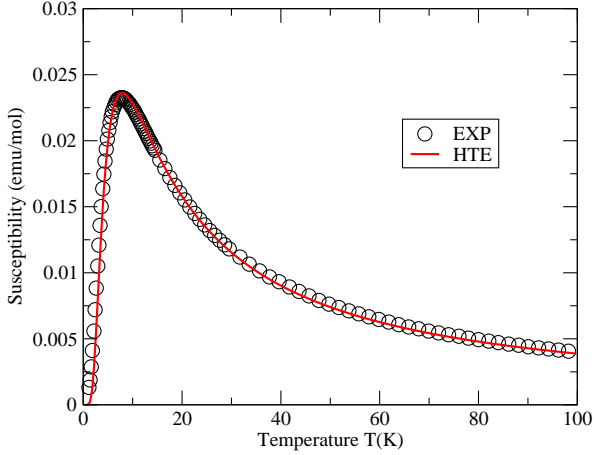}
\caption{
Susceptibility versus temperature for the compound  Cu(Hp)Cl.
Circles represent the experimental data  \cite{Hagiwara}.
The solid curve is the susceptibility evaluated directly from the HTE with 
$\mu_B=0.672\,$K/T, $J_{\perp}=14.5\,$K, $J_{\parallel}=1.0\,$K and $g=2.1$.
The conversion constant is $\chi_{{\rm HTE}}\approx 0.40615 \chi _{{\rm EXP}}$.
}
\label{fig:Cu(Hp)Clsus}
\end{center}
\end{figure}

The ladder compound Cu(Hp)Cl has been extensively studied, both 
experimentally and theoretically, including several numerical-based calculations \cite{Chaboussant1,Chaboussant2,Chaboussant3,numers1,numers5,
Hagiwara,FT1,Chiari,Mayaffre,Stone,HTE1,TBAladder1a,TBAladder1b,ying2}. 
Nevertheless, the thermodynamic quantities of the model,
such as the high field magnetization, the full temperature susceptibility, 
the magnetic specific heat, the entropy and the correlation length, are worth
close examination via the TBA and HTE methods.  
Here we show that the integrable model provides an efficient way of investigating 
the thermodynamics of this compound. 
The solid line in figure \ref{fig:Cu(Hp)Clsus} shows the zero
field magnetic susceptibility curve obtained by HTE with up to fifth
order terms.  
The rounded hump is observed around $T=8.0\,$K.  
A full fit with the experimental curve for the susceptibility 
of Hamiltonian (\ref{eq:Ham2}) suggests the values
$J_{\perp}=14.5\,$K and $J_{\parallel}=1.0\,$K with
$g=2.1$, which lies in the same phase as the conventional Heisenberg
ladder model \cite{FT1}. 
The TBA gap $\Delta \approx 10.5\,$K agrees well with the experimental gap
$\Delta =10.9\,$K \cite{Chaboussant1,Chaboussant2,Chaboussant3}. 

At this point it is appropriate to remark on the convergence of the HTE series.
The HTE in equation (\ref{eq:t-expan}) is not a series in the usual Taylor sense.
Here the expansion parameter is ${J_\parallel}/{T}$. 
For a given compound $J_\parallel$ is fixed, with ${J_\parallel}/{T}$ a small parameter
at high temperatures. 
It is important to note that the expansion coefficients $c_{ij}^{(k)}$ depend on the
temperature through the $Q$-system, which
in turn depends on the chemical potentials, see equations (\ref{eq:Q-sl(r+1)}) and
(\ref{eq:Q-sl4}).
For this more sophisticated type of series expansion it suffices to consider only
the first few terms in the physically interesting regime.
When the expansion parameter is no longer small, 
adding higher order terms does not improve the fit significantly.
In the following numerical example we show this behavior for a generic case, 
namely the particular compound Cu(Hp)Cl under consideration (figure \ref{fig:Cu(Hp)Clsus}).
For the susceptibility $\chi(T)$ the relative error in using the HTE up to
third order and up to fifth order compared to using up to ninth order terms (as an example of
higher order terms) for some sample temperatures is given in the following table.

\begin{center}
\begin{tabular}{l|llll}
$T$ & $2\,$K & $5\,$K & $10\,$K & $50\,$K \\
\hline
error $\chi_{1-3}$ rel. to $\chi_{1-9}$  & $7.33\%$        & $0.122\%$        &        $0.0003288\%$
&        $<9.2\cdot 10^{-7}\%$         \\
error $\chi_{1-5}$ rel. to $\chi_{1-9}$  & $0.56 \%$        &        $0.00147\%$ & $3.8\cdot
10^{-6}\%$  & $<5.2\cdot 10^{-10}\%$                 \\
\end{tabular}
\end{center}

In this article for comparison with experimental data we have used the HTE expansion 
with up to 5th order terms.
Technically it is not harder to include higher order HTE terms in the analysis. 
The computation of the coefficients in Appendix \ref{appendix1} has to be done only 
once, as the expansion formula gives the coefficients in terms of the abstract $Q$-system. 
For each compound or new fit only the numerical values have to be fed into the chemical 
potentials, allowing an immediate calculation of all physical properties up to the order 
of the initial free energy expansion.

In principle, the range of validity of the HTE can be extended to low
temperatures via Pad\'e series analysis.
This uses the fact that the free energy expression is
a  smooth function of its arguments.

\subsubsection{Magnetic specific heat and entropy} 
\begin{figure}
\begin{center} 
\includegraphics[width=.60\linewidth]{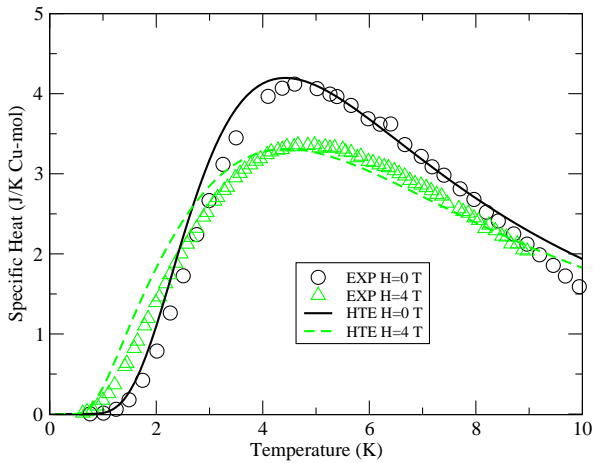}
\caption{
Specific heat versus temperature at different magnetic field values for the compound 
Cu(Hp)Cl with the same coupling constants $J_{\perp}$, $J_{\parallel}$ and $g$ as in figure \ref{fig:Cu(Hp)Clsus}. 
The experimental data \cite{Hagiwara} is given by circles for $H=0\,$T and triangles for $H=4\,$T.
The solid and dashed curves are evaluated from the HTE for these same magnetic field values.
%
The conversion constant is $C_{{\rm HTE}}\approx 4.515 C_{{\rm EXP}}$ (J/mol-Cu-K).
}
\label{fig:Cu(Hp)Clheat1-1}
\end{center}
\end{figure}

\begin{figure}
\begin{center} 
\includegraphics[width=.60\linewidth]{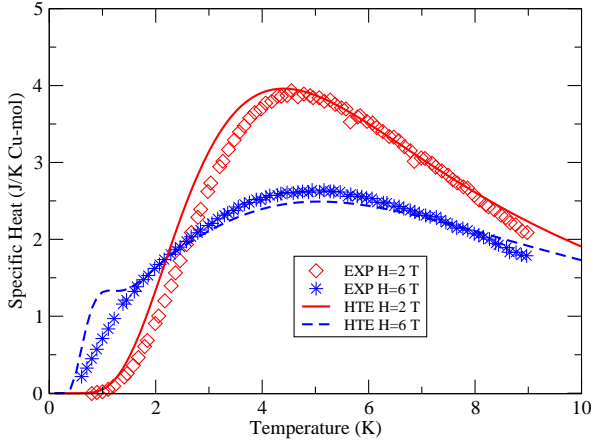}
\caption{
Specific heat  versus temperature for the compound Cu(Hp)Cl, now for the
experimental data for $H=2\,$T and $H=6\,$T \cite{Hagiwara}.
We use the same coupling constants as in figure \ref{fig:Cu(Hp)Clsus} and figure \ref{fig:Cu(Hp)Clheat1-1}. 
The solid and dashed curves are evaluated from the HTE  for these same magnetic field values.
%
%
The conversion constant is $C_{{\rm HTE}}\approx 4.515 C_{{\rm EXP}}$ (J/mol-Cu-K).
} 
\label{fig:Cu(Hp)Clheat1-2}
\end{center}
\end{figure}

\begin{figure}
\begin{center} 
\includegraphics[width=.60\linewidth]{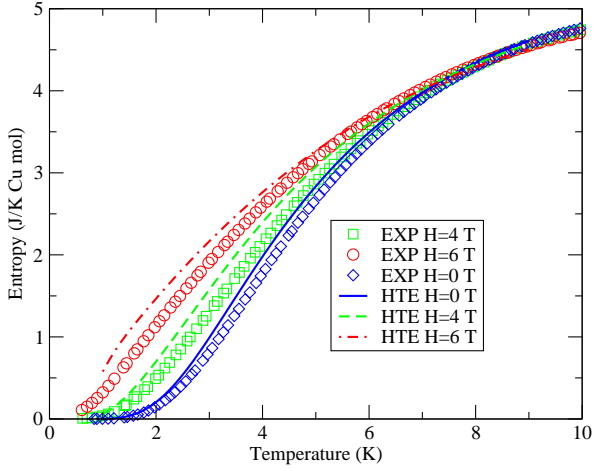}
\caption{ 
Entropy vs temperature comparison between theory and experimental data  \cite{Hagiwara} 
for the compound Cu(Hp)Cl at the external magnetic field values $H=0\,$T, $H=4\,$T and $H=6\,$T.
The coupling constants are the same as in the previous figures for Cu(Hp)Cl.
The solid, dashed and dot-dashed curves evaluated from the HTE for these same
magnetic field values.
A discrepancy with the experimental entropy data for high external magnetic field is observed, 
probably due to paramagnetic impurities.  
The conversion constant is
$S_{{\rm HTE}}\approx 4.153 S_{{\rm EXP}}$ (J/mol-Cu-K).
}
\label{fig:Cu(Hp)Clentropy}
\end{center}
\end{figure}

The solid and dashed lines in figure \ref{fig:Cu(Hp)Clheat1-1} denote the
specific heat curves at $H=0\,$T and $4\,$T obtained from the
HTE with the same coupling constants as in figure  \ref{fig:Cu(Hp)Clsus}. 
They agree well with the experimental data \cite{Hagiwara}.  
A rounded peak indicating short range ordering is observed around $T=4.5\,$K. 
At temperatures below $T=4.5\,$K there is an exponential decay 
which is believed to be due to an ordered phase. 
A similar fit for the specific heat at $H=2\,$T and $6\,$T is
observed in figure \ref{fig:Cu(Hp)Clheat1-2}.  
Another peak is found around $T=1\,$K for the magnetic field $H=6\,$T.  
This is mainly because the HTE is not convergent for this model 
if the temperature is less than $1\,$K. 
At these values the temperature scale is comparable to the coupling $J_{\parallel}=1.0\,$K.
Adding further HTE terms is not sufficient to overcome such kinds of divergency.
The entropy has been calculated directly from the free energy (\ref{eq:freeenergyHTE}) 
with up to fifth order terms.
Figure \ref{fig:Cu(Hp)Clentropy} shows agreement with the
experimental entropy curves \cite{Hagiwara}. 
A discrepancy with the fitting is observed for high magnetic field, 
which is probably caused by paramagnetic impurity effects.

\subsubsection{High field magnetization} 
\begin{figure}
\begin{center}
\includegraphics[width=.60\linewidth]{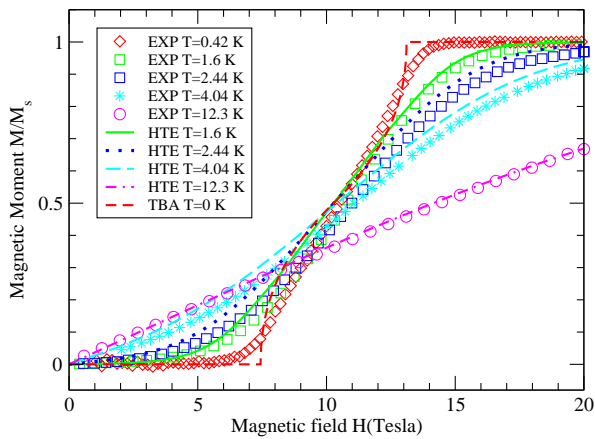}
\caption{
Magnetization vs magnetic field comparison between theory and experimental data \cite{FT1}
for the compound Cu(Hp)Cl for varying temperature.
The TBA with the same coupling constants, $J_{\perp}=14.5\,$K, $J_{\parallel}=1.0\,$K and $g=2.1$,   
predicts critical fields at zero temperature which coincide with the experimental values at low temperature. 
The magnetization curves evaluated from the HTE give satisfactory agreement with the experimental curves.
}
\label{fig:Cu(Hp)Clmag}
\end{center}
\end{figure}

{}From table $1$, we see that the parameter fitting values 
$J_{\perp}=14.5\,$K and $J_{\parallel}=1.0$K lead to
the critical field values $H_{c1}=7.4\,$T and $H_{c2}=13.1\,$T, 
which are in good agreement with the experimental values
$H_{c1}=7.5\,$T and $H_{c2}=13.0\,$T. 
The experimental high field magnetization has been measured by different 
groups \cite{FT1,Chaboussant3}. 
The zero temperature TBA magnetization curve (red dashed line)
in figure \ref{fig:Cu(Hp)Clmag} indicates the critical field values in the
proximity of the lowest temperature data at $T=0.42\,$K \cite{FT1}. 
We cannot evaluate the magnetization directly from the HTE for 
temperatures lower than $T=1.5\,$K due to the lack of convergence. 
At the higher temperatures $T=12.3\,$K and $T=4.04\,$K the HTE magnetization 
curves fits are seen to fit the experimental data well.  
At low temperatures $T=1.6\,$K and $T=2.44\,$K, a small discrepancy 
between the HTE and experimental curves is observed. 
This is mainly because the coupling constants $J_{\perp}=14.5\,$K and $J_{\parallel}=1.0\,$K 
used in fitting the experimental data of Ref.~\cite{Hagiwara} 
may have some small deviation for fitting the experimental data of Ref.~\cite{FT1}.  
It is observed that the temperature induces a spin-flip in the gapped ground state.

\subsection{\rm{(C$_5$H$_{12}$N )$_2$CuBr$_4$}}

\subsubsection{Magnetic suceptibility} 
\begin{figure}
\begin{center}
\vskip 5mm
\includegraphics[width=.60\linewidth]{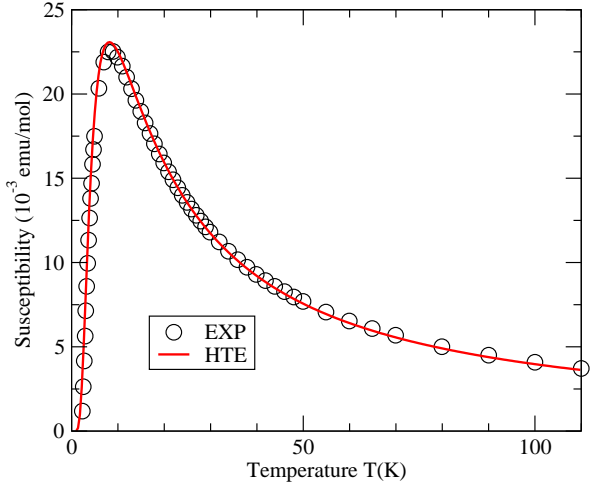}
\caption{
Susceptibility vs temperature comparison between the HTE method and the
experimental data \cite{Watson} for the compound 
(C$_5$H$_{12}$N)$_2$CuBr$_4$.  
The solid curve is the susceptibility evaluated directly from the HTE at $H=0\,$T. 
A parameter fit suggests the coupling constants $J_{\perp}=15.4\,$K and $J_{\parallel}=1.2\,$K 
with $g=2.13$ and $\mu_B=0.672$ K/T.  
The conversion constant is $\chi_{{\rm HTE}}\approx 0.40615 \chi _{{\rm EXP}}$. }
\label{fig:BPCBsus}
\end{center}
\end{figure}

The ladder compound (C$_5$H$_{12}$N)$_2$CuBr$_4$, aka BPCB \cite{Watson},  
has provided a new ground for understanding microscopic mechanisms in
low-dimensional physics. 
The ladder structure extends along the $a$-axis \cite{Watson}. 
The Cu$^{2+}$ ions are bridged via the orbital overlap of Br$^{-2}$ ions along the
intra- and inter-chain directions.
This compound was identified as a spin-$\frac12$ Heisenberg
two-leg ladder in the strong coupling regime with $J_{\perp}=13.3\,$K
and $J_{\parallel}=3.8\,$K \cite{Watson}. 
The gapped phase was observed directly from the experimental high field 
magnetization at low temperature with the typical rounded peak susceptibility shape
at $T=8\,$K. 
In fitting the HTE susceptibility to the experimental results,
we find that the integrable ladder model (\ref{eq:Ham2}) with
coupling constants $J_{\perp}=15.4\,$K, $J_{\parallel}=1.2\,$K and
$g=2.13$ describes this compound well. 
A comparison between the HTE result and experimental  
susceptibility is given in  figure \ref{fig:BPCBsus}.
A rounded peak is observed at $T\approx 8\,$K. 
The susceptibility exponentially decays as the temperature decreases, which
indicates a gapped phase. 
In this case the Troyer \etal formula \cite{Troyer}, 
$\chi \propto {\rm e}^{-\Delta/T}/\sqrt{T}$, suggests a gap of $\Delta
\approx 10\,$K, which coincides with our TBA zero temperature gap
$\Delta \approx 10.06\,$K.

\subsubsection{High field magnetization} 
\begin{figure}
\begin{center} 
\vskip 5mm
\includegraphics[width=.60\linewidth]{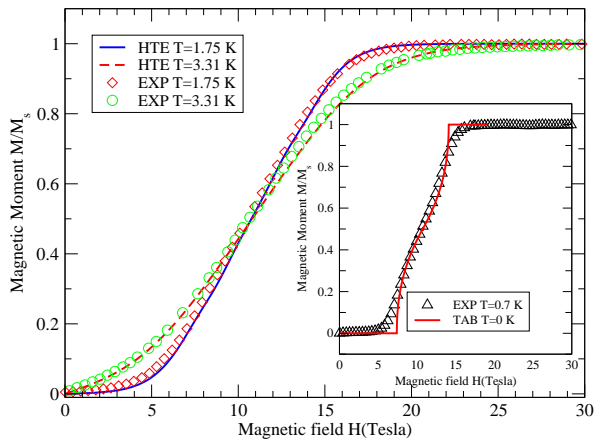}
\caption{ 
Magnetization vs magnetic field comparison between the HTE approach and the
experimental data \cite{Watson} for the compound (C$_5$H$_{12}$N)$_2$CuBr$_4$.
These curves indicate the nature of the high magnetic field quantum phase diagram.
The same parameters are used as in figure \ref{fig:BPCBsus}.
The inset shows the exact TBA magnetization curve obtained at zero temperature.
} 
\label{fig:BPCBmag}
\end{center}
\end{figure}

The high field magnetization has been measured at different temperatures \cite{Watson}. 
The experimental magnetization curve at $T=0.7\,$K reveals that the singlet gapped phase lies in
the region $H<6.6\,$T and the saturation magnetization occurs at
magnetic fields greater than $H_{c2}=14.6\,$T.  
The TBA magnetization curve is given in the inset of figure \ref{fig:BPCBmag}, where the
parameter setting is the same as in figure \ref{fig:BPCBsus}.  
This curve indicates the critical field values $H_{c1}=7\,$T and $H_{c1}=14.1\,$T which are in 
agreement with the experimental result.  
It is seen that the HTE magnetization curves at
$T=1.75\,$K (solid line) and $T=3.31\,$ K (dashed line) 
in figure \ref{fig:BPCBmag} fit the experimental curves well. 

\subsection{\rm{[Cu$_2$(C$_2$O$_2$)(C$_{10}$H$_8$N$_2$)$_2$)](NO$_3$)$_2$}}

\subsubsection{Magnetic suceptibility} 
\begin{figure}
\begin{center}
\vskip 5mm
\includegraphics[width=.60\linewidth]{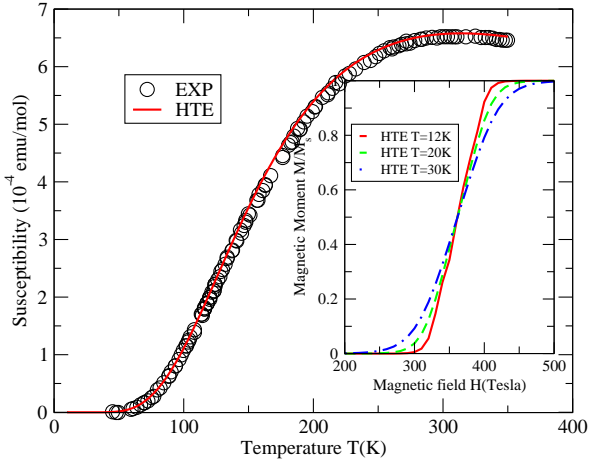}
\caption{
Susceptibility vs temperature comparison between theory (solid line) and 
experimental data (circles) \cite{SC2} for the ladder compound CuCON.
The solid line is the susceptibility evaluated directly from the HTE at $H=0\,$T. 
A parameter fit suggests the coupling constants $J_{\perp}=520\,$K and 
$J_{\parallel}=10\,$K with $g=2.14$ and $\mu_B=0.672\,$K/T. 
The inset shows the HTE magnetization versus magnetic field at different temperatures.  
The conversion constant is $\chi_{{\rm HTE}} \approx 0.40615 \chi _{{\rm EXP}}$.
}
\label{fig:CuCONsus}
\end{center}
\end{figure}

The magnetic properties of the strong coupling ladder compound
[Cu$_2$(C$_2$O$_2$)(C$_{10}$H$_8$N$_2$)$_2$)](NO$_3$)$_2$ 
(abbreviated CuCON) were investigated experimentally in Ref.~\cite{SC2}. 
The spin-spin exchange interactions between Cu$^{2+}$ ions along the leg and rung are
bridged by NO$^-_3$ and C$_2$O$^{2-}_4$, respectively and believed to
be antiferromagnetic. 
The experimental susceptibility of CuCON has a broad peak at about $T=300\,$K. 
Below this temperature, the susceptibility smoothly decreases,  indicating a large gap. 
Perturbation theory predicts the gap $\Delta \approx J_{\perp}-J_{\parallel} \approx 360\,$K. 
The HTE susceptibility evaluated from the free energy with the
coupling constants $J_{\perp}=520\,$K, $J_{\parallel}=10\,$K, $g=2.14$
and $\mu_B=0.672\,$K/T is shown in figure \ref{fig:CuCONsus} (solid line).
An overall good fit to the experimental curve is found. 
The inset of figure \ref{fig:CuCONsus} shows the magnetization
curves at different temperature.  
The TBA result predicts the critical field values $H_{c1}\approx 334\,$T and $H_{c2}\approx 389\,$T.
These high field values appear to be experimentally inaccessible.
 
\subsection{\rm{BIP-BNO}}

\subsubsection{Magnetic suceptibility} 
\begin{figure}
\begin{center}
\includegraphics[width=0.60\linewidth]{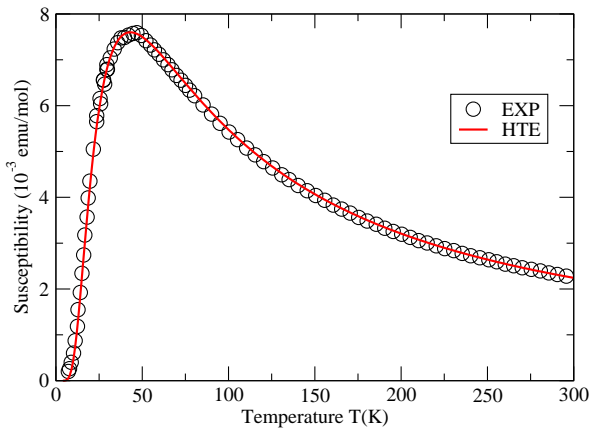}
\caption{
Susceptibility vs temperature comparison between HTE theory and experiment \cite{BIPa,BIPb} 
for the strong coupling ladder compound BIP-BNO.
The solid line is the susceptibility evaluated directly from the HTE at $H=0$T.  
A parameter fit suggests the coupling constants $J_{\perp}=83\,$K and $J_{\parallel}=7.0\,$K with
$g=2.0$ and $\mu_B=0.672\,$K/T.  
The conversion constant is $\chi_{{\rm HTE}}\approx 2 \times 0.40615 \chi _{{\rm EXP}}$ 
due to the double chain structure.  
}
\label{fig:BIP-BNOsus}
\end{center}
\end{figure}

The organic polyradical BIP-BNO has been recognised as a strong coupling 
ladder compound \cite{BIPa,BIPb}. 
The intramolecular spin-spin exchange interaction within the
BIP-BNO molecules is antiferromagnetic. 
Two intermolecular NO groups have the same crystallographic structure so that there is a twofold
symmetry along the $b$-axis. 
It can be seen to give rise to an antiferromagnetic doubled rung interaction.  
Experimental investigation of the magnetic susceptibility reveals a broad peak at
$T \approx 45\,$K \cite{BIPa,BIPb}. 
Below this temperature the susceptibility decreases steeply as the temperature decreases. 
Fitting the Troyer \etal  formula \cite{Troyer}, 
$\chi \propto {\rm e}^{-\Delta/T}/\sqrt{T}$, below $T=25\,$K
gives \cite{BIPa,BIPb} an energy gap $\Delta \approx 47\,$K. 
To fit the susceptibility and the magnetization, two sets of
coupling constants, $J_{\perp}=73\,$K, $J_{\parallel}=17\,$K and
$J_{\perp}=67\,$K, $J_{\parallel}=25\,$K have been suggested \cite{BIPa,BIPb}. 
However, 
the experimental magnetization curve at temperature $T=1.6\,$K indicates
an energy gap $\Delta \approx 38.8\,$T (roughly $\Delta \approx 52\,$K).
{}From perturbation theory, they caculated the gap $\Delta=J_{\perp}-J_{\parallel}=56\,$K or 
$42\,$K corresponding to the two sets of coupling parameters.  
There is obviously some inconsistency among these fittings. 
We find the parameters
$J_{\perp}=84\,$K and $J_{\parallel}=7.0\,$K with $g=2.0$ for the
integrable ladder model (\ref{eq:Ham2}) in fitting the susceptibility, as shown
in figure \ref{fig:BIP-BNOsus}.
Here the value for the $g$-factor is fixed in the experiment \cite{BIPa,BIPb}.
The solid HTE line shows an excellent fit with experimental susceptibility curve. 
Here the TBA gives the gap $\Delta \approx 41\,$T (roughly $\Delta \approx 55\,$K),
which is in agreement with the experimental observation of $\Delta \approx 38.8\,$T.  
We conclude that their coupling constants $J_{\perp}=73\,$K  and   $J_{\parallel}=17\,$K 
seem to be more reasonable for the Heisenberg ladder.  
In addition, we mention that the conversion constant for this compound is twice as
large as the one for the other compounds considered here, but of the same order as the one
for the spin-$1$ chain compounds \cite{BGNF}. 
This is mainly because of the doubled intermolecular structure along the rungs, 
effectively forming two overlapping spin ladders.

\section{Thermodynamics of the  spin-$(\frac{1}{2},1)$  ladder model}
\label{mixHTE}

In this section we turn to the calculation of the thermodynamic properties of the 
integrable mixed spin-$(\frac{1}{2},1)$ model (\ref{eq:Ham3}).
Various theoretical studies suggest that fractional magnetization plateaux
exist in mixed spin-$(\frac{1}{2},1)$ chains \cite{mixc1a,mixc1b,mixc2a,mixc2b,mixc3a,mixc3b,mixc4,mixc5} 
and the mixed spin ladders \cite{mixl3,mixl4,mixl5,Aristov,CWW}.
Such fractional magnetization plateaux have been found in a number of
low-dimensional magnetic systems,
including Shastry-Sutherland systems \cite{FPL,FPL2,FPL3,FP-Miyahara} and
spin ladders \cite{mixl1,mixl2a,mixl2b,FP-Sakai,FP-Hagiwara,FP-Sakai2}.

\subsection{TBA analysis and ground state properties}

\begin{figure}
\vskip 5mm
\begin{center}
\includegraphics[width=0.60\linewidth]{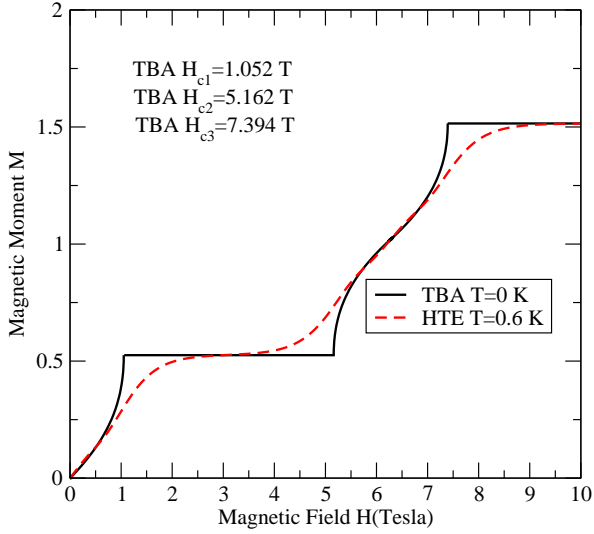}
\caption{
Magnetization versus magnetic field $H$ for the integrable model (\ref{eq:Ham3}) 
in the strong antiferromagnetic rung coupling regime.
The magnetic moment is normalised by the saturation value
$M_s=\frac{1}{2}(g_s+g_t)\mu_B$.  
The full curve is obtained from the TBA analysis.
The coupling constants used are \cite{mix} $J_{\perp}=6.0\,$K, $J_{\parallel}=0.4\,$K 
with Land\'{e} $g$-factors $g_s=2.22,\,g_t=2.09$ and $\mu_B=0.672\,$K/T.  
We predict from the TBA analysis that the one-third saturation magnetization plateau 
opens with the critical fields $H_{c1}\approx 1.052\,$T, $H_{c2}\approx 5.162\,$T and 
$H_{c3}\approx 7.394\,$T which coincide with the numerically estimated values.  
The dashed curve is evaluated from the HTE free energy (\ref{eq:mixEQTM3}). 
Both the TBA and HTE approaches are seen to yield consistent results.
In addition, the inflection point at $H=H_{{\rm IP}}\approx 6.278\,$T and 
$M \approx 1$ is observed from the curves. 
$H_{{\rm IP}}$ indicates a point of equal probability for the states
$\psi_{\frac{3}{2}}$ and $\psi_{\frac{1}{2}}^{(-)}$. 
}
\label{fig:mixmag1}
\end{center}
\end{figure}

A one-third magnetization plateau has been found to exist in the integrable 
mixed spin-$(\frac{1}{2},1)$ ladder for strong rung coupling via the
TBA approach \cite{mix}.
Gapped or gapless states appear in turn as the external magnetic field increases, 
as can be seen in figure  \ref{fig:mixmag1} where the whole magnetization curve is evaluated
numerically by solving the TBA equations in the different phases \cite{mix}.  
This approach predicts that a two-component massless quantum
magnetic phase lies in the regime $H\!<\!H_{c1}$, with a ferrimagnetic
phase appearing for $H\!>\!H_{c1}$ where the component
$\psi^{(-)}_{\frac{1}{2}}$ (recall (\ref{eq:mixbasis})) becomes a physical
ferrimagnetic ground state. 
This state forms a one third magnetization plateau which extends until the 
critical field value $H_{c2}$.  
As the magnetic field increases beyond $H_{c2}$, the state $\psi_{\frac{3}{2}}$ 
becomes involved in the ground state and the ground state becomes a mixture 
of doublet and quadruplet states.  
The two components $\psi^{(-)}_{\frac{1}{2}}$ and $\psi_{\frac{3}{2}}$
form an effective $XXZ$ model with massless excitations. 
Other components of the multiplets are gapful by virtue of both the rung
coupling and magnetic field. 
Eventually, if the magnetic field is greater than $H_{c3}$, the ground state is the 
fully polarized state. 
The critical fields are found from the TBA to be \cite{mix}
\begin{eqnarray}
H_{c1}&= &\frac{4J_{\parallel}}{[g_s+\ffrac{1}{3}(g_s-g_t)]\mu_B} \nonumber\\
H_{c2}&=&\frac{3J_{\perp}-8J_{\parallel}}{[(g_t+g_s)-\frac{1}{3}(g_s-g_t)]\mu_B}
\label{eq:mixfield}
\\
H_{c3}&=&\frac{3J_{\perp}+8J_{\parallel}}{[(g_t+g_s)-\frac{1}{3}(g_s-g_t)]\mu_B}.\nonumber
\end{eqnarray}
The magnetization in the vicinity of the critical fields exhibits  square root field dependent 
behaviour \cite{mix}.

\begin{figure}
\begin{center}
\vskip 5mm
\includegraphics[width=0.60\linewidth]{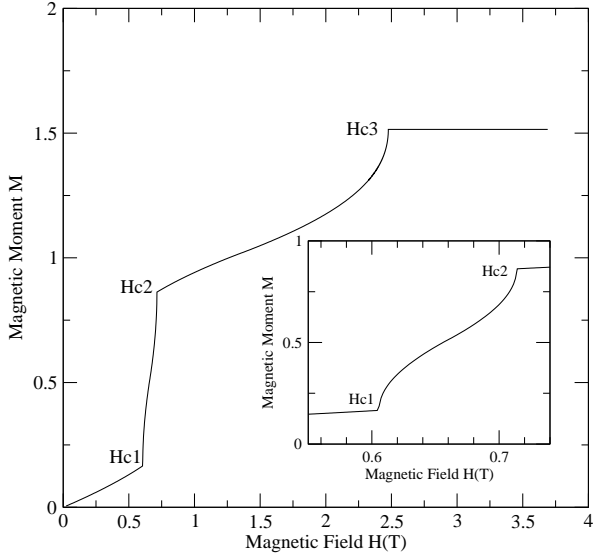}
\caption{
Magnetization versus magnetic field $H$ in the weak antiferromagnetic rung coupling 
regime \cite{mix}.
The numerical values are the same as for figure \ref{fig:mixmag1}, but now using the smaller rung
coupling constant $J_{\perp}=1.3K<J_c^{+ F}\approx 2.07\,$K. 
In this case the fractional magnetization plateau vanishes.
The TBA critical fields coincide again with the numerically estimated values in the figure.
The inset shows an enlargement of the magnetization between $H_{c1}$ and $H_{c2}$.
}
\label{fig:mixmag2}
\end{center}
\end{figure}

On the other hand, the first magnetization
plateau depends mainly on the rung coupling.
If $ \frac{4}{3}J_{\parallel}\ln 2 \!<\!J_{\perp}\! <\!J_c^{+F}$, where 
\begin{eqnarray}
J_c^{+F}&=&\frac{8J_{\parallel}}{3}\frac{[\frac{1}{2}g_t+\frac{3}{2}g_s+
\frac{1}{6}(g_s-g_t)]}{[g_s+\frac{1}{3}(g_s-g_t)]}\label{eq:Jc}
\end{eqnarray} 
this plateau disappears and the critical points $H_{c1}$ and $H_{c2}$
become close to each other.  
Figure \ref{fig:mixmag2} shows the numerical evaluation of the full magnetization curve for 
weak rung coupling $J_{\perp}=1.3\,$K.
We clearly see that the fractional magnetization plateau is closed.
This is because the rung interaction can not quench the quadruplet. 
The numerical value is $H_{c1} \approx 0.605\,$T.
This implies that for $H < H_{c1}$ the ground state is the doublet spin liquid phase. 
The quadruplet component $\psi_{\frac{3}{2}}$ is involved in the groundstate for  $H> H_{c1}$.
Therefore the critical point $H_{c1}$ indicates a quantum phase
transition from a two-state phase to a three-state phase.  
Hence for $H > H_{c1}$, two Fermi seas, $\epsilon ^{(1)}$ and $\epsilon^{(2)}$, are filled. 
However, the probability of the component $\psi_{\frac{3}{2}}$ quickly increases while the 
probability of the component $\psi^-_{-\frac{1}{2}}$ quickly decreases as the magnetic field
increases.  
It follows that the magnetization increases rapidly in the region
$H_{c1}\!<H<\!H_{c2}$,  where  the critical point $H_{c2}$ indicates a phase
transition from the three-state into the two-state phase.
Thus from the theory we predict the critical field to be  
$H_{c2}=6J_{\perp}/[(g_t+g_s)-\frac{1}{3}(g_s-g_t)]\mu_B-H_{c1} \approx 0.715\,$T.  
This again is in agreement with the numerical value of $H_{c2}=0.713\,$T. 
In the region $H_{c2}\!<\!H\!<\!H_{c3}$ the two components 
$\psi^-_{\frac{1}{2}}$ and $\psi _{\frac{3}{2}}$ compete to be in the groundstate. 
If the magnetic field is strong enough, so that $H>H_{c3}$
where $H_{c3}$ is given in (\ref{eq:mixfield}), the reference state 
$\psi _{\frac{3}{2}}$ becomes the true physical groundstate.

\subsection{HTE approach}

In order to investigate the thermodynamic properties, we again apply the HTE scheme 
which has been set up in section \ref{QTM-NLIE}. 
Explicitly, the eigenvalue of the QTM (up to constants in the chemical potentials) is given by \cite{ZT2} 
\begin{eqnarray}
T^{(1)}_1(v)&=&{\rm e}^{-\beta \mu_1}\phi _-(v-\mathrm{i})
\phi _+(v)\frac{Q_1(v+\frac12 \mathrm{i})}{Q_1(v-\frac12 \mathrm{i})}\nonumber\\
& &
+{\rm e}^{-\beta \mu_2}\phi _-(v)\phi _+(v)
\frac{Q_1(v-\frac32 \mathrm{i})Q_2(v)}{Q_1(v-\frac12 \mathrm{i})Q_2(v-\mathrm{i})}\nonumber\\
& &+{\rm e}^{-\beta \mu_3}\phi _-(v)\phi _+(v)
\frac{Q_2(v-2\mathrm{i})Q_3(v-\frac12 \mathrm{i})}{Q_2(v-\mathrm{i})Q_3(v-\frac32 \mathrm{i})}
\nonumber\\
& &
+{\rm e}^{-\beta \mu_4}\phi _-(v)\phi _+(v)\frac{Q_3(v-\frac52 \mathrm{i})Q_4(v-\mathrm{i})}
{Q_3(v-\frac32 \mathrm{i})Q_4(v-2\mathrm{i})}\nonumber\\
& &
+{\rm e}^{-\beta \mu_5}\phi _-(v)\phi _+(v)\frac{Q_4(v-3\mathrm{i})Q_5(v-\frac32 \mathrm{i})}
{Q_4(v-2\mathrm{i})Q_5(v-\frac52 \mathrm{i})}\nonumber\\
& &
+{\rm e}^{-\beta \mu_6}\phi _-(v)\phi _+(v+\mathrm{i})
\frac{Q_5(v-\frac72 \mathrm{i})}{Q_5(v-\frac52 \mathrm{i})}.
\label{eq:mixEQTM}
\end{eqnarray} 
In the above equations, the chemical potentials, $\mu_i$, $i=1,\ldots,6$, are chosen as
\begin{eqnarray}
\mu_1 &=&-\left[\ffrac34 J_{\perp} +\ffrac{1}{2} g_s\mu_Bh+\ffrac{1}{\sqrt 2} J_{\perp} 
\sqrt{1+\ffrac{1}{2}(g_sh^{'}-g_th^{'}+\ffrac{1}{2})^2}\right]\nonumber\\
\mu_2 &=&-\left[\ffrac34 J_{\perp} -\ffrac{1}{2}g_s\mu_Bh+\ffrac{1}{\sqrt 2} J_{\perp} 
\sqrt{1+\ffrac{1}{2}(g_sh^{'}-g_th^{'}-\ffrac{1}{2})^2}\right]\nonumber\\
\mu_3 &=&-(\ffrac{1}{2}g_t+g_s)\mu_BH\nonumber\\
\mu_4 &=&-\left[\ffrac34 J_{\perp} +\ffrac{1}{2}g_s\mu_Bh-\ffrac{1}{\sqrt 2} J_{\perp} 
\sqrt{1+\ffrac{1}{2}(g_sh^{'}-g_th^{'}+\ffrac{1}{2})^2}\right]\nonumber\\
\mu_5 &=&-\left[\ffrac34 J_{\perp}-\ffrac{1}{2}g_s\mu_Bh-\ffrac{1}{\sqrt 2} J_{\perp} 
\sqrt{1+\ffrac{1}{2}(g_sh^{'}-g_th^{'}-\ffrac{1}{2})^2}\right]\nonumber\\
\mu_6 &=&(\ffrac{1}{2}g_t+g_s)\mu_BH,
\label{eq:mixchem}
\end{eqnarray}
with $h^{'}=\mu_BH/J_{\perp}$.
These complicated chemical potentials make the HTE expansion of the free energy of the 
integrable model (\ref{eq:Ham3}) more involved. 
As a matter of fact, for the applications we have in mind, the third-order term in the
free energy is rather small. 
The higher order coefficents do not make a significant contribution to the free energy at 
high temeprature. 
Therefore, for simplicity,  we only consider the free energy up to third order, i.e., 
\begin{equation}
-\frac{1}{T}f(T,H)=\ln{Q_{1}^{(1)}}+c^{(1)}_{1,0}\left(\frac{J}{T}\right)+c^{(1)}_{2,0}\left(
\frac{J}{T}\right)^2+c^{(1)}_{3,0}\left(\frac{J}{T}\right)^3+\cdots
\label{eq:mixEQTM3}
\end{equation} 
with the coefficients $c^{(1)}_{1,0}$, $c^{(1)}_{2,0}$ and $c^{(1)}_{3,0}$, along with the 
characters $\{Q_1^{(a)}\}$ for $su(6)$, given in Appendix B.

The exact free energy (\ref{eq:mixEQTM3}) allows easy access to the thermodynamic
properties of the model (\ref{eq:Ham3}) via the standard thermodynamic relations 
given in (\ref{eq:Thermo-Relat}). 
Indeed, this approach makes it very easy  to predict novel critical behaviour 
resulting from varying the strength of the rung and leg interactions. 
In figure \ref{fig:mixmag1}, the magnetization curve (dashed line) at temperature $T=0.6\,$K, 
obtained from (\ref{eq:mixEQTM3}), indicates that the ground state remains in a gapless 
phase until the magnetic field exceeds the critical value $H_{c1} \approx 1.5\,$T. 
The one-third magnetization plateau is also observed under the same
parameter setting as that used for the TBA analysis. 
This magnetization plateau vanishes at the critical value $H_{c2}\approx 4.8\,$T. 
An ordered ferrimagnetic phase is formed in the magnetization plateau region. 
As the field increases beyond the critical point $4.8\,$T, the magnetization increases
almost linearly until the whole system is fully polarized at the
critical value $H_{c2}\approx 7.8\,$T. 
This phase diagram is consistent with the zero temperature TBA result, also
shown in figure \ref{fig:mixmag1}. 
This behaviour is reminiscent  of  the integrable spin-$\frac32$ chain \cite{BGNF}.

\begin{figure}
\begin{center}
\includegraphics[width=.650\linewidth]{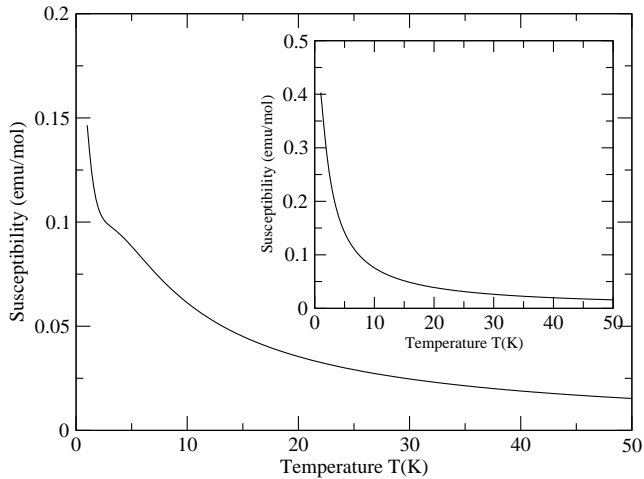}
\caption{ 
Susceptibility vs temperature for the model (\ref{eq:Ham3}) with the same 
parameters as in figure \ref{fig:mixmag1}.  
The susceptibility curve indicates the antiferromagnetic behaviour in the
absence of an energy gap. 
But in this case, the mid-plateau causes a cusp-like feature in the susceptibility curve at
$T\approx 2.5\,$K, similar to what is observed for  the integrable spin-$\frac32$ chain \cite{BGNF}. 
The inset shows the susceptibility curve with a weaker rung interaction 
$J_{\perp}=1.3\,$K, see figure \ref{fig:mixmag2}. 
The mid-plateau is closed and no cusp-like behaviour is observed in the susceptibility curve.
}
\label{fig:mixsus}
\end{center}
\end{figure}

The susceptibility curve for this model is given in figure \ref{fig:mixsus}, 
with the same parameters as in figure \ref{fig:mixmag1}. 
A gapless antiferromagnetic phase is observed from the susceptibility curve. 
However, the ordered antiferrimagnetic phase results in a cusp-like behaviour 
in the susceptibility. 
If the magnetization plateau is wider, the cusp-like  effect becomes more obvious. 
The inset shows the susceptibility for the model with a weaker rung interaction, where the
same parameter settings as figure \ref{fig:mixmag2} are used. 
As predicted from the TBA analysis, the magnetization plateau is closed. 
Thus the susceptibility shows standard antiferromagnetic behaviour without
the cusp-like behaviour.

\subsection{Organic ferrimagnetic mixed spin-$(\frac{1}{2},1)$ ladder}

\begin{figure}
\begin{center}
\includegraphics[width=0.60\linewidth]{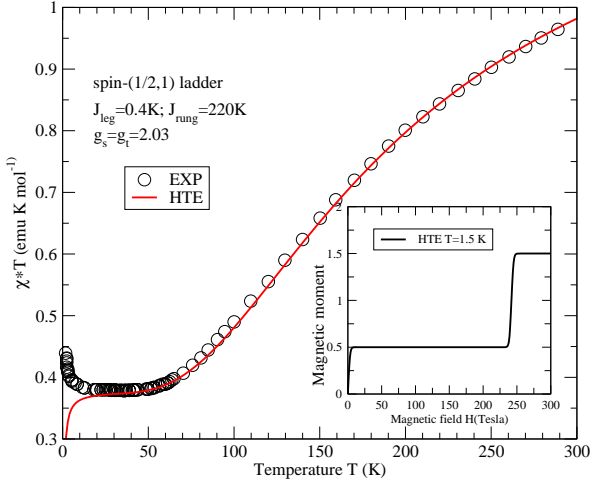}
\caption{
Susceptibility vs temperature for the mixed spin-$(\frac{1}{2},1)$ ladder PNNBNO.
Circles represent the experimental data \cite{mixl1}.
The solid line denotes the susceptibility $\chi \cdot T$ evaluated directly from the HTE free energy 
(\ref{eq:mixEQTM3}) with $J_{\perp}=220\,$K and $J_{\parallel}=0.4\,$K with
$\mu_B=0.672\,$K/T and $g_s=g_t=2.03$.
The conversion constant is $\chi_{{\rm HTE}}\approx 2\times 0.40615\,  \chi _{{\rm EXP}}$ 
due to the double-chain structure. 
The inset shows  the magnetization curve predicted from HTE. 
A large mid-plateau forms a ferrimagnetic trimerization phase.
}
\label{fig:PNNBNOsus}
\end{center}
\end{figure}

To the best of our knowledge, candidates for mixed spin-$(\frac{1}{2},1)$ compounds appear to be
very rare due to their more complicated structure. 
Nevertheless, some organic ferrimagnets may have such mixed structures. 
One compound that we are aware of is the organic ferrimagnet PNNBNO \cite{mixl1,mixl2a,mixl2b},
which has been recognised as a ladder compound with alternating
spin-$\frac{1}{2}$ and spin-$1$ units, i.e., as two coupled alternating
mixed spin chains. 
PNNBNO forms a ferrimagnetic ladder with antiferromagnetic rung and leg interaction. 
The structure along each leg consists of an alternating spin-$\frac{1}{2}$ and 
spin-$1$ chain, which is different from our theoretical model (\ref{eq:Ham3}), which
has uniform spin-$\frac{1}{2}$ on one leg and uniform spin-1 on the other.
However, at high temperature, the thermodynamic properties are entirely 
dominated by the strong antiferromagnetic rung interaction. 
The integrable model (\ref{eq:Ham3}) should therefore be expected to
adequately describe the compound PNNBNO at high temperatures. 
In figure \ref{fig:PNNBNOsus}, we give the susceptibility curve evaluated from the
exact HTE free energy (\ref{eq:mixEQTM3}). 
Figure \ref{fig:PNNBNOsus} demonstrates that at high temperatures $T>10\,$K, the theoretical HTE 
prediction fits the experimental results well. 
We see that the value $\chi T$ derived from the HTE
remains  constant in the temperature range of $10-50\,$K, while it
increases for temperatures over $50\,$K. 
This constant range reveals the mid-plateau effect -- the ground state 
of the ladder lies in a ferrimagnetic state state with $S^z=\frac{1}{2}$, 
see the inset of figure \ref{fig:PNNBNOsus}. 
When $T>50\,$K, the curves indicate antiferromagnetic behaviour. 
For low temperatures, the HTE $\chi T$ values decrease as the temperature tends to zero. 
This indicates antiferromagnetic behaviour for the integrable mixed spin-$(\frac{1}{2},1)$ ladder. 
However, the experimental data for the susceptiblity suggests
ferromagnetic behaviour at low temperature due to the ferrimagnetic
structure along the legs.

\begin{figure}
\begin{center} 
\vskip 5mm
\includegraphics[width=.60\linewidth]{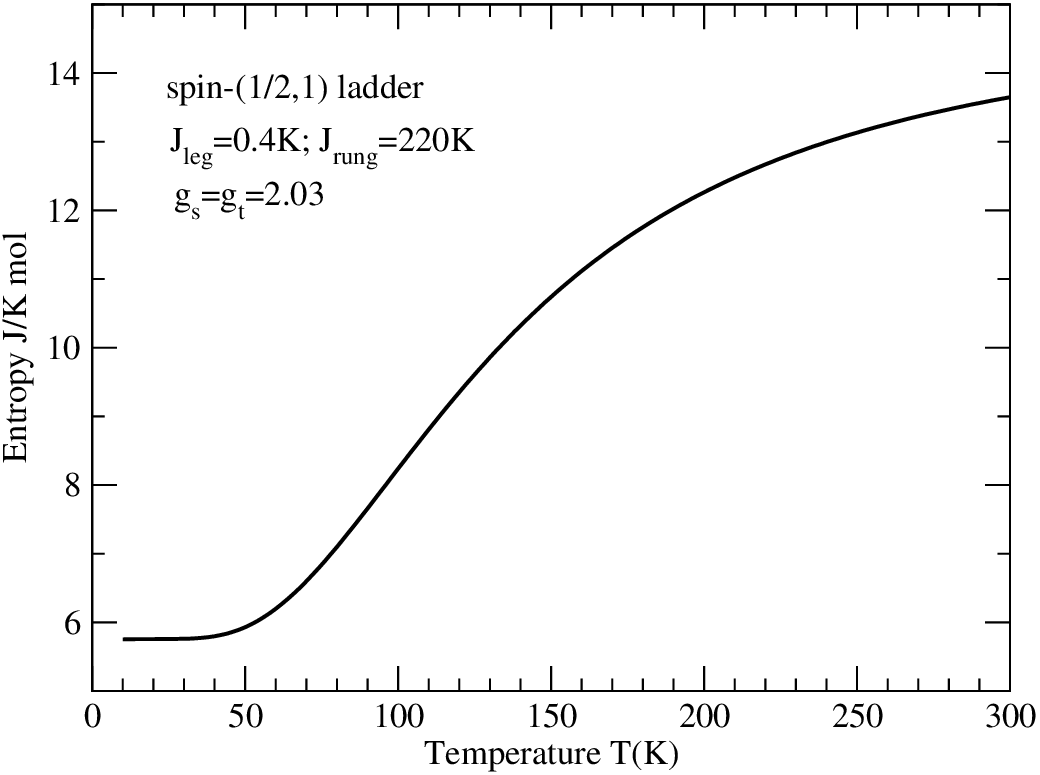}
\caption{ 
Entropy vs temperature for the the mixed spin-$(\frac{1}{2},1)$
ladder PNNBNO at high temperature. 
The curve evaluated from the HTE with the same constants as in figure \ref {fig:PNNBNOsus} for
zero field.  
In the temperature region $10-50\,$K, the entropy remains 
constant at the value $S=5.7569$ K/J-mol, which is consistent with the theoretical
value of $NK_b\ln2 \approx 5.76$ K/J-mol \cite{mixl1}.  
It indicates a pure ferrimagnetic trimerization phase with total spin $S^z=\frac12$.  
The conversion constant is $S_{{\rm HTE}}\approx 2 \times 4.153 S_{{\rm EXP}}$ (J/mol-Cu-K),  
due to the double-chain structure.
}
\label{fig:PNNBNOentropy}
\end{center}
\end{figure}

In addition, we predict the high temperature entropy at zero field in figure \ref{fig:PNNBNOentropy}.  
It remains constant, i.e., at $S=5.7569\,$K/J-mol in the temperature range $10-50\,$K. 
The entropy increases as the quadruplet enters into the ground state
from the ferrimagnetic phase (trimerization phase) for $T>50\,$K. 
In the ferrimagnetic phase, only the doublet component $S^z=\frac{1}{2}$ occupies the ground state. 
The theoretical prediction for the entropy in this phase is $NK_B\ln2 \approx 5.76$ \cite{mixl1}, 
which is in excellent agreement with our HTE value $S=5.7569\,$K/J-mol. 
Below $T \approx 10\,$K the entropy decreases as the temperature tends to zero.

\section{Magnetization plateaux for the spin-orbital model}
\label{sec:SO}

In this section we discuss extending the analysis to the integrable spin-orbital model.

\subsection{Hamiltonians}
The transition metal oxides have been extensively studied in condensed
matter physics due to their close connections to superconductivity. 
In particular, the electronic and spin properties of the D-electron shells of the transition metal ions 
lead to rich and novel phase transitions.  
In many metal oxide compounds \cite{SOcomp1,SOcomp2,SOcomp3a,SOcomp3b,SOcomp4}, the
orbital order has been observed besides charge and spin fluctuations. 
As a consequence, a great deal of 
interest \cite{SOtheo1,SOtheo2a,SOtheo2b,SOtheo3,SOtheo4,SOtheo5a,SOtheo5b,SOtheo6,SOtheo6a,yinga,yingb} 
in spin-orbital models with both spin and orbit double degeneracies arises
in this context. 
In the absence of Hund rule coupling, a simple spin-orbital model Hamiltonian is \cite{SOtheo5a,SOtheo5b}
\begin{equation}
{\cal H}_{{\rm
S-O}}=\sum_{j=1}^{L}\left(2\vec{S}_j \cdot \vec{S}_{j+1}+x\right)\left(2\vec{T}_j \cdot \vec{T}_{j+1}+y\right)
\label{eq:SP-Ham1} 
\end{equation}
where $x$ and $y$ are arbitrary constants, $\vec{S}$ is a spin-$\frac12$ operator
and $\vec{T}$ is the orbital pseudospin-$\frac12$ operator.
The model has $su(2)\otimes su(2)$ symmetry.

The phase diagram of this model has been discussed in 
Refs.~\cite{SOtheo5a,SOtheo5b,SOtheo6,SOtheo6a,SOtheo7}. 
In the $(x,y)$-plane, there are five phases associated with different symmetries of the model. 
In phase I both spin and orbital degrees of freedom are in fully polarized ferromagnetic states.
In phase II the orbital degrees of freedom are in the fully polarized ferromagnetic
state while the spin degrees of freedom are in the antiferromagnetic ground state and
vice versa in phase III.
Phase IV is a gapped phase.
Phase V is a gapless region surrounding the origin in the $(x,y)$-plane. 
The point $(\frac{1}{2},\frac{1}{2})$ on the IV-V phase boundary has $su(4)$ symmetry. 
The simplest $su(4)$ spin-orbital model is given by the permutation operator. 
It has a rotation symmetry in the spin $\vec{S}$ and $\vec{T}$ spaces.  
In general, this symmetry is broken by Hund's rule coupling or
other anisotropic hybridizations. 
In this section, we consider two types of symmetry breaking interactions: 
(1) spin-orbital Ising coupling \cite{yinga,yingb}, and (2) orbital zero-field splitting \cite{SOtheo8}, 
where the Hamiltonians are respectively given by
\begin{eqnarray}
{\cal H}_{{\rm S-O-I}} &=& {\cal H}_{{\rm S-O-C}}+J_z\sum_{j=1}^LS^z_jT^z_j
-\mu_BH\sum_{j=1}^L\left(g_s S^z+g_t T^z\right)
\label{eq:SP-Ham1-1} \\
{\cal H}_{{\rm S-O-S}} &=& {\cal H}_{{\rm S-O-C}}+\Delta_z\sum_{j=1}^LT^z_j
-\mu_BH\sum_{j=1}^L\left(g_s S^z+g_t T^z\right)
\label{eq:SP-Ham1-2}
\end{eqnarray}
with
\begin{equation}
{\cal H}_{{\rm
S-O-C}}=J\sum_{j=1}^{L}\left(2\vec{S}_j\cdot \vec{S}_{j+1}+
\frac{1}{2}\right)\left(2\vec{T}_j\cdot \vec{T}_{j+1}+\frac{1}{2}\right).
\label{eq:SP-Ham1-0} 
\end{equation}
In the above equations, $g_s$ and $g_t$ denote Land\'{e} factors for spin and orbit. 
We see that for Model 1  (\ref{eq:SP-Ham1-1}) the $su(4)$
symmetry is broken into $su(2)\otimes su(2)$ due to the spin-orbital Ising coupling. 
For Model 2 (\ref{eq:SP-Ham1-2}) the $su(2)$ symmetry and rotation symmetry for the orbit are
broken down by the presence of zero-field splitting caused by a crystalline field effect. 
We shall see that these different kinds of anisotropy can trigger different phase transitions 
in the presence of an external  magnetic field.

\subsection{Ground state properties and fractional magnetization plateaux}

\subsubsection{Model 1} 
In analogy to section \ref{sec:TBA}, the model  (\ref{eq:SP-Ham1-1}) can be diagonalised by 
the fundamental basis $\st[S^z,T^z]$, with  
$\st[1]=\st[\frac{1}{2},\frac{1}{2}]$,
$\st[2]=\st[\frac{1}{2},-\frac{1}{2}]$,
$\st[3]=\st[-\frac{1}{2},\frac{1}{2}]$,
$\st[4]=\st[-\frac{1}{2},-\frac{1}{2}]$,
with energy \cite{yinga,yingb}
\begin{eqnarray}
{\cal E} &=&-J\sum^{M_1}_{i=1}\frac{1}{(v_i^{(1)})^2+\frac{1}{4}}
-\frac{1}{2}\mu_BH(g_s+g_t)N_1 \nonumber\\
& &-\frac{1}{2}[J_z+\mu_BH(g_s-g_t)]N_2-\frac{1}{2}[J_z-\mu_BH(g_s-g_t)]N_3\nonumber\\
& &+\frac{1}{2}\mu_BH(g_s+g_t)N_4
\end{eqnarray}
where $N_{i},\, i=1,2,3,4$ are the occupation numbers of  state $\st[i]$ and
the parameter $ v_i^{(1)}$ satisfy the Bethe equations (\ref{eq:BE}).

There is no spin singlet state as a result of the different spin and orbit Land\'{e} factors. 
The states $\st[2]$ and $\st[3]$ have nonzero magnetic moment. 
The Ising coupling $J_z$ energetically favours the states $\st[2]$ and $\st[3]$. 
For very strong Ising coupling, these two states may form an effective $su(2)$ gapless phase. 
Following the TBA analysis given in section  \ref{sec:TBA}, for $J_z< 4J\ln 2$, we find that the states $\st[1]$
and $\st[4]$ are involved in the ground state in the absence of a magnetic field.  
In the presence of a  magnetic field, the spin and orbit degrees of freedom
undergo quite different magnetic ordering due to the Ising coupling. 
In the strong coupling regime $J_z \gg 4J\ln 2$, the states $\st[1]$ and
$\st[4]$ are not involved in the ground state.  
When the magnetic field is applied, the spin and orbital degrees of freedom are both in an
antiferromagnetic state but with opposite magnetic moment.  
Thus the ground state problem is reduced to solving the TBA equation
\begin{equation}
\epsilon ^{(1)}(v)=-2\pi J a_1(v)+G_1-\frac{1}{2\pi}\int _{-Q}^{Q}\frac{2\, \epsilon^{(1)-}(k)}{1+(v-k)^2}dk
\label{eq:SOQ1}
\end{equation} 
where the driving term $G_1=(g_s-g_t)\mu_BH$ and $Q$ denotes  the Fermi boundary. 
The density $\rho^{(1)}(v)$ is determined by 
\begin{equation}
\rho^{(1)}(v)=\frac{1}{2\pi}\frac{1}{v^2+\frac{1}{4}}-\frac{1}{2\pi}
\int _{-Q}^{Q}\frac{2\, \rho^{(1)}(k)}{1+(v-k)^2}dk.\label{eq:SOQbethe}
\end{equation}

\begin{figure}
\vskip 5mm
\begin{center}
\includegraphics[width=.60\linewidth]{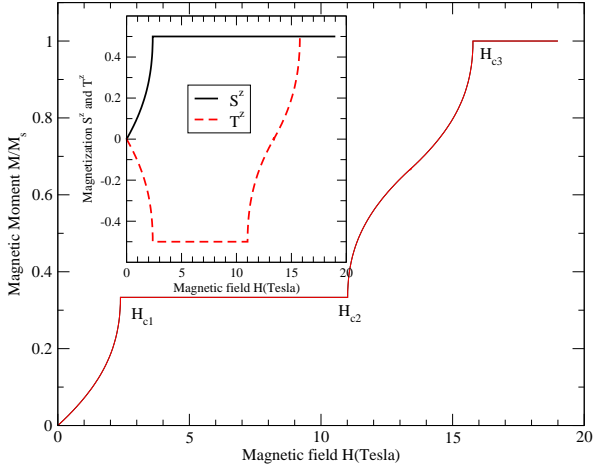}
\caption{
Magnetization versus magnetic field $H$ in units of
saturation magnetization $M_s=\frac{1}{2}(g_s+g_t)\mu_B$ 
for the spin orbital model (\ref{eq:SP-Ham1-1}).  
This figure illustrates the magnetization plateau obtained from the model 
via the TBA with $J_z=18\,$K and $J=0.4\,$K
with $g_s=2$ and $g_t=1$. 
The inset shows the magnetization for the spin and orbital degrees of freedom.
}
\label{fig:SO-Jz}
\end{center}
\end{figure}

If $g_s>g_t$, due to Zeeman splitting, the energy level of state $\st[3]$ becomes higher while state
$\st[2]$ becomes lower as the magnetic field increases. 
As a consequence, the magnetization for the spin degrees increases whereas the orbital magnetization
decreases. 
If the magnetic field is strong, i.e.,  $H>H_{c1}$, with the critical value
\begin{equation}
H_{c1}=\frac{4J}{\mu_B(g_s-g_t)}
\end{equation}
all sites are occupied by state $\st[2]$, with no state $\st[3]$ appearing in the ground state.  
Thus a  magnetization plateau with magnetic moment $M=\frac{1}{2}\mu_B(g_s-g_t)$ opens up.  
The spin and orbital degrees of freedom become fully polarized, but the orbital moment direction  
is antiparallel to the magnetic field. 
Explicitly, the total magnetization is $M=\mu_Bg_sS^z+\mu_Bg_tT^z$, with  
the spin  and orbit magnetizations given by 
\begin{eqnarray}
S^z=\frac{1}{2}-\int_{-Q}^Q\rho^{(1)}(v)dv,\,\,\,\,\,T^z=-\frac{1}{2}+\int_{-Q}^Q\rho^{(1)}(v)dv.
\end{eqnarray}
The spin and orbit magnetization and the total magnetization may be obtained by 
numerically solving (\ref{eq:SOQ1}) and (\ref{eq:SOQbethe}).
The results as presented in figure \ref{fig:SO-Jz}.

On the other hand, the magnetic field can induce the state $\st[1]$
into the ground state. 
The magnetic plateau vanishes at the critical point
\begin{equation}
H_{c2}=\frac{J_z-8J}{2\mu_Bg_t}
\end{equation}
where the phase transition from a ferrimagnetic phase (with both spin and
orbital degrees of freedom fully polarized) into an antiferromagnetic phase occurs.
In this antiferromagnetic phase the spin degrees are fully-polarized while the orbital
degrees of freedom are in an antiferromagnetic state.  
We notice that in the regime $H_{c2} \leq H<\frac{J_z}{2\mu_Bg_t}$ the driving term in
(\ref{eq:SOQ1}) is $G_1=\frac12 {J_z}-\mu_Bg_tH$. 
Correspondingly, the magnetization for the spin and orbital degrees of freedom is given by
$S^z=\frac{1}{2}$ and $T^z=-\frac{1}{2}+\int_{-Q}^Q\rho^{(1)}(v)dv$. 
In the regime
$\frac{J_z}{2\mu_Bg_t} \leq H<H_{c3}$, the driving term is
$G_1=-\frac12{J_z}+\mu_Bg_tH$, thus the magnetization is changed to  
$T^z=\frac{1}{2}-\int_{-Q}^Q\rho^{(1)}(v)dv$ for the orbital degrees of freedom. 
As the magnetic field increases, the state $\st[1]$ occupies more and more sites in
the ground state.  
For strong magnetic fields, i.e., $H>H_{c3}$,
both the spin and orbital degrees of freedom are fully polarized.
The critical field $H_{c3}$ is given by
\begin{equation}
H_{c3}=\frac{J_z+8J}{2\mu_Bg_t}.
\end{equation}
The critical behaviour in the vicinity of the critical points is a Pokrovsky-Talapov-type 
transition as indicated in figure \ref{fig:SO-Jz}.

\subsubsection{Model 2} 
In   the above fundamental basis, the nested algebaic Bethe Ansatz gives the 
energy for  model 2 in the form
\begin{eqnarray}
{\cal E} &=&-J\sum^{M_1}_{i=1}\frac{1}{(v_i^{(1)})^2+\frac{1}{4}}
-\frac{1}{2}\mu_BH(g_s+g_t)N_1\nonumber\\
& &-[\Delta _z+\frac{1}{2}\mu_BH(g_s-g_t)]N_2 +\mu_BH(g_s-g_t)N_3\nonumber\\
& &-[\Delta _z-\frac{1}{2}\mu_BH(g_s-g_t)]N_4.
\end{eqnarray}
Here $N_{i},\, i=1,2,3,4$ are the occupation  numbers of state $\st[i]$ and
the parameters $v_i^{(1)}$ satisfy the Bethe equations (\ref{eq:BE}).

\begin{figure}
\vskip 5mm
\begin{center}
\includegraphics[width=.60\linewidth]{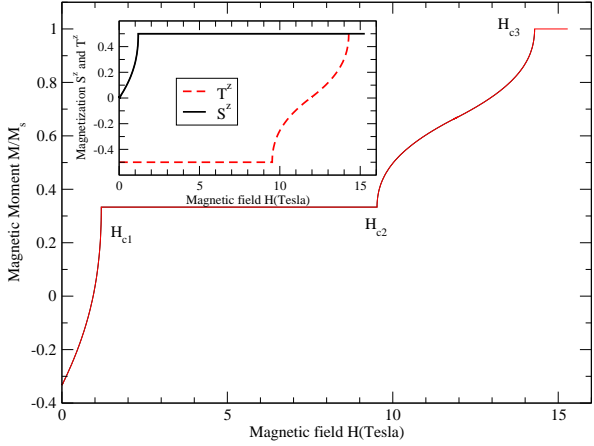}
\caption{
Magnetization versus magnetic field $H$ in units of
saturation magnetization $M_s=\frac{1}{2}(g_s+g_t)\mu_B$ for the spin
orbital model (\ref{eq:SP-Ham1-2}).  
This figure  exhibits the magnetization plateau obtained through the TBA with
$\Delta_z=8\,$K and $J=0.4\,$K with $g_s=2$ and $g_t=1$. 
The inset shows the magnetization for the spin and orbital degrees of freedom.}
\label{fig:SO-delta}
\end{center}
\end{figure}

The orbit splitting $\Delta_z$ energetically favours the states $\st[2]$ and $\st[4]$.  
For a very large orbital splitting, these two states may form an effective $su(2)$ gapless phase. 
Performing the TBA analysis we find that for $\Delta_z>2J\ln 2$ the states $\st[1]$ and
$\st[3]$ are gapful in the absence of a magnetic field.  
Therefore, in this case the ground state problem is reduced to the $su(2)$ TBA equation
(\ref{eq:SOQ1}) with the driving term $G_1=g_s\mu_BH$.  
The density $\rho^{(1)}(v)$ is determined by (\ref{eq:SOQbethe}).  
In this mixed state, the orbital degrees of freedom can be in a fully polarized state,  while the
spin degrees are in an  antiferromagnetic phase. 
Again we assume $g_s>g_t$, then the energy level of state $\st[4]$ becomes higher
and the energy level of state $\st[2]$ becomes lower as the magnetic field increases. 
As a consequence, the magnetization of the spin degrees increases.  
If the magnetic field is larger than the critical value $H_{c1}$,  given by
\begin{equation}
H_{c1}=\frac{4J}{\mu_Bg_s}
\end{equation}
the state $\st[2]$ occupies all sites, while the state $\st[4]$ is not involved in the ground state.  
Here a magnetization plateau with magnetic moment $M=\frac{1}{2}\mu_B(g_s-g_t)$ opens.  
The spin and orbital degrees of freedom are both fully polarized, as seen in the inset of figure \ref{fig:SO-delta}.  
Explicitly, in the region $H<H_{c1}$, the total
magnetization is $M=\mu_Bg_sS^z+\mu_Bg_tT^z$, 
with  the spin and orbit magnetizations  given by
\begin{eqnarray}
S^z=\frac{1}{2}-\int_{-Q}^Q\rho^{(1)}(v)dv,\,\,\,\,\,T^z=-\frac{1}{2}.
\end{eqnarray}
The values of the spin and orbit magnetization and the total magnetization may be again obtained by
numerically solving (\ref{eq:SOQ1}) and (\ref{eq:SOQbethe}), 
the results of which are shown in figure \ref{fig:SO-delta}.

For this model the magnetic plateau vanishes for magnetic fields greater than $H_{c2}$, 
due to the involvement of state $\st[1]$.
This critical field has the value 
\begin{equation}
H_{c2}=\frac{\Delta_z-4J}{\mu_Bg_t}.
\end{equation}
At the further critical point $H_{c2}$ a phase transition from a ferrimagnetic
phase into an antiferromagnetic phase occurs. 
In this phase, the spin state is fully polarized,  whereas the orbit is in an antiferromagnetic state. 
As the magnetic field increases, state $\st[2]$ becomes less involved in the ground state.  
For a strong magnetic field, i.e., $H>H_{c3}$, both the spin and orbit degrees are fully polarized. 
Thus the state $\st[1]$ becomes the  true physical ground state. 
The critical point $H_{c3}$ is given by
\begin{equation}
H_{c2}=\frac{\Delta_z+4J}{\mu_Bg_t}.
\end{equation}
These phase transition points for model  2 (\ref{eq:SP-Ham1-2}) are
clearly indicated in figure \ref{fig:SO-delta}.

We conclude this section by remarking that the thermodynamic
properties of the spin orbital models are easily accessible from the
HTE,  along the lines discussed earlier in section \ref{QTM-NLIE}. 
As yet there appears to be a lack of relevant real materials, 
so we do not discuss this further.

\section{Concluding remarks}
\label{sec:concl}

In this article we have reviewed a systematic approach for investigating the
thermal and magnetic properties of the integrable two-leg spin ladder models. 
The thermodynamic Bethe Ansatz (TBA) and High Temperature Expansion (HTE) methods 
used  to analyse the ground state properties at zero temperature and the thermodynamics at
finite temperatures have been described in detail. 
The TBA approach outlined in section 3 is best suited to the calculation of 
physical properties at zero temperature.
On the other hand, the HTE method outlined in section 4 provides an exact 
expansion for the free energy.
Although the HTE method involves sophisticated mathematical calculations, 
the resulting expansion for the free energy, e.g., equations (\ref{eq:freeenergyHTE}) 
and (\ref{eq:mixEQTM3}), is relatively simple.
In practice it suffices to consider only the first few terms in the expansion and the 
relevant thermal and magnetic properties follow almost trivially as derivatives
to the free energy.
In this way the integrable spin-$\frac{1}{2}$ ladder model defined in section 2.1 
can be used to examine the physics of a 
number of strong coupling ladder compounds which have been extensively 
studied in the literature. 
The two-leg ladder compounds discussed in section 5 are 
(5IAP)$_2$CuBr$_4\cdot 2$H$_2$O \cite{5IAP}, 
Cu$_{2}$(C$_5$H$_{12}$N$_2$)$_2$Cl$_4$
\cite{compound1,Chaboussant1,Chaboussant2}, 
(C$_5$H$_{12}$N)$_2$CuBr$_4$ \cite{Watson}, 
BIP-BNO \cite{BIPa,BIPb} and
[Cu$_2$(C$_2$O$_2$)(C$_{10}$H$_8$N$_2$)$_2$)](NO$_3$)$_2$ \cite{SC2}.
The critical magnetic fields derived from the TBA equations are in good
agreement with the experimental results for the compounds considered. 
The exact results obtained for the susceptibility, the specific heat, the magnetization and  
the entropy obtained from the HTE method provide excellent overall agreement with 
the known experimental data.

In addition to these results, the ground state and thermodynamic properties of the
integrable mixed spin-$(\frac{1}{2},1)$ ladder model defined in section 2.2  
have been investigated via the TBA and HTE in section 6.
The results have been used to examine the magnetic properties of the ferrimagnetic 
mixed spin ladder compound PNNBNO \cite{mixl1,mixl2a,mixl2b}. 
Two integrable spin orbital models based on an underlying 
$su(4)$ symmetry were discussed in section 7.
It was seen that both the single-ion anisotropy and the zero-field orbital splitting 
can trigger a fractional magnetization plateau with respect to different Land\'{e}
factors for the spin and orbital degrees of freedom. 
The phase diagrams for the two spin orbital models were also discussed.

It is evident from these results that integrable models can provide an alternative way 
to investigate the thermal and magnetic properties of a number of low-dimensional 
ladder compounds. 
For arbitrary two-leg spin-$(s,S)$ ladders with strong rung coupling, 
where spins of value $s$ and $S$ reside on different legs, the rung interaction terms
 $J_{\perp}\sum_{j=1}^L\vec{S}_{j}{\cdot}\vec{s}_j$ and
 $\tilde{J}_{\perp}\sum_{j=1}^L(\vec{S}_{j}{\cdot}\vec{s}_j)^2$ may
 overwhelm the spin-spin intrachain interactions and dominate the low
 temperature behaviour. 
Therefore it is not surprising that the physics of the spin-$(s,S)$ ladder compounds 
may be captured by the integrable spin-$(s,S)$ ladder models based on $su(n)$ symmetry, 
where $n=(2s+1)(2S+1)$. 
For instance, the integrable spin-$1$ ladder with sophisticated rung interaction may
 trigger three gapped phases with magnetization plateaux at
 $M/M_s=0,\frac{1}{2},1$, which are also found to exist in spin chains
 \cite{AFF3,Sutherland,Sato}. 
However, the spin-$1$ ladder compound 
BIP-TENO \cite{BIP-TENO1a,BIP-TENO1b,BIP-TENO2,BIP-TENO3} exhibits a
magnetization plateau at $\frac{1}{4}$ which is induced by
frustration in the next neighbour interactions along the legs. 
Obviously,  the integrable spin-$1$ ladder model does not favor the $\frac{1}{4}$
magnetization plateau. 
Nevertheless, it may be used to analyse the fractional magnetization plateau 
for a known spin-$1$ alternating chain compound \cite{FP-Hagiwara}. 
In fact, there are relatively few of the more exotic spin ladder compounds.
The predictive power of this approach has yet to be fully exploited.

\begin{figure}[t]
\begin{center}
\vskip 5mm
\includegraphics[width=0.65\linewidth]{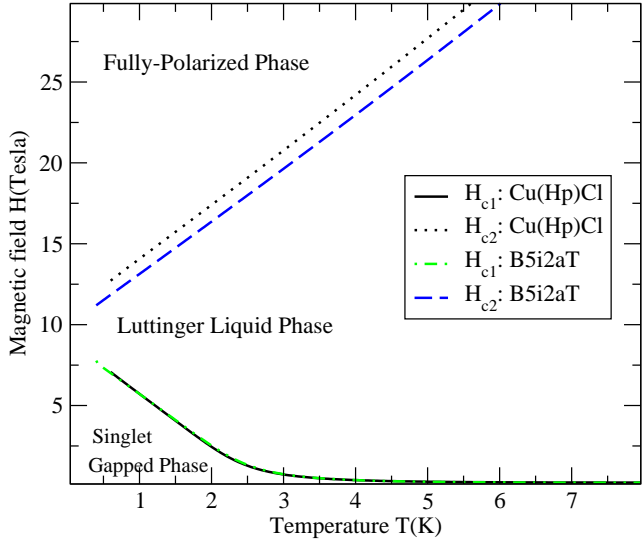}
\caption{
Phase diagrams for the compounds B$5$i$2$aT and Cu(Hp)Cl with spin couplings 
used in sections \ref{sec:5IAP} and \ref{sec:CuHpCl}.
Similar phase diagrams are found using the HTE approach for the other spin
ladder compounds discussed in section \ref{sec:Exp_comp}.  
In general the critical point $H_{c2}$ approaches the critical point $H_{c1}$ as 
$J_{\parallel}$ decreases.  
For $J_{\parallel}=0$ the two critical points merge at zero temperature.}
\label{Fig5}
\end{center}
\end{figure}

Finally, we emphasize that the gapped phase induced by
strong rung interaction in the spin ladder models is the field-induced
singlet ground state which is similar to the phase induced by a large
single-ion anisotropy for the integrable spin chains \cite{BGN,BGNF}. 
The transition is of the universal Pokrovsky-Talapov type.  
The universal nature of the critical fields is seen in the full $H-T$ phase diagram.
For example, the phase diagram for the compounds B$5$i$2$aT and
Cu(Hp)Cl is shown in Fig.~\ref{Fig5}. 
It is anticipated that this phase diagram may be further extended into the low
temperature region using the recently derived nonlinear integral equations for the 
thermodynamics of the $su(4)$ model \cite{DA06}.
In general thermal fluctuations increase the saturation field and 
diminish the energy gap.
The slopes of the critical curves indicate  that  the estimated values of $H_{c1}$ and $H_{c2}$ at 
$T=0$ K coincide with the TBA results given in table 1.
The field-induced gapped phase is significantly different from the Haldane valence bond
solid state or the dimerization in the spin-Peierls transition. 
The former does not exhibit long range order spin-spin correlations, however the latter may.

The HTE scheme also provides a means of calculating correlation functions for integrable
models.
In particular, there are multiple integral formulae \cite{GKS04-3} 
(see also, Refs.~\cite{GKS04-1,GKS04-2,GS05,GHS05,BGKS06}) for 
the density matrix of the one-dimensional Heisenberg model  
in a magnetic field for any finite temperature.
The evaluation of the multiple integral formulae is of interest because 
any correlation functions for the model can be written in terms of the density matrix. 
The HTE method has been applied \cite{TS,ZTp} for
their formulae for short ranged case and proved to be useful to 
evaluate them. 
In the case of zero magnetic field, there is a different way \cite{FK02,F03} 
to calculate the HTE of correlation functions. 
It will be worthwhile to explore correlation functions of the integrable ladder models.

A particularly exciting development for quantum science in general is that fundamental  
interacting spin systems, traditionally realised in the context of condensed-matter based systems
like those discussed in this article, may also be realised by trapping atoms in optical lattices.
This atom-optics based setting \cite{Bloch} offers a potentially cleaner environment to 
mimic and study condensed matter physics phenomena.
It offers, for example, a way to construct a complete toolbox for realising various kinds of
spin models \cite{toolbox,beyond}. 
It also provides a way of realising low-dimensional quantum systems, for which 
the spatial confinement leads to enhanced dynamics and correlations.
Integrable models, and the methods reviewed in this article for calculating their
physical properties, will clearly continue to have a bearing on these developments.
This is because they treat genuine interacting quantum many-body systems for which
perturbative approaches and mean-field theories often fail.
The methods discussed here may be generalised to spin systems with open boundary 
conditions or impurities \cite{boundary}.
This is particularly important for small-sized systems in which quantum fluctuations may be 
strong enough that boundary effects are no longer negligible at low temperature.  
In general obtaining an exact systematic low-temperature expansion for integrable models in analogy 
with the HTE remains an outstanding open problem.
So too does the development of exact expansion methods for the integrable Bose and Fermi gases.


\section*{Acknowledgements}

This work has been supported by the Australian Research Council through
the Discovery and Linkage International programs and the Japanese Society for the
Promotion of Science (JSPS) Bilateral Joint Project 
``Solvable models and their thermodynamics in statistical mechanics and field theory".
Z. Tsuboi was partially supported by a JSPS Grant-in-Aid for Scientific Research (no. 16914018). 
N. Oelkers has also been supported by DAAD during the early stages of this work.
We thank A. Foerster, M. Orend\'{a}\v{c}, M. Shiroishi, Z.-J. Ying and H.-Q. Zhou for helpful discussions.
The authors thank M. Takahashi and M. Shiroishi for their kind hospitality at the 
Institute for Solid State Physics at the University of Tokyo where part of this work was undertaken.




\newcommand{\JMP}{{J. Math. Phys.} }
\newcommand{\PRL}{{Phys. Rev. Lett.} }
\newcommand{\APNY}{{Ann. Phys. N.Y.\/} }

\clearpage

\appendix
\section{$su(4)$ HTE coefficients}
\label{appendix1}

The first five coefficients in the HTE for the 
free energy (\ref{eq:freeenergyHTE}) of the integrable spin ladder (\ref{eq:Ham2}) 
are explicitly given by
\begin{eqnarray}
 c^{(1)}_{1,0} &=& \frac{2Q^{(2)}_{1}}{{Q^{(1)}_{1}}^2}  \nonumber\\
c^{(1)}_{2,0} &=& \frac{3 Q^{(2)}_{1}}{{Q^{(1)}_{1}}^2} - 
\frac{6 {Q^{(2)}_{1}}^2}{{Q^{(1)}_{1}}^4} 
+ \frac{3Q^{(3)}_{1}}{{Q^{(1)}_{1}}^3}  \nonumber \\
 c^{(1)}_{3,0}&=&\frac{10 Q^{(2)}_{1}}{3 {Q^{(1)}_{1}}^2} - 
                \frac{18 {Q^{(2)}_{1}}^2}{{Q^{(1)}_{1}}^4} + 
                \frac{80 {Q^{(2)}_{1}}^3}{3 {Q^{(1)}_{1}}^6} + 
                \frac{8 Q^{(3)}_{1}}{{Q^{(1)}_{1}}^3} - \frac{24 
                Q^{(2)}_{1} Q^{(3)}_{1}}{{Q^{(1)}_{1}}^5} + \frac{4 
                Q^{(4)}_{1}}{{Q^{(1)}_{1}}^4}  \nonumber \\
 c^{(1)}_{4,0} &=& \frac{35 
                Q^{(2)}_{1}}{12 {Q^{(1)}_{1}}^2} - 
                \frac{205 {Q^{(2)}_{1}}^2}{6 {Q^{(1)}_{1}}^4} + 
                \frac{120 {Q^{(2)}_{1}}^3}{{Q^{(1)}_{1}}^6} - 
                \frac{140 {Q^{(2)}_{1}}^4}{{Q^{(1)}_{1}}^8} + \frac{55 
                Q^{(3)}_{1}}{4 {Q^{(1)}_{1}}^3} - \frac{100 Q^{(2)}_{1} 
                Q^{(3)}_{1}}{{Q^{(1)}_{1}}^5} \nonumber \\
      && + \frac{180 {Q^{(2)}_{1}}^2 
                Q^{(3)}_{1}}{{Q^{(1)}_{1}}^7} - 
                \frac{45 {Q^{(3)}_{1}}^2}{2 {Q^{(1)}_{1}}^6} + \frac{15 
                Q^{(4)}_{1}}{{Q^{(1)}_{1}}^4} - \frac{40 Q^{(2)}_{1} 
                Q^{(4)}_{1}}{{Q^{(1)}_{1}}^6}  \nonumber \\
 c^{(1)}_{5,0}&=& \frac{21 
                Q^{(2)}_{1}}{10 {Q^{(1)}_{1}}^2} - 
                \frac{50 {Q^{(2)}_{1}}^2}{{Q^{(1)}_{1}}^4} +
                 \frac{320 {Q^{(2)}_{1}}^3}{{Q^{(1)}_{1}}^6} 
                 - \frac{840 {Q^{(2)}_{1}}^4}{{Q^{(1)}_{1}}^8} +
                  \frac{4032 {Q^{(2)}_{1}}^5}{5 {Q^{(1)}_{1}}^{10}} 
                  + \frac{37 
                Q^{(3)}_{1}}{2 {Q^{(1)}_{1}}^3}\nonumber \\
               &&  - \frac{249 Q^{(2)}_{1} 
                Q^{(3)}_{1}}{{Q^{(1)}_{1}}^5} + \frac{1020 {Q^{(2)}_{1}}^2 
                Q^{(3)}_{1}}{{Q^{(1)}_{1}}^7} - \frac{1344 {Q^{(2)}_{1}}^3 
                Q^{(3)}_{1}}{{Q^{(1)}_{1}}^9} - 
                \frac{120 {Q^{(3)}_{1}}^2}{{Q^{(1)}_{1}}^6}\nonumber \\
               && + \frac{378 
                Q^{(2)}_{1} {Q^{(3)}_{1}}^2}{{Q^{(1)}_{1}}^8} + \frac{34 
                Q^{(4)}_{1}}{{Q^{(1)}_{1}}^4} - \frac{210 Q^{(2)}_{1} 
                Q^{(4)}_{1}}{{Q^{(1)}_{1}}^6} + \frac{336 {Q^{(2)}_{1}}^2 
                Q^{(4)}_{1}}{{Q^{(1)}_{1}}^8} - \frac{72 Q^{(3)}_{1} 
                Q^{(4)}_{1}}{{Q^{(1)}_{1}}^7}  \nonumber. 
\end{eqnarray}
The $Q$ functions are defined in (\ref{eq:Q-sl4}).


\section{$su(6)$ HTE coefficients}
\label{appendix2}

The first three coefficients in the HTE for the 
free energy (\ref{eq:mixEQTM3}) of the integrable mixed spin ladder (\ref{eq:Ham3})
are explicitly given by
\begin{eqnarray}
c_{1,0}^{(1)}&=&2 \frac{Q^{(2)}_1}{{Q^{(1)}_1}^2} \nonumber\\
c_{2,0}^{(1)}&=&3 \frac{Q^{(2)}_1}{{Q^{(1)}_1}^2}- 6 \frac{{Q^{(2)}_1}^2}{{Q^{(1)}_1}^4} 
+3 \frac{{Q^{(3)}_1}}{{Q^{(1)}_1}^3} \nonumber\\
c_{3,0}^{(1)}&=& \frac{10}{3}\frac{Q^{(2)}_1}{{Q^{(1)}_1}^2}
-18\frac{{Q^{(2)}_1}^2}{{Q^{(1)}_1}^4}	 
+\frac{80}{3}\frac{{Q^{(2)}_1}^3}{{Q^{(1)}_1}^6}
+8\frac{{Q^{(3)}_1}}{{Q^{(1)}_1}^3}
-24\frac{{Q^{(2)}_1}{Q^{(3)}_1}}{{Q^{(1)}_1}^5}+ 4 \frac{Q^{(4)}_1}{{Q^{(1)}_1}^4}.
\nonumber 
\end{eqnarray}
In turn the characters of the antisymmetric representations of $su(6)$ are 
explicitly written in terms of the chemical potentials (\ref{eq:mixchem}) as
\begin{eqnarray}
Q_1^{(1)} &=& {\rm e}^{-\beta \mu_1}+{\rm e}^{-\beta \mu_2}+{\rm e}^{-\beta \mu_3}
+{\rm e}^{-\beta \mu_4}+{\rm e}^{-\beta \mu_5}+{\rm e}^{-\beta \mu_6} \nonumber \\
Q_1^{(2)} &=& {\rm e}^{-\beta(\mu_1+\mu_2)}+{\rm e}^{-\beta(\mu_1+\mu_3)}+{\rm e}^{-\beta(\mu_1+\mu_4)}
+{\rm e}^{-\beta(\mu_1+\mu_5)}+{\rm e}^{-\beta(\mu_1+\mu_6)}\nonumber \\
&& + {\rm e}^{-\beta(\mu_2+\mu_3)} + {\rm e}^{-\beta(\mu_2+\mu_4)}+{\rm e}^{-\beta(\mu_2+\mu_5)}
+{\rm e}^{-\beta(\mu_2+\mu_6)}+{\rm e}^{-\beta(\mu_3+\mu_4)}\nonumber \\
&& + {\rm e}^{-\beta(\mu_3+\mu_5)} + {\rm e}^{-\beta(\mu_3+\mu_6)}+{\rm e}^{-\beta(\mu_4+\mu_5)}
+{\rm e}^{-\beta(\mu_4+\mu_6)}+{\rm e}^{-\beta(\mu_5+\mu_6)}
\nonumber \\
Q_1^{(3)} &=& {\rm e}^{-\beta(\mu_1+\mu_2+\mu_3)}+ {\rm e}^{-\beta(\mu_1+\mu_2+\mu_4)}+
 {\rm e}^{-\beta(\mu_1+\mu_2+\mu_5)}+ {\rm e}^{-\beta(\mu_1+\mu_2+\mu_6)}\nonumber \\
 && +{\rm e}^{-\beta(\mu_2+\mu_3+\mu_4)}+ {\rm e}^{-\beta(\mu_2+\mu_3+\mu_5)}+
 {\rm e}^{-\beta(\mu_2+\mu_3+\mu_6)}+ {\rm e}^{-\beta(\mu_2+\mu_4+\mu_5)}\nonumber \\
 && +{\rm e}^{-\beta(\mu_2+\mu_4+\mu_6)}+ {\rm e}^{-\beta(\mu_2+\mu_5+\mu_6)}+
 {\rm e}^{-\beta(\mu_3+\mu_4+\mu_5)}+ {\rm e}^{-\beta(\mu_3+\mu_4+\mu_6)}\nonumber \\
&& +{\rm e}^{-\beta(\mu_3+\mu_5+\mu_6)}+ {\rm e}^{-\beta(\mu_4+\mu_5+\mu_6)} \nonumber \\
Q_1^{(4)} &=&{\rm e}^{-\beta(\mu_1+\mu_2+\mu_3+\mu_4)}+{\rm e}^{-\beta(\mu_1+\mu_2+\mu_3+\mu_5)}
+{\rm e}^{-\beta(\mu_1+\mu_2+\mu_3+\mu_6)}\nonumber \\
&&+{\rm e}^{-\beta(\mu_1+\mu_2+\mu_4+\mu_5)}+{\rm e}^{-\beta(\mu_1+\mu_2+\mu_4+\mu_6)}
+{\rm e}^{-\beta(\mu_1+\mu_2+\mu_5+\mu_6)}\nonumber \\
&&+{\rm e}^{-\beta(\mu_1+\mu_3+\mu_4+\mu_5)}+{\rm e}^{-\beta(\mu_1+\mu_3+\mu_4+\mu_6)}
+{\rm e}^{-\beta(\mu_1+\mu_3+\mu_5+\mu_6)}\nonumber \\
&&+{\rm e}^{-\beta(\mu_1+\mu_4+\mu_5+\mu_6)}+{\rm e}^{-\beta(\mu_2+\mu_3+\mu_4+\mu_5)}
+{\rm e}^{-\beta(\mu_2+\mu_3+\mu_4+\mu_6)}\nonumber \\
&&+{\rm e}^{-\beta(\mu_2+\mu_3+\mu_5+\mu_6)}+{\rm e}^{-\beta(\mu_2+\mu_4+\mu_5+\mu_6)}
+{\rm e}^{-\beta(\mu_3+\mu_4+\mu_5+\mu_6)} \nonumber \\
Q_1^{(5)} &=&{\rm e}^{-\beta (\mu_1+\mu_2+\mu_3+\mu_4+\mu_5)}+
{\rm e}^{-\beta (\mu_1+\mu_2+\mu_3+\mu_4+\mu_6)}
+{\rm e}^{-\beta (\mu_2+\mu_3+\mu_4+\mu_5+\mu_6)} \nonumber\\
Q_1^{(6)} &=&{\rm e}^{-\beta (\mu_1+\mu_2+\mu_3+\mu_4+\mu_5+\mu_6)}
={\rm e}^{{3J_{\perp}}/{T}}.\nonumber
\end{eqnarray}

%
\section{Location of Bethe roots and zeros of the largest QTM eigenvalue}
\label{appendfig}

In this Appendix the location of Bethe roots and zeros of the largest QTM eigenvalue
are illustrated for finite systems.

\begin{figure}[h]
\vskip 10mm
\begin{minipage}{\linewidth}
\includegraphics[width=1.\linewidth]{roots_n_zeroes_no_chem_pot_2.eps}
\end{minipage}
\caption{
(left) Location of roots $v_k^{(a)} (a = 1, 2, 3)$ of the $su(4)$ 
Bethe equations (\ref{eq:qtm_BAE4}) in the complex plane. 
The parameters are $N=12$, $M_1=M_2=M_3=N/2=6$ and $u_N = -0.05$, 
with no chemical potentials ($\mu_i=0$ for $i=1,2,3,4$).
Symbols \color{black} $\bigcirc$, $\color{red} \Box$, $\color{blue} \triangle$
denote colours 1, 2, 3, respectively. 
Note that the roots of each colour form 6 one-strings.
(right) Location of the zeros of the corresponding 
largest QTM eigenvalue $\widetilde{T}^{(a)}_{1}(v)$ 
($a = 1, 2, 3$) in the complex plane. 
The symbols $\color{black} \bigcirc$, $\color{red} \Box$, $\color{blue} \triangle$
denote colours 1, 2, 3, respectively.
Note the `physical strip' around the real axis which is free of zeros.
}
\label{fig:extra1}
\end{figure}

\begin{figure}[t]
\vskip 5mm
\begin{minipage}{\linewidth}
\includegraphics[width=1.\linewidth]{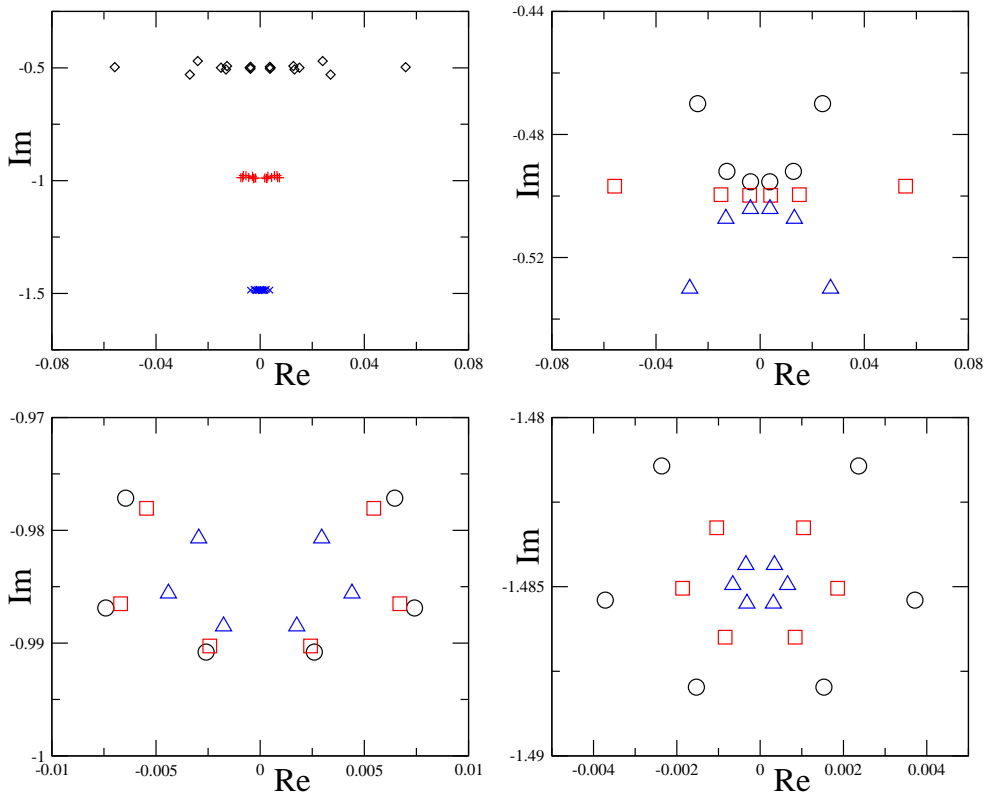}
\end{minipage}
\caption{
(top left) Location of roots $v_k^{(a)} (a = 1, 2, 3)$ of the $su(4)$ Bethe equations 
(\ref{eq:qtm_BAE4}) in the complex plane corresponding
to the largest eigenvalue of the QTM for magnetic fields $H = 5\,$T, $9\,$T and $13\,$T. 
Symbols $\color{black} \diamond$, $\color{red} +$, $\color{blue} \times$
denote the roots of colour 1, 2, 3, respectively.
The parameters are $N=12$ and $T = 1.59\,$K, with chemical potentials given by
equation (\ref{eq:chem}), i.e., 
$\mu_1 = -13.3\,$K, $\mu_2 = -1.4112\,H$, $\mu_3 = 0$ and $\mu_4=-\mu_2$.
These values correspond to the compound $\mathrm{(5IAP)_{2}CuBr_{4}\cdot 2H_{2}O}$ 
(see figure \ref{fig:B5i2aTSZ}) with $J_\perp=13.3\,$K.
The remaining figures are more detailed versions of this figure.
(top right) Location of the Bethe roots $v_k^{(1)}$.
Symbols $\color{black} \bigcirc$, $\color{red} \Box $, $\color{blue} \triangle$
denote magnetic fields of $H = 5\,$T, $9\,$T and $13\,$T, respectively.
(bottom left) Location of the Bethe roots $v_k^{(2)}$.
(bottom right) Location of the Bethe roots $v_k^{(3)}$.
}
\label{fig:extra2}
\end{figure}

\begin{figure}[tp]
\vskip 5mm
        \begin{minipage}{\linewidth}
        \includegraphics[width=1.\linewidth]{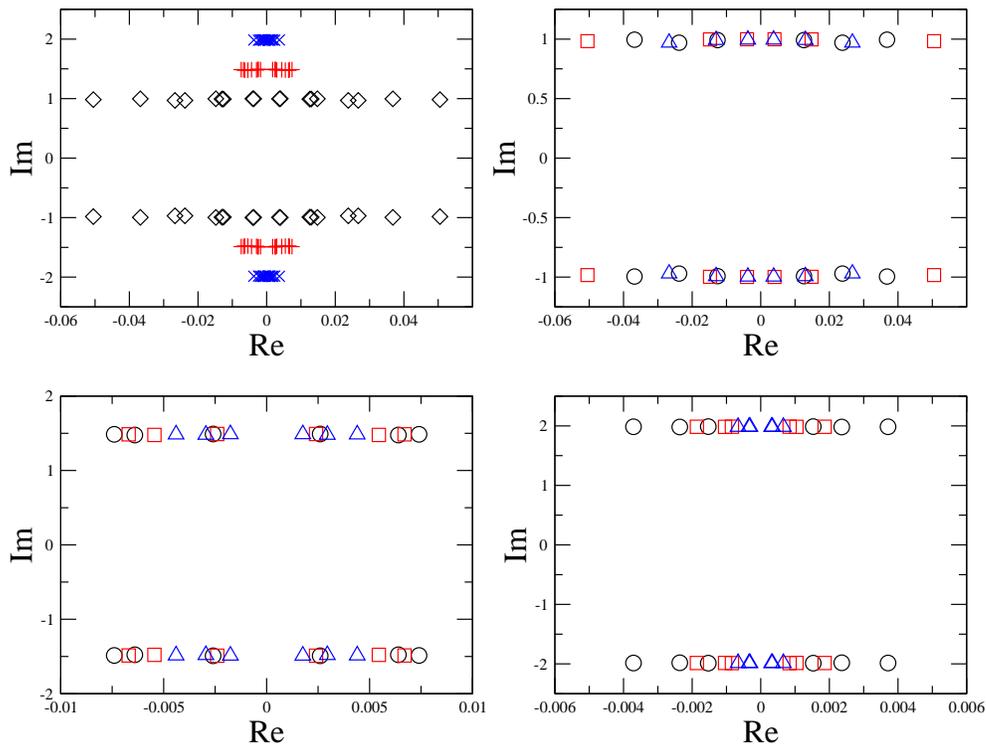}
        \end{minipage}
\caption{
(top left) Location of the zeros of the largest QTM eigenvalue $\widetilde{T}^{(a)}_{1}(v)$ 
($a = 1, 2, 3$) in the complex plane corresponding to the Bethe roots in  figure \ref{fig:extra2} 
for the same numerical parameters. 
This figure is an overlay of the other three figures and corresponds to the top left of
 figure \ref{fig:extra2}.
Symbols $\color{black} \diamond$, $\color{red} +$, $\color{blue} \times$
denote the zeros for colours 1, 2, 3, respectively.
Note that the `physical strip' around the real axis is free of zeros  
for each of the magnetic fields, in accord with Conjecture 4.1
The remaining panels show the location of zeros of $\widetilde{T}^{(a)}_{1}(v)$ for the 
corresponding Bethe root distributions in  figure \ref{fig:extra2}.
Symbols $\color{black} \bigcirc$, $\color{red} \Box $, $\color{blue} \triangle$
denote magnetic fields $H=5\,$T, $9\,$T and $13\,$T, respectively.
(top right) Colour $a=1$. (bottom left) Colour $a=2$. (bottom right) Colour $a=3$.
}
\label{fig:extra3}
\end{figure}

\end{document}